\documentclass[twocolumn,amsmath,amssymb,nofootinbib,eqsecnum,tighten,floatfix]{revtex4}

\usepackage{graphicx}
\usepackage{dcolumn} 
\usepackage{bm, bbm}      
\usepackage{curves}
\usepackage{epic}
\usepackage{wasysym}
\usepackage{pslatex} 
\usepackage{latexsym}
\usepackage[mathcal]{euscript}
\usepackage{epsf,times}
\usepackage{subfigure}

\newcommand\A{\mathcal{A}}
\newcommand\B{\mathcal{B}}

\newcommand\N{\mathcal{N}}

\newcommand\T{\mathcal{T}}

\def\openone{\leavevmode\hbox{\small1\kern-4.2pt\normalsize1}}

\newcommand{\beq}{\begin{equation}}
\newcommand{\eeq}{\end{equation}}
\newcommand{\bea}{\begin{eqnarray}}
\newcommand{\eea}{\end{eqnarray}}

\begin{document}


\title{
Topological order in a three-dimensional toric code at finite temperature
}

\author{
Claudio Castelnovo$^1$, 
and
Claudio Chamon$^2$
       }
\affiliation{
$^1$ 
Rudolf Peierls Centre for Theoretical Physics, 
University of Oxford, Oxford, OX1 3NP, UK
\\ 
$^2$ Physics Department, Boston University, Boston, MA 02215, USA, 
            }

\date{\today}

\begin{abstract}
We study topological order in a toric code in three spatial
dimensions, or a 3+1D $\mathbb{Z}_2$ gauge theory, at finite
temperature. We compute exactly the topological entropy of the
system, and show that it drops, for any infinitesimal temperature, to
half its value at zero temperature. The remaining half of the entropy
stays constant up to a critical temperature $T_c$, dropping to zero
above $T_c$. These results show that topologically ordered phases
exist at finite temperatures, and we give a simple interpretation of
the order in terms of fluctuating strings and membranes, and how
thermally induced point defects affect these extended
structures. Finally, we discuss the nature of the topological order
at finite temperature, and its quantum and classical aspects.
\end{abstract}

\maketitle
%
%

\section{\label{sec: Introduction}
Introduction
        }
Some quantum systems are characterized by a type of order which cannot
be captured by a local order parameter that signals broken symmetries,
but instead the order is topological in nature.~\cite{topo refs} 
One of the ways in which this topological order manifests itself is in a 
ground state degeneracy that cannot be lifted by any local perturbation, and 
that depends on the genus of the surface in which the system is
defined. Recently, there have been efforts to find characterizations
of topological order other than ground state degeneracies, in
particular exploring the entanglement in the ground state
wavefunction.~\cite{Levin2006,Kitaev2006}

At zero temperature, topological order can be detected using the von
Neumann entanglement entropy, more precisely a topological
contribution to it that can be separated from the boundary
contribution by appropriate subtractions of different bipartitions of
the system.~\cite{Levin2006,Kitaev2006} Because the pure state density
matrix is constructed from the ground state, it was argued
in Ref.~\onlinecite{Levin2006} that topological order is a property of the
wavefunction, and not of the Hamiltonian, at absolute zero temperature. 

An interesting question is what happens with topological order at finite
temperature. The question is relevant because thermal fluctuations, no
matter how small, are present in any laboratory system. To address
this issue, it was proposed in Ref.~\cite{Castelnovo2007} to use the
topological entropy as a probe of topological order, but to compute it
using an equilibrium mixed state density matrix
$\hat\rho={Z^{-1}}e^{-\beta \hat H}$. It becomes clear that, as
opposed to zero temperature for which one can do away with the full
information contained in the Hamiltonian and just use the ground state
wavefunction, topological order, if present at finite temperature,
must be a property of the Hamiltonian.

The topological entropy was computed exactly for the 2D Kitaev
model~\cite{Kitaev2003} at finite temperature $T$, and it was shown
that the infinite system size limit and the $T\to 0$ limit do not
commute, and that at finite $T$ the topological entropy vanishes in
the thermodynamic limit. Thus, it was argued that the topological
order in the 2D system was 
fragile.~\cite{Dennis2002,Nussinov2006,Castelnovo2007} 

Here we show that the situation in 3D is rather different, using 
the 3D version of Kitaev's model as an example.~\cite{Hamma2005a} 
In contrast to 2D, topological order
survives up to a phase transition at a finite temperature $T_c$. 
The order can be probed through a non-vanishing topological
entropy, as well as understood from a simple cartoon picture that we
present in the paper, using the fact that in 3D strings can move
around point defects (as opposed to 2D).

We prove in this paper that the von Neumann entropy of a subsystem 
$\A$ of a ${\mathbb Z}_2$ gauge model such as Kitaev's toric code, in any 
number of dimensions, can be always decomposed into two additive 
contributions from each of the two gauge structures 
(magnetic and electric):
~\cite{footnote:separability} 
\bea
S^{\ }_{\textrm{VN}}(\A;T) 
&=& 
S^{(S)}_{\textrm{VN}}(\A ; T/\lambda^{\ }_{A}) 
+ 
S^{(P)}_{\textrm{VN}}(\A ; T/\lambda^{\ }_{B}) 
,
\eea
where $S^{(S)}_{\textrm{VN}}$ and $S^{(P)}_{\textrm{VN}}$ are the
separable contributions from the stars and plaquettes of the
model, and $\lambda_A$ and $\lambda_B$ the associated coupling
constants for these two structures. 
Consequently, the same additive separability holds for the
topological entropy, which is a sum of two independent contributions:
\bea
S^{\ }_{\textrm{topo}}(T) 
&=& 
S^{(S)}_{\textrm{topo}}(T/\lambda^{\ }_{A}) 
+
S^{(P)}_{\textrm{topo}}(T/\lambda^{\ }_{B}) 
.
\eea
One of the contributions,
$S^{(S)}_{\textrm{topo}}$, evaporates for any infinitesimal temperature in the
thermodynamic limit, just as in 2D, but the other one,
$S^{(P)}_{\textrm{topo}}$, remains constant up to a finite temperature phase 
transition at $T_c = 1.313346(3) \lambda_B$, that occurs for the 3D case: 
\begin{equation}
S^{\rm 3D}_{\textrm{topo}}(T) =
\begin{cases}
2\ln 2 & \text{$T=0$} \\
\ln 2 & \text{$0< T < T_c$} \\
0 & \text{$T > T_c$} 
.
\end{cases}
\end{equation}

As a consequence of these results, we argue that topological order can
be well defined at finite temperatures in 3D.~\cite{footnote:Nussinov2008}
This finding raises the following interesting question: is the finite $T$
order classical or quantum?  Perhaps another way to ask the question
is the following: Which kind of information can be robustly stored
using the isolated topological sectors in phase space that cannot be
connected by local moves ($2^3$ such states in 3D): classical (bits)
or quantum (qubits) information?  While we cannot argue that the
system does not realize a full quantum memory, we can at the least
argue that it can store probabilistic information (pbits --
probabilistic bits~\cite{pbit refs}) in the form of a quantum
superposition of states in the different topological sectors, where
the square amplitudes for all states in a given sector (a probability)
does not fluctuate in the thermodynamic limit if the coupling to a
thermal bath is local. However, the relative phases for all these
amplitudes could be scrambled. This weak type of quantum superposition
is not discernible from a classical probability distribution.

Finally, this example shows that the notion of classical topological
order, suggested for hard constrained models in
2D,~\cite{Castelnovo2006} is well defined in 3D without resorting to
any hard constraints.
%
%

\section{\label{sec: the model}
The model
        }
Consider a three-dimensional version of Kitaev's toric code,~\cite{Hamma2005a}
defined on 
a simple cubic lattice of size $N = L \times L \times L$, with 
periodic boundary conditions and spin-$1/2$ degrees of freedom 
$\vec{\sigma}^{\ }_{i}$ living on the bonds, $i = 1,\ldots,3N$ 
($\sigma^{x}_{i}$, $\sigma^{\textrm{y}}_{i}$ and 
$\sigma^{z}_{i}$ being the three Pauli matrices). 
Let us label the centers of each single square plaquette in the lattice 
with $p = 1,\ldots,3N$, and each site of the cubic lattice with 
$s = 1,\ldots,N$. 

\begin{figure}[ht]
\vspace{0.4 cm}
\begin{center}
\includegraphics[width=0.7\columnwidth]{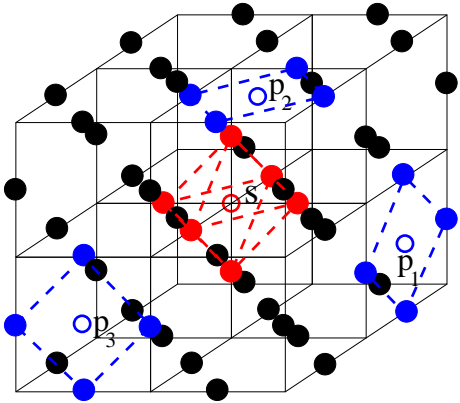}
\end{center}
\caption{
\label{fig: Kitaev 3D}
(Color online) -- 
Illustration of the Kitaev model in 3D, with explicit examples of a star 
operator $A^{\ }_{s} = \prod \sigma^{x}_{i}$ at the 
lattice site $s$, and of three plaquette operators 
$B^{\ }_{p} = \prod \sigma^{z}_{i}$ at the plaquette-dual sites 
$p_1$, $p_2$ and $p_3$. 
The $\sigma$ spin index $i$ labels respectively the $6$ (red) spins around 
$s$ and the $4$ (blue) spins around $p$ (connected by dashed lines). 
}
\end{figure}

Let us define the plaquette and star operators on the lattice 
\beq
B^{\ }_{p} = \prod^{\ }_{i\in p} \sigma^{z}_{i} 
\qquad \qquad 
A^{\ }_{s} = \prod^{\ }_{i\in s} \sigma^{x}_{i} 
\label{eq: B,A ops}
\eeq
as illustrated in Fig.~\ref{fig: Kitaev 3D}. 
The Hamiltonian of the model can then be written in terms of these operators 
as 
\beq
H = 
- \lambda^{\ }_{A} \sum^{\ }_{s} A^{\ }_{s} 
- \lambda^{\ }_{B} \sum^{\ }_{p} B^{\ }_{p} 
\label{eq: Kitaev Ham}
\eeq
where $\lambda^{\ }_{A}$ and $\lambda^{\ }_{B}$ are two real, positive 
constants. 

Notice that all star and plaquette operators commute, but they are not all 
independent. While only the product of all star operators equals 
the identity, therefore leaving $N-1$ independent star operators, the 
product of the plaquette operators around each cubic unit cell gives the 
identity, therefore introducing $N-1$ constraints in the $3N$ total 
plaquette operators (the product of all but one cube is 
equivalent to that same cube, so we have one less constraint). 
Moreover, three additional constraints come from the fact that the 
product of all plaquette operators along any crystal plane in the cubic 
lattice (i.e., $\langle x,y \rangle$, $\langle x,z \rangle$, or 
$\langle y,z \rangle$) yields the identity, and we are finally left with 
$2N-2$ independent plaquette operators. 

\begin{figure}[ht]
\vspace{0.4 cm}
\begin{center}
\includegraphics[width=0.7\columnwidth]{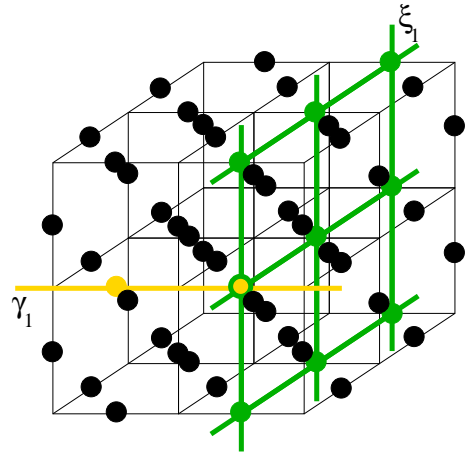}
\end{center}
\caption{
\label{fig: non local ops}
(Color online) -- 
Two examples of the non-local operators needed to distinguish between 
the degenerate GS of the 3D Kitaev model. 
}
\end{figure}

The ground state (GS) manifold of the system is identified by 
having all plaquette and star quantum numbers equal to $+1$, and it is 
$2^{3N - (N-1) - (2N-2)}_{\ } = 2^{3}_{\ }$ dimensional, assuming periodic 
boundary conditions in all three directions. 
Similarly to the 2D case, one can notice that this degeneracy has a 
topological nature, and the different sectors are distinguished by 
three non-local operators 
\bea
\Gamma^{\ }_{1} 
&=& 
\prod^{\ }_{i \in \gamma^{\ }_{1}} \sigma^{z}_{i}
\qquad 
\Gamma^{\ }_{2} 
= 
\prod^{\ }_{i \in \gamma^{\ }_{2}} \sigma^{z}_{i}
\qquad 
\Gamma^{\ }_{3} 
= 
\prod^{\ }_{i \in \gamma^{\ }_{3}} \sigma^{z}_{i}
\label{eq: topo ops z}
\eea
or 
\bea
\Xi^{\ }_{1} 
&=& 
\prod^{\ }_{i \in \xi^{\ }_{1}} \sigma^{x}_{i}
\qquad 
\Xi^{\ }_{2} 
= 
\prod^{\ }_{i \in \xi^{\ }_{2}} \sigma^{x}_{i}
\qquad 
\Xi^{\ }_{3} 
= 
\prod^{\ }_{i \in \xi^{\ }_{3}} \sigma^{x}_{i}
\label{eq: topo ops x}
\eea
that are diagonal in the $\sigma^{z}_{\ }$ and $\sigma^{x}_{\ }$ basis, 
respectively. Here the $\gamma^{\ }_{i}$ can be any winding paths along 
the edges of the cubic lattice in each of the three crystal directions 
($x$, $y$, or $z$), 
and the $\xi^{\ }_{i}$ can be any winding planes perpendicular to each of 
the crystal directions and passing through the midpoints of the corresponding 
edges of the cubic lattice (i.e., crystal planes in the dual lattice whose 
sites sit at the centers of the elementary cubic cells). 
Two examples are shown in Fig.~\ref{fig: non local ops} for clarity. 

In the $\sigma^{z}_{\ }$ basis and in the topological sector where 
all the $\Gamma^{\ }_{i}$ equal $+1$, 
the GS wavefunction of the system can be written as 
\beq
\vert GS \rangle 
= 
\frac{1}{\vert G \vert^{1/2}_{\ }} 
  \sum^{\ }_{g \in G} 
    g \vert 0 \rangle, 
\label{eq: Kitaev GS}
\eeq
where $\vert 0 \rangle$ is any state in the sector, say the state with all the 
$\sigma^{z}_{i} = +1$, and $G$ is the Abelian group generated by 
all products of star operators (of dimension $\vert G \vert = 2^{N-1}_{\ }$). 
In the $\sigma^{x}_{\ }$ basis and in the topological sector where 
all the $\Xi^{\ }_{i}$ equal $+1$, 
the GS wavefunction of the system can be written as in 
Eq.~(\ref{eq: Kitaev GS}), where now 
$\vert 0 \rangle$ is any state in the sector, say the state with all the 
$\sigma^{x}_{i} = +1$, and $G$ is the Abelian group generated by 
all products of plaquette operators (of dimension $\vert G \vert = 2^{2(N-1)}_{\ }$). 

\begin{figure}[ht]
\vspace{0.4 cm}
\begin{center}
\includegraphics[width=0.7\columnwidth]{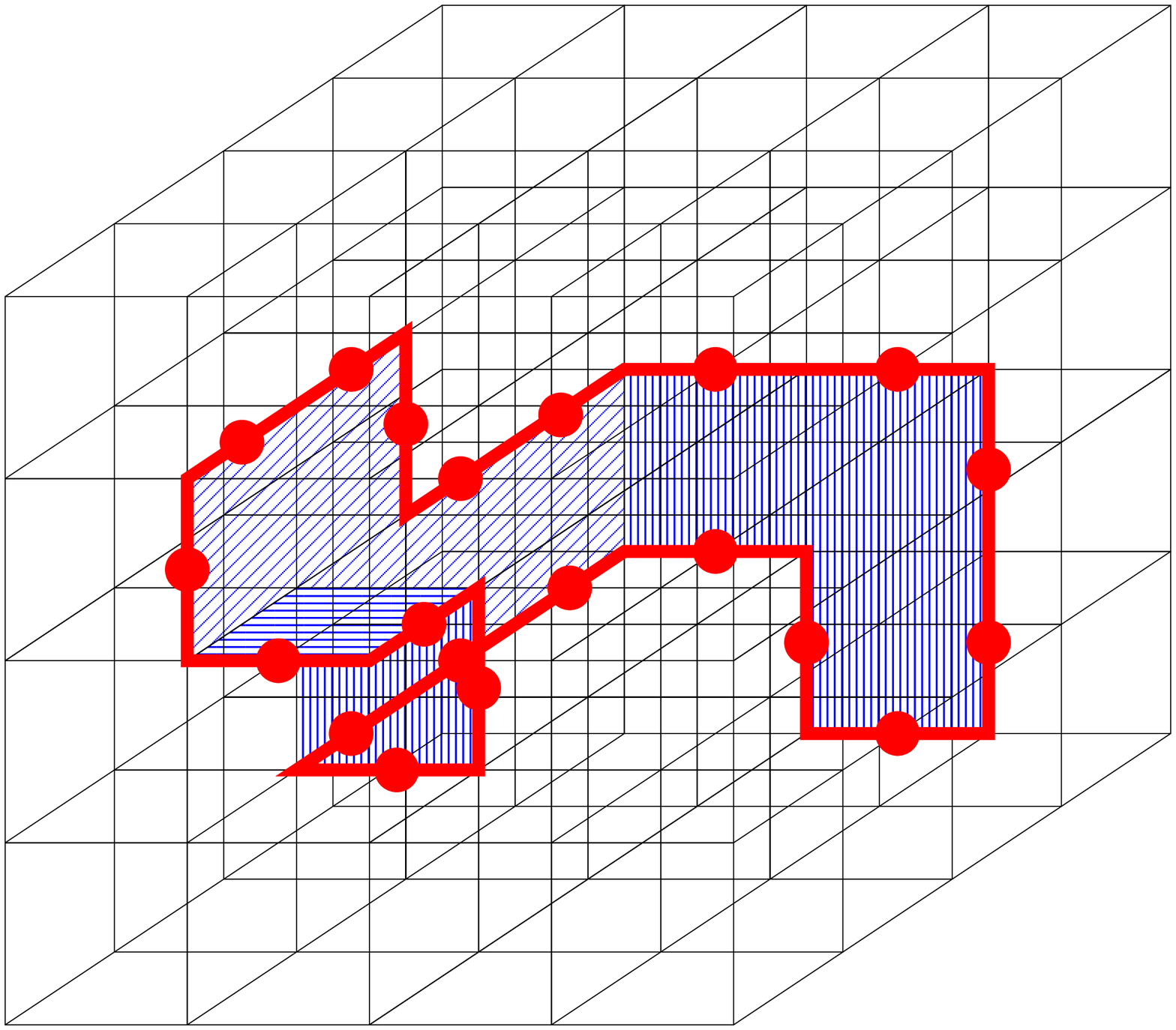}
\\ 
\includegraphics[width=0.7\columnwidth]{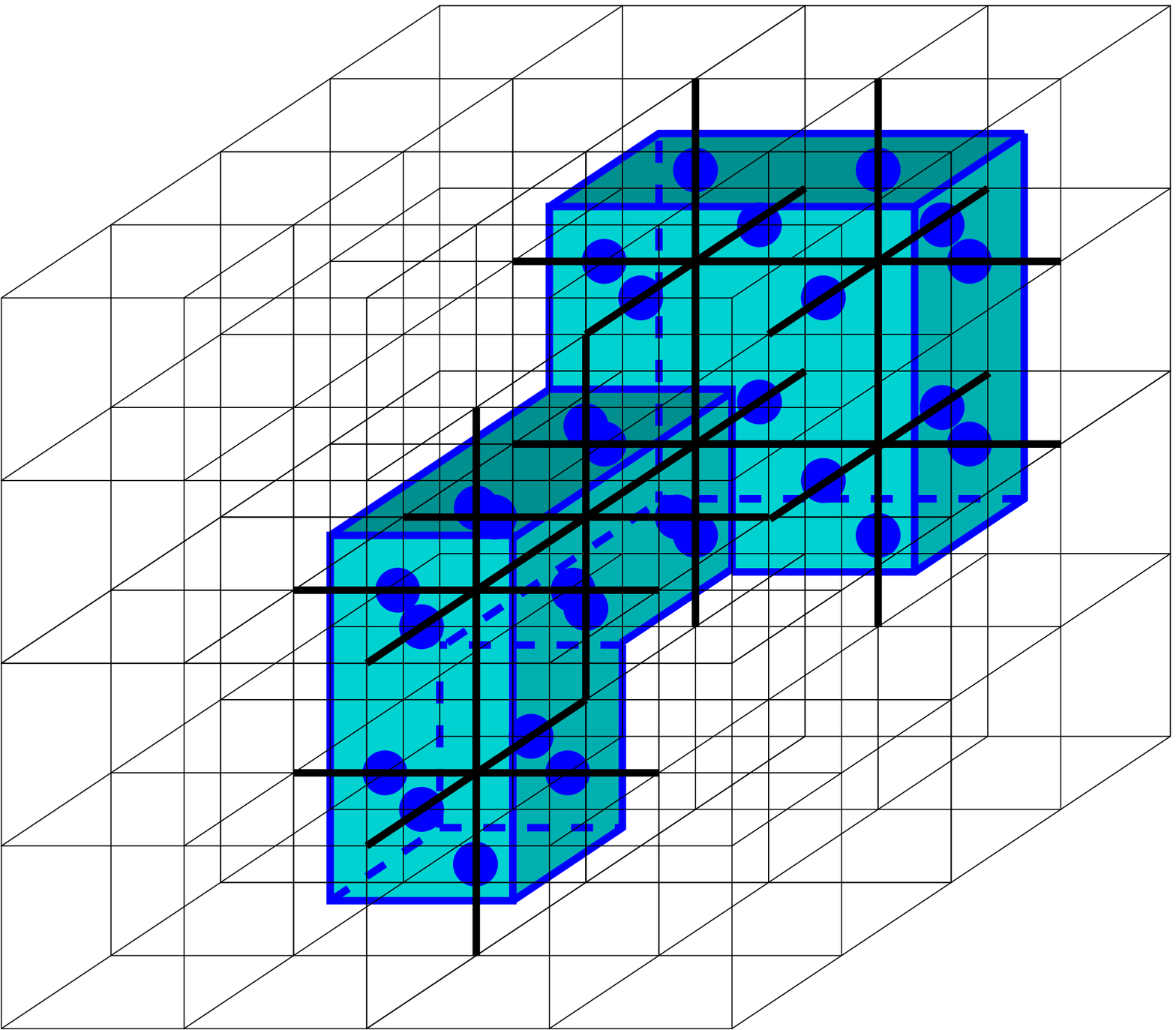}
\end{center}
\caption{
\label{fig: gauge structures}
(Color online) -- 
Two examples of the underlying structures of the 3D Kitaev model: 
the closed $\sigma^{z}_{\ }$ \emph{loops} along the edges of the 
cubic lattice, which satisfy 
$\prod^{\ }_{\textrm{loop}} \sigma^{z}_{i} = 1$, 
and the closed $\sigma^{x}_{\ }$ \emph{membranes} in the 
body-centered dual lattice, satisfying 
$\prod^{\ }_{\textrm{membrane}} \sigma^{x}_{i} = 1$. 
}
\end{figure}

Notice the two \emph{different} underlying structures in the system: 
the closed $\sigma^{z}_{\ }$ \emph{loops} along the edges of the 
cubic lattice, which satisfy 
$\prod^{\ }_{\textrm{loop}} \sigma^{z}_{i} = 1$ identically, 
and the closed $\sigma^{x}_{\ }$ \emph{membranes} in the 
body-centered dual lattice (locally perpendicular to the edges of the 
original lattice), 
satisfying $\prod^{\ }_{\textrm{membrane}} \sigma^{x}_{i} = 1$ 
identically (see Fig.~\ref{fig: gauge structures}). 
%
%
%

\section{\label{sec: zero temperature gauge}
The topological entropy at zero temperature
        }
\begin{figure*}[ht]
\vspace{0.6 cm}
\begin{center}
\includegraphics[width=1.6\columnwidth]{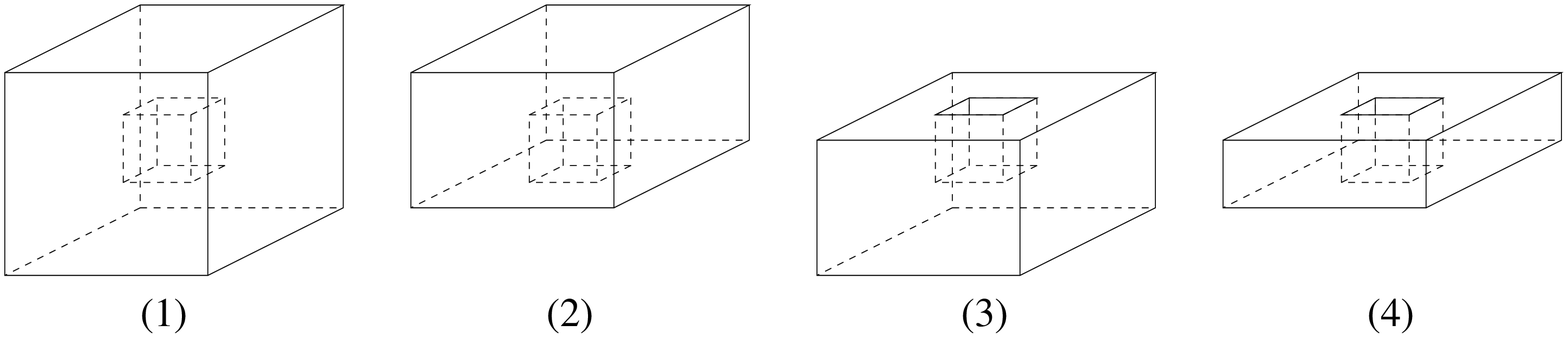}
\\ 
\vspace{0.5 cm}
\includegraphics[width=1.6\columnwidth]{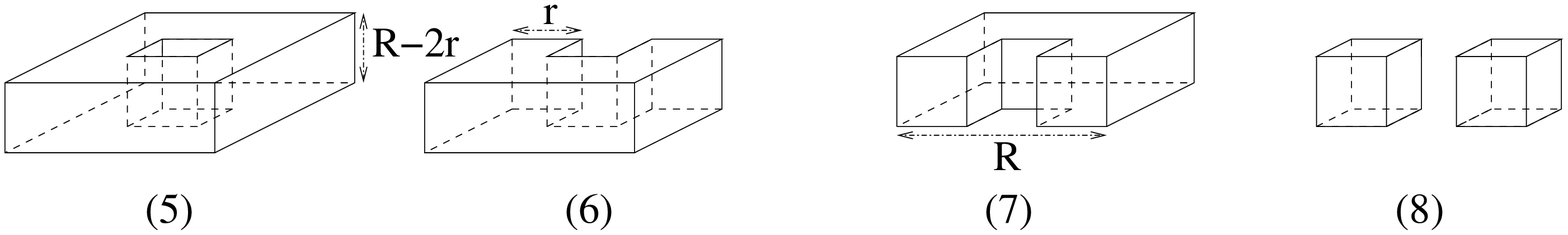}
\end{center}
\caption{
\label{fig: bipartition schemes}
Illustration of the two bipartition schemes used for the 3D Kitaev model: 
`spherical' (Top), and `donut-shaped' (Bottom). 
}
\end{figure*}

Let us first compute the zero-temperature topological entropy of the system, 
using a three-dimensional version of the bipartition scheme proposed by 
Levin and Wen~\cite{Levin2006} in two dimensions. 
Notice, however, that in 3D a bipartition can be topologically non-trivial with respect to closed loops 
but not with respect to closed membranes -- e.g., a donut --, and 
vice versa -- e.g., a spherical shell. Thus, there is no unique way to generalize 
the 2D case. 
Two equally valid options are illustrated in 
Fig.~\ref{fig: bipartition schemes}, based on a `spherical' (1-4) and 
a `donut-shaped' (5-8) bipartition scheme, respectively. 

In the $\sigma^{z}_{\ }$ basis~\cite{footnote-basis}, 
where $G$ is generated by the star 
operators, the calculation of the entanglement entropy 
$S^{\ }_{\textrm{VN}}$ proceeds as in the 2D 
case,~\cite{Hamma2005b,Castelnovo2006,Castelnovo2007}. 
Using the group property of $G$ in Eq.~(\ref{eq: Kitaev GS}), one can show 
that 
\bea
S^{\ }_{\textrm{VN}}(\A) 
&=& 
- \ln \frac{d^{\ }_{\A} d^{\ }_{\B}}{\vert G \vert}
, 
\label{eq: S_VN general}
\eea
where $d^{\ }_{\A}$ is the dimension of the subgroup $G^{\ }_{\A} \subset G$ 
containing all the elements of $G$ that act as the identity on $\B$, 
$G^{\ }_{\A} = \{g\in G \;\vert\; g=g^{\ }_{\A}\otimes \openone^{\ }_{\B}\}$, 
and similarly for subsystem $\B$. 
As in the 2D case, these subgroup dimensions depend on the number 
$\N^{(s)}_{\A}$ ($\N^{(s)}_{\B}$) of star operators acting solely on spins 
in $\A$ ($\B$), and on the number $m^{\ }_{\A}$ ($m^{\ }_{\B}$) of connected 
components of $\A$ ($\B$): 
\bea
d^{\ }_{\A} 
&=& 
2^{\N^{(s)}_{\A}+m^{\ }_{\B}-1}_{\ } 
\label{eq: d_A 1}
\\ 
d^{\ }_{\B} 
&=& 
2^{\N^{(s)}_{\B}+m^{\ }_{\A}-1}_{\ } 
, 
\label{eq: d_B 1}
\eea
The $m^{\ }_{\B}$ contribution to $d^{\ }_{\A}$, and vice versa the 
$m^{\ }_{\A}$ contribution to $d^{\ }_{\B}$, 
come from the so-called collective operations, i.e., elements of the groups 
$G^{\ }_{\A}$ ($G^{\ }_{\B}$) that cannot be expressed as products of 
star operators in $\A$ ($\B$). 
In the 3D case, such collective operations correspond to non-contractible 
closed membranes. In this respect, bipartitions 1 and 8 are special in 
that subsystems $1\B$ and $8\A$ are composed of two separate connected components 
($m^{\ }_{1\B} = m^{\ }_{8\A} = 2$), 
while all other subsystems have only one component. 

We can then compute the topological entropy $S^{\ }_{\textrm{topo}}$ of the 
system in the $\sigma^{z}_{\ }$ basis 
from either the spherical or the donut-shaped bipartition scheme, 
\bea
S^{(z)}_{\textrm{topo}} 
&=& 
\lim^{\ }_{r,R \to \infty} 
  \left[
    - S^{1\A}_{\textrm{VN}} 
    + S^{2\A}_{\textrm{VN}} 
    + S^{3\A}_{\textrm{VN}} 
    - S^{4\A}_{\textrm{VN}} 
  \right] 
= 
\ln 2 
\nonumber \\ 
S^{(z)}_{\textrm{topo}} 
&=& 
\lim^{\ }_{r,R \to \infty} 
  \left[
    - S^{5\A}_{\textrm{VN}} 
    + S^{6\A}_{\textrm{VN}} 
    + S^{7\A}_{\textrm{VN}} 
    - S^{8\A}_{\textrm{VN}} 
  \right] 
= 
\ln 2 
, 
\label{eq: S_topo 1}
\eea
where we used the fact that all $\N^{(s)}_{\ }$ 
contributions cancel out exactly. 
In fact, if we define 
$\N^{(s)}_{i\A\B} = \N^{(s)}_{\ } - \N^{(s)}_{i\A} - \N^{(s)}_{i\B}$ 
to be the number of star operators acting simultaneously on $\A$ and 
$\B$, $\N^{(s)}_{\ } = N$ being the total number of star operators in 
the system, one can show that 
%
%
\bea
&&
\N^{(s)}_{1\A}
+ 
\N^{(s)}_{1\B} 
+ 
\N^{(s)}_{4\A} 
+ 
\N^{(s)}_{4\B} 
\nonumber\\ 
&&
\qquad
- 
\left[
\N^{(s)}_{2\A} 
+ 
\N^{(s)}_{2\B} 
+ 
\N^{(s)}_{3\A} 
+ 
\N^{(s)}_{3\B} 
\right]
\nonumber\\ 
&& 
\qquad 
= 
2N - \N^{(s)}_{1\A\B} - \N^{(s)}_{4\A\B} 
- 2N + \N^{(s)}_{2\A\B} + \N^{(s)}_{3\A\B} 
, 
\nonumber \\ 
&& 
\qquad 
= 0 
. 
\eea
%
%
This result relies on the fact that the total boundary in bipartitions 1 and 
4 is the same -- with the same multiplicity, and with precisely the same edge 
and corner structure -- as in bipartitions 2 and 3, by construction. 
Therefore, 
$\N^{(s)}_{1\A\B} + \N^{(s)}_{4\A\B} = \N^{(s)}_{2\A\B} + \N^{(s)}_{3\A\B}$. 
Similarly for bipartitions 5-8. 

Let us also compute the topological entropy in the $\sigma^{x}_{\ }$ 
basis,~\cite{footnote-basis}
as it will be useful when we consider the finite temperature case. 
The group $G$ is now generated by the plaquette operators, which are highly 
redundant and require more involved calculations to obtain the von Neumann 
entropy $S^{\ }_{\textrm{VN}}$. 
In fact, while Eq.~(\ref{eq: S_VN general}) still holds, one needs to count 
the number of independent plaquette generators of subgroups $G^{\ }_{\A}$ 
and $G^{\ }_{\B}$ in order to obtain the equivalent of 
Eqs.~(\ref{eq: d_A 1}) and~(\ref{eq: d_B 1}). 
Notice that the collective operations are now given by closed 
loops, and only bipartitions 4 and 5 allow for non-trivial (i.e., 
non-contractible) loops. 

As we discussed before, $\vert G \vert = 2^{2(N-1)}_{\ }$. 
This arises from counting all independent generators of $G$ as the total 
number of plaquettes in $G$ (all possible generators), minus the number of 
independent constraints. These are all but one of the cubic unit 
cells, plus three crystal planes. 
Similar arguments apply to the bipartitions 1-8. 
Notice that in all of the bipartitions, 
subsystem $\A$ does \emph{not} contain any entire 
crystal plane, while subsystem $\B$ \emph{always} contains all three crystal 
planes. 
Taking advantage of this simplification, in the following it will be 
understood that $G^{\ }_{\B}$ has three less 
independent generators with respect to $G^{\ }_{\A}$. 

Let us proceed case by case. 
For bipartitions where both $\A$ and $\B$ have only one connected 
component without handles, 
such as bipartitions 2,3,6, and 7 in Fig.~\ref{fig: bipartition schemes}, 
the group $G^{\ }_{\A}$ (equivalently $G^{\ }_{\B}$) 
is generated by all the plaquette operators acting solely on $\A$, subject to 
the constraints given by all cubic unit cells entirely contained in $\A$. 
There are no collective operations in this case, and one obtains 
\bea
d^{\ }_{\A} 
&=& 
2^{\N^{(p)}_{\A}-\N^{(c)}_{\A}}_{\ } 
\\ 
d^{\ }_{\B} 
&=& 
2^{\N^{(p)}_{\B}-\N^{(c)}_{\B}}_{\ } 
, 
\eea
where $\N^{(p)}_{\A}$ is the number of plaquette operators acting on spins 
in $\A$, $\N^{(c)}_{\A}$ is the number of cubic unit cells in $\A$, 
and similarly for $\B$. 

\begin{figure}[ht]
\vspace{0.6 cm}
\begin{center}
\hspace{0.2 cm}
\includegraphics[width=0.72\columnwidth]{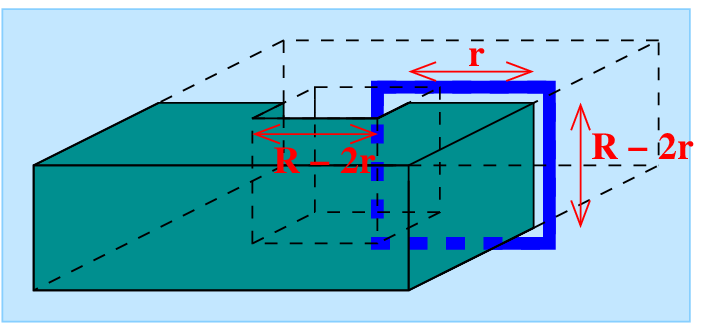}
\\ 
\vspace{0.5 cm}
\includegraphics[width=0.93\columnwidth]{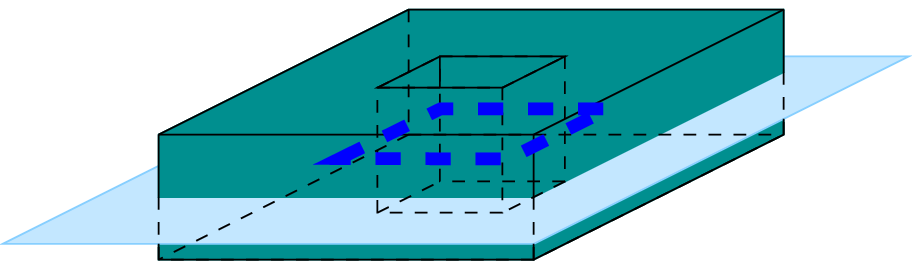}
\end{center}
\caption{
\label{fig: collective ops 1}
\label{fig: collective ops 2}
(Color online) -- 
Illustration of the collective operations in bipartitions 4 and 5 in the 
$\sigma^{x}_{\ }$ basis, acting on subsystem $\B$ (Top) and 
subsystem $\A$ (Bottom), respectively. 
}
\end{figure}

Consider then the case of bipartition 4 (equivalently, 5). 
Although both $\A$ and $\B$ are still connected, the presence of a handle 
allows now for collective operations. 
Take a crystal plane perpendicular to the largest surface of 
subsystem $\A$, and draw it so that it bisects the 
donut into two identical U-shaped portions 
[see Fig.~\ref{fig: collective ops 1} (Top)]. 
The intersection of this plane with $\A$ gives two rectangles of size 
$r \times (R-2r)$, a distance $R-2r$ apart. 
Now take the product of all plaquettes belonging to one of the rectangles 
plus those at its boundary. The resulting operation acts on $\B$ alone, yet 
it cannot be constructed from plaquettes in $\B$ because the 
``outer boundary'' of the rectangle cannot be the \emph{sole} boundary of a 
surface in $\B$. 
Notice that this collective operation can be deformed at will and moved along 
the donut by appropriate products of plaquettes in $\B$, therefore there 
is only one independent such operation. 
Similar arguments apply if we repeat the construction starting from a plane 
parallel to the largest surface of the subsystem $\A$, again chosen so as to 
bisect the donut. This yields another independent collective operation 
acting now on $\A$ [see Fig.~\ref{fig: collective ops 2} (Bottom)]. 
As a result, 
\bea
d^{\ }_{\A} 
&=& 
2^{\N^{(p)}_{\A}-\N^{(c)}_{\A}+n^{\ }_{\A}}_{\ } 
\\ 
d^{\ }_{\B} 
&=& 
2^{\N^{(p)}_{\B}-\N^{(c)}_{\B}+n^{\ }_{\B}}_{\ } 
, 
\eea
where $n^{\ }_{\A}=1$ and $n^{\ }_{\B}=1$ are the number of collective 
operations in $\A$ and $\B$, respectively. 

Finally, one can show that there are no collective operations in the 
$\sigma^{x}_{\ }$ basis in bipartitions 1 and 8. 
In fact, all closed loops are contractible to a point 
both in $\A$ and in $\B$ in these bipartitions. 
However, the disconnected nature of subsystem $\B$ in bipartition 1 
(equivalently, subsystem $\A$ in bipartition 8), requires special care in the 
counting of the independent generators of $G^{\ }_{\A}$ (respectively, 
$G^{\ }_{\B}$). 
As in the previous cases, all plaquettes in $\A$ belong to 
$G^{\ }_{\A}$, and all cubic unit cells in $\A$ act as independent constraints 
towards the counting of the independent generators of $G^{\ }_{\A}$. 
However, in bipartition 1, there is a class of closed membranes in 
$\A$ that cannot be assembled as a product of cubic cells in $\A$. 
This is the case, for example, of the closed cubic membranes in $\A$ that 
surround entirely the inner component of $\B$. 
Any two such membranes can be obtained one from the other 
via multiplication by cubic unit cells in $\A$. 
Thus, they only give rise to one additional constraint in the counting of the 
independent generators. 
In general, the number of such constraints is given 
by $m^{\ }_{\B}-1$, where $m^{\ }_{\B}$ is the number of connected components 
of $\B$. 
Similarly for bipartition 8 and subsystem $\B$, one obtains $m^{\ }_{\A}-1$ 
additional constraints, where $m^{\ }_{\A}$ is the number of connected 
components of $\A$. 

Combining all of the above considerations into a general expression for the 
dimensions of subgroups $G^{\ }_{\A}$ and $G^{\ }_{\B}$ in the 
$\sigma^{x}_{\ }$ basis, one obtains 
\bea
d^{\ }_{\A} 
&=& 
2^{\N^{(p)}_{\A} 
   - \N^{(c)}_{\A} 
   + n^{\ }_{\A} - (m^{\ }_{\B}-1) 
   - m^{(\textrm{c.p.})}_{\A}}_{\ } 
\\ 
d^{\ }_{\B} 
&=& 
2^{\N^{(p)}_{\B} 
   - \N^{(c)}_{\B} 
   + n^{\ }_{\B} - (m^{\ }_{\A}-1) 
   - m^{(\textrm{c.p.})}_{\B}}_{\ } 
, 
\label{eq: d_A in general}
\eea
where $m^{(\textrm{c.p.})}_{\A}$ ($m^{(\textrm{c.p.})}_{\B}$) is the number 
of crystal planes (c.p.) entirely contained in $\A$ ($\B$). 
Recall that all bipartitions of interest have 
$m^{(\textrm{c.p.})}_{\A} = 0$ and $m^{(\textrm{c.p.})}_{\B} = 3$. 

We can then use Eq.~(\ref{eq: S_VN general}) 
to compute the topological entropy of the system using the spherical 
and the donut-shaped bipartition schemes in the $\sigma^{x}_{\ }$ 
basis, 
\bea
S^{(x)}_{\textrm{topo}} 
&=& 
\lim^{\ }_{r,R \to \infty} 
  \left[
    - S^{1\A}_{\textrm{VN}} 
    + S^{2\A}_{\textrm{VN}} 
    + S^{3\A}_{\textrm{VN}} 
    - S^{4\A}_{\textrm{VN}} 
  \right] 
\nonumber\\ 
&=& 
\left(- 1 + 2 \right) \ln 2 
= 
\ln 2 
\nonumber \\ 
S^{(x)}_{\textrm{topo}} 
&=& 
\lim^{\ }_{r,R \to \infty} 
  \left[
    - S^{5\A}_{\textrm{VN}} 
    + S^{6\A}_{\textrm{VN}} 
    + S^{7\A}_{\textrm{VN}} 
    - S^{8\A}_{\textrm{VN}} 
  \right] 
\nonumber\\ 
&=& 
\left( 2 - 1 \right) \ln 2 
= 
\ln 2, 
\label{eq: S_topo 2}
\eea
where we used the fact that all $\N^{(p)}_{\ }$ and $\N^{(c)}_{\ }$ 
contributions cancel out exactly. 
In fact, if we define 
$\N^{(p)}_{i\A\B} = \N^{(p)}_{\ } - \N^{(p)}_{i\A} - \N^{(p)}_{i\B}$ 
to be the number of plaquette operators acting simultaneously on $\A$ and 
$\B$, $\N^{(p)}_{\ } = 3N$ being the total number of plaquette operators in 
the system, and we define 
$\N^{(c)}_{i\A\B} = \N^{(c)}_{\ } - \N^{(c)}_{i\A} - \N^{(c)}_{i\B}$ 
to be the number of cubic unit cells simultaneously encompassing spins in 
$\A$ and in $\B$, $\N^{(c)}_{\ } = N$ being the total number of cubic unit 
cells in the system, 
one can show that 
\begin{widetext}
\bea
&&
\left( \N^{(p)}_{1\A} - \N^{(c)}_{1\A} \right) 
+ 
\left( \N^{(p)}_{1\B} - \N^{(c)}_{1\B} \right) 
+ 
\left( \N^{(p)}_{4\A} - \N^{(c)}_{4\A} \right) 
+ 
\left( \N^{(p)}_{4\B} - \N^{(c)}_{4\B} \right) 
\nonumber\\ 
&&
\qquad\qquad\qquad\qquad\qquad 
- 
\left[
\left( \N^{(p)}_{2\A} - \N^{(c)}_{2\A} \right) 
+ 
\left( \N^{(p)}_{2\B} - \N^{(c)}_{2\B} \right) 
+ 
\left( \N^{(p)}_{3\A} - \N^{(c)}_{3\A} \right) 
+ 
\left( \N^{(p)}_{3\B} - \N^{(c)}_{3\B} \right) 
\right]
\nonumber\\ 
&& 
\qquad\qquad\qquad\qquad\qquad 
= 
4N - \N^{(p)}_{1\A\B} - \N^{(p)}_{4\A\B} 
+ \N^{(c)}_{1\A\B} + \N^{(c)}_{4\A\B} 
- 4N + \N^{(p)}_{2\A\B} + \N^{(p)}_{3\A\B} 
- \N^{(c)}_{2\A\B} - \N^{(c)}_{3\A\B} , 
\nonumber \\ 
&& 
\qquad\qquad\qquad\qquad\qquad 
= 0 
. 
\eea
\end{widetext}
This result relies on the fact that the total boundary in bipartitions 1 and 
4 is the same -- with the same multiplicity, and with precisely the same edge 
and corner structure -- as in bipartitions 2 and 3, by construction. 
Therefore, 
$\N^{(p)}_{1\A\B} + \N^{(p)}_{4\A\B} = \N^{(p)}_{2\A\B} + \N^{(p)}_{3\A\B}$ 
and 
$\N^{(c)}_{1\A\B} + \N^{(c)}_{4\A\B} = \N^{(c)}_{2\A\B} + \N^{(c)}_{3\A\B}$. 
Similarly for bipartitions 5-8. 

Clearly, both bipartition schemes capture the topological nature of the 
system, and provide an equally valid measure of the topological entropy. 
In 2D the choice of bipartitions 1-4 in Ref.~\onlinecite{Levin2006} is such 
that bipartition 1 is topologically equivalent to bipartition 4 upon exchange 
of subsystem $\A$ with subsystem $\B$, while bipartitions 2 and 3 are actually 
topologically invariant upon the same exchange. 
Hence, because the von Neumann entropy for the ground state is symmetric
under the exchange of $\A$ and 
$\B$, the topological contribution measured in the 2D scheme is bound to 
be double counted, namely $S^{\ }_{\textrm{topo}} = 2 \ln D = \ln D^{2}_{\ }$, 
where $D$ is the so called quantum dimension of the 
system.~\cite{Kitaev2006,Levin2006} 
In 3D, both the scheme 1-4 and the scheme 5-8 isolate the topological 
contribution to the entanglement entropy without double counting. 
Notice that all the bipartitions are topologically invariant under the 
exchange of $\A$ and $\B$, except for bipartitions 1 and 8. 
If we want to recover the symmetry of the 2D scheme, a possible solution 
is to define 
\bea 
S^{\ }_{\textrm{topo}} 
&=& 
\lim^{\ }_{r,R \to \infty} 
  \left[
    - S^{1\A}_{\textrm{VN}} 
    + S^{2\A}_{\textrm{VN}} 
    + S^{3\A}_{\textrm{VN}} 
    - S^{4\A}_{\textrm{VN}} 
\right. 
\nonumber \\ 
&& 
\left. 
\qquad\;\;\,
    - S^{5\A}_{\textrm{VN}} 
    + S^{6\A}_{\textrm{VN}} 
    + S^{7\A}_{\textrm{VN}} 
    - S^{8\A}_{\textrm{VN}} 
  \right] 
\nonumber \\ 
&=& 
\ln D^2 , 
\eea
with $D = 2$. 
As we will see in the following, the symmetric 1-8 choice is actually 
required if we are interested in studying the finite temperature case, since 
the von Neumann entropy is no longer invariant upon exchange of $\A$ and $\B$, 
and a non-topologically-symmetric choice of bipartitions would lead to 
different results depending on whether we work with subsystem $\A$ or 
subsystem $\B$.~\cite{footnote: mutual info} 
%
%

\section{\label{sec: finite temperature}
The finite temperature behavior
        }
In this section we study the behavior of the entanglement and topological 
entropies at finite temperature, via a generalization of the approach 
used for the 2D Kitaev model in Ref.~\onlinecite{Castelnovo2007}. 

\begin{figure}[ht]
\vspace{0.6 cm}
\begin{center}
\includegraphics[width=0.7\columnwidth]{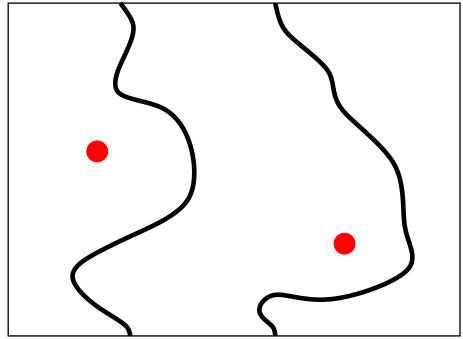}
\end{center}
\caption{
\label{fig: 2D defect picture}
(Color online) -- 
Qualitative illustration of the disruptive effect of two defects 
(solid red dots) 
in the 2D Kitaev model on a torus: two winding loops (black wavy lines) 
on either side of a defect (solid circle) \emph{read off} opposite eigenvalues 
of the corresponding winding loop operator. 
}
\end{figure}

A qualitative picture of the effect of thermal fluctuations can be 
argued by comparison with the two dimensional case. 
There the information about the 
topological sectors is stored in the eigenvalues of winding loop operators, 
namely products of spin operators along winding loops. 
On a torus, there are infinitely many choices for such winding loop operators, 
but the absence of magnetic and electric charges (i.e., plaquettes and stars 
with eigenvalue $-1$) in the gauge structure at zero temperature reduces them 
to only two independent ones: the two non-contractible winding loops on the 
torus. 
Any other can be obtained from these two via multiplication by an appropriate 
set of plaquette or star operators, which have eigenvalue $+1$ at $T=0$. 
Clearly the presence of order $1$ (deconfined) thermal defects destroys immediately all 
topological information stored in the system, since the eigenvalues of two 
loops on opposite sides of a defect are no longer consistent with each other 
(see Fig.~\ref{fig: 2D defect picture}).
%
%

Let us now consider the case of the Kitaev model in 3D. 
First of all, we need to discuss the two gauge structures separately, since 
they are no longer identical as in 2D. 
If we work in the $\sigma^{x}$ basis, then the topological 
information is stored in the eigenvalues of winding membrane operators, given 
by the product of all $\sigma^{x}$ operators belonging to a closed 
winding surface locally perpendicular to the bonds of the sites it crosses 
(see Fig.~\ref{fig: non local ops}). 
All possible choices of these membranes yield the same result at zero 
temperature since the corresponding operators can be obtained one from the 
other by products of sets of star operators, which have all eigenvalue $+1$ 
in the GS. 
Thermal defects in this case play exactly the same role as in 2D, since two 
membranes on opposite sides of a defect \emph{read off} opposite eigenvalues 
of the corresponding winding membrane operator.

\begin{figure}[ht]
\vspace{0.6 cm}
\begin{center}
\includegraphics[width=0.8\columnwidth]{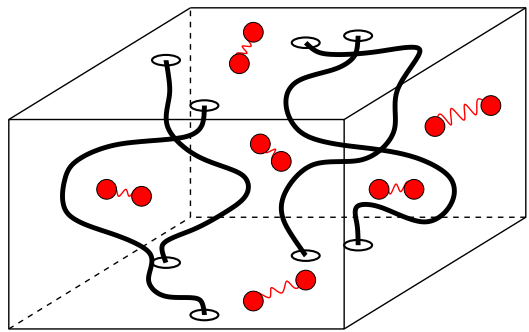}
\end{center}
\caption{
\label{fig: 3D defect picture}
(Color online) -- 
Qualitative illustration of the reason why the topological information stored 
in the underlying $\sigma^{z}_{\ }$ loop structure of the 3D Kitaev model 
is robust to thermal fluctuations: even in presence of sparse defects 
(solid red circles), any two winding loops (black wavy lines), with equal 
winding numbers, can be smoothly deformed one into the other without crossing 
any defects. 
(The wiggly lines represent qualitatively the confining strings between 
defect `pairs' discussed in the text.) 
}
\end{figure}

On the other hand, the situation is quite different for the loop operators 
defined in the $\sigma^{z}$ basis. There the topological information is stored 
in winding loop operators -- as in the 2D case -- but they are 
now embedded in 3D. 
Clearly, localized defects have no disruptive effects on the topological 
information because any two winding loops (with equal winding numbers) 
can be smoothly deformed one into the other 
\emph{without crossing any defects} at low enough temperatures 
(see Fig.~\ref{fig: 3D defect picture}). 
This is indeed the case here, where we learn from 3D lattice gauge theory 
that defective plaquettes are confined at low temperatures. 
They are created in quadruplets by a single spin flip operation, 
and they can be pairwise separated only at the cost of creating a string of 
defective plaquettes in between the two pairs.~\cite{Wegner1971,Kogut1979} 
Therefore, the winding loop operators will keep carrying the same quantum 
information in presence of a low density of defects. If we were to read 
out the topological information from the system, we would be getting the 
correct result as long as the chosen loop does not 
pass directly \emph{through} a defect.

However, can this information be accessed by means of the same expectation 
values of loop operators that are used at zero temperature, 
Eq.~(\ref{eq: topo ops z}) and Eq.~(\ref{eq: topo ops x})? 
The answer to this question is negative, as it was recently shown using gauge 
theory arguments in Ref.~\onlinecite{Nussinov2008}. A simple reason 
as to why naively choosing a given loop operator and
looking at its expectation value alone does not capture the order below
$T_c$ is that, typically, winding loops will pass through at least one defect
in the thermodynamic limit (the probability of a loop not crossing any 
defect scales as $(1-\rho_{\textrm{def}})^L$, where $\rho_{\textrm{def}}$ is 
the equilibrium density of defects at a given temperature, and $L$ is the 
linear size of the system). However, only those loops that
avoid the defects contain the topological information. (Recall that in
2D the eigenvalues of loop operators, even when they do not pass
through defects, differ on two sides of one defect, in contrast to the
situation in 3D.) 
This implies that the average expectation value of loop operators is bound 
to vanish exponentially in system size for any finite density of defects, 
i.e., for any finite temperature, independently of the nature of the 
system. As we show in the following, the topological entropy
of the system is capable of capturing these physical differences, and
it accurately reflects the topological properties of the different
phases.

The physical meaning of the distinct sectors can be understood as
follows. Consider preparing the system in a coherent superposition of
different topological sectors at zero temperature. Raise the
temperature to some value $T < T_c$, and then lower it again back to
zero. If defects are confined, transitions between different loop
sectors are forbidden throughout the process. We are thus bound to
obtain a final state where the probability (magnitude of amplitude
square) of finding the final state in each loop sector is the same as
in the initial state. In this sense, the loop sectors are protected
from thermal fluctuations at low temperatures, and topological order
survives at finite temperature ($T<T_c$).

That the system does not change sectors during the time that it is in
thermal equilibrium with the bath is a {\it dynamical} problem (broken
ergodicity). This can be understood by contrasting the time scales for
mixing sectors if defects are confined or deconfined. Deconfined
thermal defects are free to randomly walk across the system, and
induce transitions between different topological sector by means of
creation, system-spanning propagation, and annihilation processes.
The characteristic time for a sector-changing process scales therefore
as some power of the system size, $\tau_{\textrm{deconfined}} \sim
L^{\alpha}$. In contrast, confined defects will have to overcome an
energy barrier of the order of $L$ to be able to wind around the
system and induce a change in the topological sector. As a result,
their characteristic time scale is instead exponential in system size,
$\tau_{\textrm{confined}} \sim e^{c L}$. Even for rather small
systems, confined defects would require time scales larger than the
age of the universe to transition between sectors.

An even more interesting situation occurs when both $\mathbb{Z}_2$
gauge defect types are confined, so that the $\sigma^x$ and $\sigma^z$
topological sectors are both protected. This case is briefly discussed
in Appendix~\ref{app: confined-confined}, and it is related to error
recovery that was argued to be realizable for example in a 4D toric
code.~\cite{Dennis2002} What we argue here based on the finite
temperature studies is that the system can be {\it self-correcting}:
if the system is prepared in a given superposition at zero temperature
and its temperature is raised and again lowered to zero without ever
going above $T_c$, the system returns to the same original quantum
state (a ``boomerang'' effect).

The protection holds at low temperatures, but it is bound
to vanquish as the density of defective plaquettes with eigenvalue
$-1$ grows with temperature: once enough defects are in place, one can
no longer deform paths around them. Therefore, we expect a loss of
topological information as temperature is increased, via a topological
phase transition at finite temperature.

In analogy with 3D lattice gauge theory, we expect this transition to occur 
when plaquette defects deconfine at high enough temperature. This is captured 
by the expectation value of Wilson loop operators, which is exponentially 
suppressed with the length of the loop (perimeter law) at low temperatures, 
while it is suppressed with the area of the minimal enclosed surface 
(area law) at high temperatures.%
~\cite{Wegner1971,Kogut1979,Savit1980,Caselle2002}
In our notation, the transition temperature is set by the energy 
scale $\lambda_B$, and the transition is expected to occur at the critical point of 
the 3D lattice gauge theory.

The topological entropy is a non-local order parameter that detects
the presence of topological order in a system. Any loss of topological
information, e.g., whenever some topological sectors become
ill-defined, should have a measurable effect on such entropy. Indeed,
we show below that this is the case, and that the qualitative picture
inferred from the arguments above is confirmed by an exact calculation
of the topological entropy at finite temperature.
%
%

\subsection{\label{sec: density matrix}
The density matrix
           }
Let us work for convenience in the $\sigma^{x}_{\ }$ 
(tensor product) basis, 
where the Hilbert space $\mathcal{H}$ is spanned by the whole set of 
orthonormal states $\vert \alpha \rangle$, 
labeled by the configurations $\alpha$ of a classical Ising model 
on the bonds of a 3D simple cubic lattice (the value $\pm 1$ 
of each Ising variable corresponds to the eigenvalue of the 
$\sigma^{x}_{\ }$ operator at the same site). 

Define $G$ to be the group generated by all plaquette operators 
$B^{\ }_{p} = \prod^{\ }_{i \in p} \sigma^{z}_{i}$. 
Recall that any two elements of the group differing by products of 
plaquettes around closed membranes are in fact the same element 
(i.e., they are defined modulo the identities 
$\prod^{\ }_{\textrm{closed membrane}} B^{\ }_{p} = \openone$), where we are 
assuming periodic boundary conditions, and full crystal planes are therefore 
closed membranes as well. 
Recall also that $\vert G \vert = 2^{2N-2}_{\ }$, where 
$N$ is the number of sites in the simple cubic lattice. 
Every two elements of the 
group commute with each other, and 
$g^{2}_{\ } = \openone$, $\forall\,g\in G$. 
For later convenience, let us label with $\alpha=0$ the 
fully magnetized state $\sigma^{x}_{\ }=+1$. 

The equilibrium properties of the system at finite temperature are 
captured by the density matrix 
\bea
\rho(T) 
&=& 
\frac{1}{Z} \: e^{-\beta \hat H} 
\nonumber \\ 
&=& 
\frac{\sum^{\ }_{\alpha,\beta} 
      \langle\alpha\vert 
        e^{-\beta H}_{\ } 
      \vert\beta\rangle
      \;
      \vert\alpha\rangle \langle\beta\vert 
     }
     {\sum_{\alpha} 
      \langle\alpha\vert 
        e^{-\beta H} 
      \vert\alpha\rangle
     } 
. 
\label{eq: rho(T) def f.c.}
\eea
For convenience of notation, let us rewrite the 
Hamiltonian~(\ref{eq: Kitaev Ham}) as 
\bea
H 
&=& 
-\lambda^{\ }_{B} P 
-\lambda^{\ }_{A} S 
\nonumber \\
P &=& \sum^{\ }_{p} B^{\ }_{p} 
\nonumber \\
S &=& \sum^{\ }_{s} A^{\ }_{s}. 
\nonumber
\eea

Notice that 
$S \vert\alpha\rangle = M^{\ }_{s}(\alpha) \vert \alpha \rangle$, 
where $M^{\ }_{s}(\alpha)$ is the net 
``star magnetization'', i.e., the difference between the number of stars with 
eigenvalue $+1$ and with eigenvalue $-1$ in the state $\vert\alpha\rangle$. 
The action of any group element $g$ is to flip plaquettes, which 
cannot change the sign of any star operator since they commute, and 
therefore 
$S g \vert\alpha\rangle = M^{\ }_{s}(\alpha) g \vert \alpha \rangle$, 
$\forall\,g \in G$. 

Thus, the denominator of Eq.~(\ref{eq: rho(T) def f.c.}) becomes 
\bea
\sum_{\alpha} 
  \langle\alpha\vert 
    e^{-\beta H} 
  \vert\alpha\rangle
&=&
\sum_{\alpha} 
  e^{\beta \lambda^{\ }_{A} M^{\ }_{s}(\alpha)}_{\ } 
  \langle\alpha\vert 
    e^{\beta \lambda^{\ }_{B} P} 
  \vert\alpha\rangle
. 
\eea
Upon expanding 
\beq
e^{\beta\lambda^{\ }_{B} P} 
=
\prod_{p} 
  \left[
    \cosh\beta\lambda^{\ }_{B} 
    +
    \sinh\beta\lambda^{\ }_{B}\: B^{\ }_{p} 
  \right], 
\eeq
as follows from the definition $P = \sum^{\ }_{p} B^{\ }_{p}$ and from the 
fact that $B^{2}_{p} \equiv \openone$, $\forall\,p$, one can explicitly 
compute the last term 
\beq
\langle \alpha \vert 
  e^{\beta\lambda^{\ }_{B} P}_{\ } 
\vert \alpha \rangle 
= 
\langle \alpha \vert 
  \prod^{\ }_{p}
    \left[
      \cosh\beta\lambda^{\ }_{B} 
      + 
      \sinh\beta\lambda^{\ }_{B}\: B^{\ }_{p} 
    \right] 
\vert \alpha \rangle. 
\label{eq: exp S f.c.}
\eeq

All non-vanishing contributions in Eq.~(\ref{eq: exp S f.c.}) come from 
products of plaquette operators that reduce to the identity (i.e., products 
around closed membranes). The above equation is therefore independent 
of $\alpha$, which we set to the reference state $0$ 
in the following.

The set of all possible closed membranes in a periodic 3D simple cubic 
lattice is in one-to-two correspondence with all possible configurations 
of an Ising model on the dual simple cubic lattice (the membranes are, say, 
the antiferromagnetic domain boundaries), provided we allow for both periodic 
and antiperiodic boundary conditions in all three directions. 
In this language, the sum of all non-vanishing contributions can be written 
as 
\beq
\langle 0 \vert 
  e^{\beta\lambda^{\ }_{B} P}_{\ } 
\vert 0 \rangle 
= 
\frac{1}{2} 
\sum^{\ }_{\mathcal{C}} 
  \left[
    \cosh\beta\lambda^{\ }_{B} 
  \right]^{3N-N^{\ }_{\textrm{AF}}(\mathcal{C})}_{\ } 
  \left[
    \sinh\beta\lambda^{\ }_{B}
  \right]^{N^{\ }_{\textrm{AF}}(\mathcal{C})}_{\ } , 
\eeq
where $\mathcal{C}$ is a generic configuration of the 3D Ising model 
with any type of boundary conditions, $3N$ is the total number of 
nearest-neighbor (nn) bonds and $N^{\ }_{\textrm{AF}}(\mathcal{C})$ is the 
number of antiferromagnetic nn bonds. 
The factor $1/2$ comes from the $\mathbb{Z}^{\ }_{2}$ symmetry: a given 
membrane configuration corresponds to two equivalent but distinct Ising 
configurations. 
For convenience, let us introduce the simplified notation 
$c = \cosh\beta\lambda^{\ }_{B}$, 
$s = \sinh\beta\lambda^{\ }_{B}$ and 
$t = s/c = \tanh\beta\lambda^{\ }_{B}$, 
and define $J>0$ such that $e^{-2 J}_{\ } = t$ 
(recall that $\lambda^{\ }_{B} > 0$). 
The above expression can then be further simplified to 
\bea
2 \;
\langle 0 \vert 
  e^{\beta\lambda^{\ }_{B} P}_{\ } 
\vert 0 \rangle 
&=& 
c^{3N}_{\ }
\sum^{\ }_{\mathcal{C}} \: 
  t^{N^{\ }_{\textrm{AF}}(\mathcal{C})}_{\ }
\nonumber \\ 
&=& 
c^{3N}_{\ }
\sum^{\ }_{\mathcal{C}} \: 
  e^{-2 J N^{\ }_{\textrm{AF}}(\mathcal{C})}_{\ }
\nonumber \\ 
&=& 
c^{3N}_{\ }
\sum^{\ }_{\mathcal{C}} \: 
  \exp 
    \left(
      J \sum^{\ }_{\langle i,j \rangle} S^{\ }_{i} S^{\ }_{j} - 3NJ 
    \right)  
\nonumber \\ 
&=& 
(sc)^{3N/2}_{\ }
\sum^{\ }_{\mathcal{C}} \: 
  \exp
    \left( 
      J \sum^{\ }_{\langle i,j \rangle} S^{\ }_{i} S^{\ }_{j} 
    \right) 
\nonumber \\ 
&\equiv& 
(sc)^{3N/2}_{\ }
Z^{\textrm{tot}}_{J} 
, 
\label{eq: 3D Ising f.c.}
\eea
%
%
where $Z^{\textrm{tot}}_{J}$ is the partition function of an Ising model on 
a simple cubic lattice of size $N = L \times L \times L$ with reduced 
ferromagnetic coupling constant $J$, summed over all possible choices of 
(periodic or antiperiodic) boundary conditions.~\cite{footnote: Ising duality} 

We can now move on to compute the numerator of 
Eq.~(\ref{eq: rho(T) def f.c.}), 
\bea
&&
\sum^{\ }_{\alpha,\beta} \: 
  \langle\alpha\vert 
    e^{-\beta H}_{\ } 
  \vert\beta\rangle
  \;
  \vert\alpha\rangle \langle\beta\vert 
=
\nonumber \\ 
&& 
\qquad 
= 
\sum^{\ }_{\alpha,\beta} 
  e^{\beta \lambda^{\ }_{A} M^{\ }_{s}(\alpha)}_{\ } 
  \langle\alpha\vert 
    e^{\beta \lambda^{\ }_{B} P}_{\ } 
  \vert\beta\rangle
  \;
  \vert\alpha\rangle \langle\beta\vert 
\nonumber \\ 
&& 
\qquad 
= 
\sum^{\ }_{g\in G} \sum^{\ }_{\alpha} 
  e^{\beta \lambda^{\ }_{A} M^{\ }_{s}(\alpha)}_{\ } 
  \langle\alpha\vert 
    e^{\beta \lambda^{\ }_{B} P}_{\ } g 
  \vert\alpha\rangle
  \;
  \vert\alpha\rangle \langle\alpha\vert g 
, 
\eea
where we used the fact that all matrix elements 
$
\langle\alpha\vert 
  e^{\beta \lambda^{\ }_{B} P}_{\ } 
\vert\beta\rangle
$ 
vanish identically unless $\vert\beta\rangle = g \vert\alpha\rangle$, 
$\exists g \in G$. 
Once again, the expectation value 
$
\langle\alpha\vert 
  e^{\beta \lambda^{\ }_{B} P}_{\ } g 
\vert\alpha\rangle
$ 
is independent of $\alpha$, and the above expression simplifies to 
\beq
\sum_{g\in G} \sum^{\ }_{\alpha} 
  e^{\beta \lambda^{\ }_{A} M^{\ }_{s}(\alpha)}_{\ } 
  \langle0\vert 
    e^{\beta \lambda^{\ }_{B} P}_{\ } g 
  \vert0\rangle
  \;
  \vert\alpha\rangle \langle\alpha\vert g 
. 
\eeq

The expectation value can be computed explicitly 
by expanding the exponential 
\bea
&& 
\langle 0 \vert 
  e^{\beta\lambda^{\ }_{B} P}_{\ } g 
\vert 0 \rangle 
= 
\nonumber \\ 
&& 
\qquad
= 
\langle 0 \vert 
  \prod^{\ }_{p}
    \left[
      \cosh\beta\lambda^{\ }_{B} 
      + 
      \sinh\beta\lambda^{\ }_{B}\: B^{\ }_{p} 
    \right] 
  \prod^{\ }_{p^{\prime}_{\ }\in g} B^{\ }_{p^{\prime}_{\ }} 
\vert 0 \rangle 
. 
\nonumber \\
\label{eq: exp S g f.c.}
\eea
Here, the notation 
$\prod^{\ }_{p^{\prime}_{\ }\in g} B^{\ }_{p^{\prime}_{\ }}$ 
represents the decomposition of $g$ in terms of the group 
generators $\{ B^{\ }_{p} \}$. 
Clearly this decomposition is highly non-unique, since the group elements are 
defined modulo the identities 
$\prod^{\ }_{\textrm{closed membrane}} B^{\ }_{p} = \openone$, and 
Eq.~(\ref{eq: exp S g f.c.}) needs to be handled with care. 

As before, all non vanishing contributions 
come from products of plaquette operators that reduce to the identity. 
In this case, however, there are two options for every operator 
$B^{\ }_{p^{\prime}_{\ }}$: 
(i) it can be multiplied out directly by 
$\sinh\beta\lambda^{\ }_{B}\: B^{\ }_{p}$, with $p=p^{\prime}_{\ }$ (recall 
that $B^{2}_{p^{\prime}_{\ }} = \openone$); 
or 
(ii) it can be completed to an identity by an appropriate product of 
$B^{\ }_{p}$ terms so that $B^{\ }_{p^{\prime}_{\ }} \prod B^{\ }_{p}$ 
forms a closed membrane. 
Notice that in the second case the product over $p$ may \emph{not} include 
$p^{\prime}_{\ }$ itself. 

All this can be expressed in more elegant terms in the Ising language defined 
previously. Case (i) corresponds to the two spins across the bond 
$p^{\prime}_{\ }$ being ferromagnetically aligned in the Ising model, 
and contributing a Boltzmann factor $\sinh\beta\lambda^{\ }_{B}$. 
Case (ii) corresponds to the two spins across $p^{\prime}_{\ }$ being 
antiferromagnetically aligned, and contributing a Boltzmann factor 
$\cosh\beta\lambda^{\ }_{B}$. 
Notice that the correlations between the different $p^{\prime}_{\ }$ are 
automatically taken care of in the Ising language, and we obtain 
\bea
2 \;
\langle 0 \vert 
  e^{\beta\lambda^{\ }_{B} P}_{\ } g 
\vert 0 \rangle 
&=& 
(sc)^{3N/2}_{\ }
\sum^{\ }_{\mathcal{C}} \: 
  \exp 
    \left( 
      J \sum^{\ }_{\langle i,j \rangle} 
       \eta^{\ }_{ij}(g) S^{\ }_{i} S^{\ }_{j} 
    \right) 
\nonumber \\ 
&\equiv& 
(sc)^{3N/2}_{\ }
Z^{\textrm{tot}}_{J}(g), 
\label{eq: 3D Ising random f.c.}
\eea
where 
\beq
\eta^{\ }_{ij}(g) 
= 
\left\{ 
  \begin{array}{ll}
    +1  & \;\; \textrm{if} \; \langle i,j \rangle \notin g
    \\ 
    -1 & \;\; \textrm{if} \; \langle i,j \rangle \in g . 
  \end{array}
\right. 
\label{eq: random bond g f.c.} 
\eeq
Recall that a bond in the Ising model corresponds to a plaquette in the 
original system, and $\langle i,j \rangle \in g$ means that the 
corresponding plaquette operator appears in the decomposition of 
$g$. 

In order to derive Eq.~(\ref{eq: 3D Ising random f.c.}), let us define 
$N^{\ }_{\textrm{F}}(\mathcal{C} \vert g)$ 
($N^{\ }_{\textrm{AF}}(\mathcal{C} \vert g)$) to be the 
number of bonds with ferromagnetically (antiferromagnetically) aligned spins
in the 
subset of bonds corresponding to $g$ of a given Ising configuration 
$\mathcal{C}$. 
Define as well 
$N^{\ }_{\textrm{F}}(\mathcal{C} \vert \overline{g})$ 
($N^{\ }_{\textrm{AF}}(\mathcal{C} \vert \overline{g})$) to be 
the number of bonds with ferromagnetic (antiferromagnetic) spin alignment
within bonds in the subset 
complementary to $g$. 
Clearly, 
$
N^{\ }_{\textrm{F/AF}}(\mathcal{C}) 
= 
N^{\ }_{\textrm{F/AF}}(\mathcal{C} \vert g) 
+ 
N^{\ }_{\textrm{F/AF}}(\mathcal{C} \vert \overline{g})
$. 

We can then rewrite Eq.~(\ref{eq: exp S g f.c.}) in the Ising language as 
\bea
&& 
2 \;
\langle 0 \vert 
  e^{\beta\lambda^{\ }_{B} P}_{\ } g 
\vert 0 \rangle 
= 
\sum^{\ }_{\mathcal{C}} \: 
  c^{N^{\ }_{\textrm{F}}(\mathcal{C} \vert \overline{g})}_{\ } 
  s^{N^{\ }_{\textrm{AF}}(\mathcal{C} \vert \overline{g})}_{\ }
  s^{N^{\ }_{\textrm{F}}(\mathcal{C} \vert g)}_{\ } 
  c^{N^{\ }_{\textrm{AF}}(\mathcal{C} \vert g)}_{\ }
\nonumber \\ 
&& \qquad 
= 
\sum^{\ }_{\mathcal{C}} \: 
  c^{N^{\ }_{\textrm{F}}(\mathcal{C})}_{\ } 
  s^{N^{\ }_{\textrm{AF}}(\mathcal{C})}_{\ }
  t^{N^{\ }_{\textrm{F}}(\mathcal{C} \vert g)}_{\ } 
  t^{-N^{\ }_{\textrm{AF}}(\mathcal{C} \vert g)}_{\ }
\nonumber \\ 
&& \qquad 
= 
c^{3N}_{\ }
\sum^{\ }_{\mathcal{C}} \: 
  e^{-2J \left( 
           N^{\ }_{\textrm{AF}}(\mathcal{C})+
	   N^{\ }_{\textrm{F}}(\mathcal{C} \vert g)-
           N^{\ }_{\textrm{AF}}(\mathcal{C} \vert g)
	 \right)}_{\ }
\nonumber \\ 
&& \qquad 
= 
c^{3N}_{\ }
\sum^{\ }_{\mathcal{C}} \: 
  \exp 
    \left[ 
      J \left( 
          \sum^{\ }_{\langle i,j \rangle} S^{\ }_{i} S^{\ }_{j} - 3N 
          - 2 \sum^{\ }_{\langle i,j \rangle \in g} 
          S^{\ }_{i} S^{\ }_{j}
        \right) 
    \right] 
\nonumber \\ 
&& \qquad 
= 
(sc)^{3N/2}_{\ }
\sum^{\ }_{\mathcal{C}} \: 
  \exp 
    \left( 
      J \sum^{\ }_{\langle i,j \rangle} 
        \eta^{\ }_{ij}(g) S^{\ }_{i} S^{\ }_{j} 
    \right) 
\nonumber 
. 
\eea

In the following, it is convenient to introduce the convention that a 
bond $\langle ij \rangle$ belongs to or is inside a partition $\A$ of the 
system ($\langle ij \rangle \in \A$) if all the spins on the corresponding 
plaquette operator belong to $\A$, and the bond does not belong or is 
outside $\A$ ($\langle ij \rangle \notin \A$) otherwise. 
(Similarly, we will refer to a cubic unit cell in or not in $\A$ if its 
six composing plaquettes are all in $\A$ or not.) 

In conclusion, the numerator of Eq.~(\ref{eq: rho(T) def f.c.}) can be mapped 
onto the partition function of a 3D random-bond Ising model on a simple 
cubic lattice, where the randomness is controlled by the choice of 
$g$. Again, summation over all possible boundary conditions is 
understood. 

Substituting Eq.~(\ref{eq: 3D Ising f.c.}) and 
Eq.~(\ref{eq: 3D Ising random f.c.}) into Eq.~(\ref{eq: rho(T) def f.c.}) 
gives 
\beq
\rho(T) 
= 
\sum_{g \in G} 
\frac{Z^{\textrm{tot}}_{J}(g)} 
     {Z^{\textrm{tot}}_{J}(\openone)} 
\;
\sum^{\ }_{\alpha} 
  \frac{e^{\beta\lambda^{\ }_{A}M^{\ }_{s}(\alpha)}_{\ }} 
       {Z^{\ }_{s}} 
\; 
    \vert\alpha\rangle \langle\alpha\vert g 
, 
\label{eq: rho(T) f.c.}
\eeq
where $J=-(1/2) \ln[\tanh(\beta\lambda^{\ }_{B})]$, 
$
Z^{\ }_{s} 
= 
\sum^{\ }_{\alpha} 
  e^{\beta\lambda^{\ }_{A}M^{\ }_{s}(\alpha)}_{\ }
$ 
is the partition function of a non-interacting Ising system in a magnetic 
field of reduced strength $\beta\lambda^{\ }_{A}$, 
and $Z^{\textrm{tot}}_{J}(\openone) \equiv Z^{\textrm{tot}}_{J}$. 

In the limit of $T \to 0$ ($\beta \to \infty$), $J \to 0^{+}_{\ }$, 
all $g$ are equally weighed, 
\beq
Z^{\textrm{tot}}_{0}(g) = Z^{\textrm{tot}}_{0}(\openone) 
\qquad 
\forall\,g \in G 
, 
\eeq
and only the states with maximal star magnetization $M^{\ }_{s}(\alpha) = N$, 
i.e., those that are eigenstates of the star operators with eigenvalue $+1$ 
everywhere, survive: 
\beq
\frac{e^{\beta\lambda^{\ }_{A}M^{\ }_{s}(\alpha)}_{\ }} 
     {Z^{\ }_{s}} 
\;\to\; 
\frac{1}{2^{3}_{\ } \, \vert G \vert} \:
\delta^{\ }_{M^{\ }_{s}(\alpha),N} 
\label{eq: T = 0 limit of alpha term}
. 
\eeq
Such states are of the form $g \vert 0^{\ }_{k} \rangle$, where 
$k=1,\,\ldots,\,2^3$ labels the states obtained from $\vert 0 \rangle$ by 
the action of the non-local $\Gamma$ operators in Eq.~(\ref{eq: topo ops z}). 
Namely, the states $\vert 0^{\ }_{k} \rangle$ are of the form 
$\Gamma^{m_1}_{1} \Gamma^{m_2}_{2} \Gamma^{m_3}_{3} \vert 0 \rangle$, 
for all possible choices of $m_1,m_2,m_3 = 0,1$. 
The factor $1 / 2^{3}_{\ } \, \vert G \vert$ in the above equation appears 
because there are precisely $2^{3}_{\ } \, \vert G \vert$ states with 
maximal star magnetization. 
Thus, one recovers the density matrix of the zero-temperature Kitaev model, 
prepared with equal probability across all topological sectors~\cite{Hamma2005b}
\bea
\rho(T=0) 
&=& 
\frac{1}{2^{3}_{\ }} 
\sum^{2^3}_{k=1} \;
\frac{1}{\vert G \vert} 
\sum_{g,g^{\prime}_{\ } \in G} 
\;
g \vert0^{\ }_{k}\rangle 
  \langle0^{\ }_{k}\vert gg^{\prime}_{\ } 
. 
\label{eq: rho(T=0) f.c.}
\eea

In the limit $T \to \infty$ ($\beta \to 0$), $J \to \infty$, 
all $g$ are exponentially suppressed except for 
$g = \openone$, while all states $\alpha$ become equally 
weighed. 
In this case one obtains the mixed-state density matrix 
\bea
\rho(T\to\infty) 
&=& 
\frac{1}{2^{3N}_{\ }} \:
\sum^{\ }_{\alpha} \; 
  \vert\alpha\rangle \langle\alpha\vert 
\label{eq: rho(T=inf) f.c.}
\eea
of a non-interacting Ising model defined on the bonds of a simple cubic 
lattice. 

Clearly from Eq.~(\ref{eq: rho(T) f.c.}), one expects something to happen 
in the system when the value of the temperature $T$, i.e., the parameter 
$J$, is such that the 3D Ising model described by $Z^{\textrm{tot}}_{J}$ 
becomes critical. 
In order to understand how this relates to the presence of topological 
order at zero temperature, we need to proceed with the calculations and 
compute the von Neumann entropy and the topological entropy as a function 
of temperature. 
%
%

\subsection{\label{sec: S_VN f.c.}
The von Neumann entropy
           }
Let us consider a generic bipartition of the original system $\mathcal{S}$ 
into subsystems $\A$ and $\B$ ($\mathcal{S} = \A \cup \B$). 
The von Neumann (entanglement) entropy of partition $\A$ is given by 
\beq
S^{\A}_{\textrm{VN}}
\equiv 
-\textrm{Tr} \left[ \rho^{\ }_{\A} \ln \rho^{\ }_{\A} \right] 
= 
- \lim^{\ }_{n \to 1} 
  \partial^{\ }_{n} \textrm{Tr} \left[ \rho^{n}_{\A} \right] 
, 
\label{eq: S_VN f.c.}
\eeq
where $\rho^{\ }_{A} = \textrm{Tr}^{\ }_{\B} \rho$ is the reduced density 
matrix obtained from the full density matrix $\rho$ by tracing out the 
degrees of freedom in subsystem $\B$. 
Similarly for $S^{\B}_{\textrm{VN}}$, and 
$S^{\A}_{\textrm{VN}} = S^{\B}_{\textrm{VN}}$ holds if $\rho$ is a pure state 
density matrix.

In order to compute the von Neumann entropy~(\ref{eq: S_VN f.c.}) from the 
finite-temperature density matrix~(\ref{eq: rho(T) f.c.}), we first obtain 
the reduced density matrix of the system using an approach similar to the 
one in Ref.~\onlinecite{Hamma2005b}, 
%
%
\bea
\rho^{\ }_{\A}(T) \!\!\! 
&=& \!\!\!
\sum_{g \in G} 
\frac{Z^{\textrm{tot}}_{J}(g)} 
     {Z^{\textrm{tot}}_{J}(\openone)} 
\, 
\sum^{\ }_{\alpha} 
\frac{e^{\beta\lambda^{\ }_{A}M^{\ }_{s}(\alpha)}_{\ }}
     {Z^{\ }_{s}} 
\; 
\vert \alpha^{\ }_{\A} \rangle 
  \langle \alpha^{\ }_{\A} \vert g^{\ }_{\A} 
\;
\langle \alpha^{\ }_{\B} \vert 
  g^{\ }_{\B} 
\vert \alpha^{\ }_{\B} \rangle 
\nonumber \\ 
&=& \!\!\!
\sum_{g \in G^{\ }_{\A}} 
\frac{Z^{\textrm{tot}}_{J}(g)} 
     {Z^{\textrm{tot}}_{J}(\openone)} 
\, 
\sum^{\ }_{\alpha} 
\frac{e^{\beta\lambda^{\ }_{A}M^{\ }_{s}(\alpha)}_{\ }}
     {Z^{\ }_{s}} 
\; 
\vert \alpha^{\ }_{\A} \rangle 
  \langle \alpha^{\ }_{\A} \vert g^{\ }_{\A} 
, 
\label{eq: rho_A(T) f.c.}
\eea
%
%
where we used the generic tensor decomposition 
$
\vert \alpha \rangle 
= 
\vert \alpha^{\ }_{\A} \rangle 
\otimes 
\vert \alpha^{\ }_{\B} \rangle
$, 
$g = g^{\ }_{\A} \otimes g^{\ }_{\B}$, 
and the fact that 
$
\langle \alpha^{\ }_{\B} \vert 
  g^{\ }_{\B} 
\vert \alpha^{\ }_{\B} \rangle
= 
1
$ 
if $g^{\ }_{\B} = \openone^{\ }_{\B}$ and zero otherwise. 
As in the previous section, we denoted by 
$
G^{\ }_{\A} 
= 
\{ 
g \in G 
\;\vert\; 
g^{\ }_{\B} = \openone^{\ }_{\B}
\}
$ 
the subgroup of $G$ given by all operations $g$ that act trivially on $\B$ 
(similarly for $G^{\ }_{\B}$). 

Notice that a plaquette operator $B^{\ }_{p}$ can either act solely on spins 
in partition $\A$ (represented in the following by the notation $p \in \A$), 
solely on spins in partition $\B$ ($p \in \B$), or simultaneously on spins 
belonging to $\A$ and $\B$ (which we will refer to as 
\emph{boundary plaquette operators}, and represent by $p \in \A\B$). 

Recall from Sec.~\ref{sec: zero temperature gauge} that a complete set of 
generators for the subgroup $G^{\ }_{\A}$ can be constructed by taking: 
(i) All plaquette operators that act solely on $\A$, i.e., 
$\{ B^{\ }_{p} \;\vert\; p \in \A\}$ 
($\N^{(p)}_{\A} = \vert\{ B^{\ }_{p} \;\vert\; p \in \A\}\vert$). 
(ii) All possible (independent) collective operators constructed from 
plaquettes in $\B$ and at the boundary, but acting solely on $\A$; 
as illustrated in Sec.~\ref{sec: zero temperature gauge}, the number of such 
collective operators equals the number 
$n^{\ }_{\A}$ of non-contractible loops in subsystem $\A$. 
And by (iii) accounting for all constraints given by the independent closed 
membranes in $\A$. That is,  all $\N^{(c)}_{\A}$ cubic unit cells in $\A$, 
all possible $(m^{\ }_{\B} - 1)$ additional closed membranes if $\B$ is 
disconnected, and all independent entire crystal planes inside $\A$ 
($m^{\textrm{(c.p.)}}_{\A} = 0,1,2,3$). 
Again, for all bipartitions of interest in our study 
$m^{\textrm{(c.p.)}}_{\A} = 0$ and $m^{\textrm{(c.p.)}}_{\B} = 3$, 
and for simplicity we will restrict to this specific case. 

The cardinalities of the subgroups $G^{\ }_{\A}$  and $G^{\ }_{\B}$ are 
thus given by 
\begin{subequations}
\bea
d^{\ }_{\A} 
&\equiv& 
\vert G^{\ }_{\A} \vert 
= 
2^{\N^{(p)}_{\A} 
   - \N^{(c)}_{\A} 
   + n^{\ }_{\A}  
   - (m^{\ }_{\B} - 1)}_{\ }
\\ 
d^{\ }_{\B} 
&\equiv& 
\vert G^{\ }_{\B} \vert 
= 
2^{\N^{(p)}_{\B} 
   - \N^{(c)}_{\B} 
   + n^{\ }_{\B}  
   - (m^{\ }_{\A} - 1) 
   - 3}_{\ }
. 
\eea
\end{subequations}
In particular, $n^{\ }_{\A} = n^{\ }_{\B} = 1$ in bipartitions 4,5 and zero 
otherwise; and $m^{\ }_{\A} = 2$ in bipartition 8, $m^{\ }_{\B} = 2$ in 
bipartition 1, and they equal 1 in all other cases. 

Let us then use Eq.~(\ref{eq: rho_A(T) f.c.}) to compute the trace of the 
$n$-th power of $\rho^{\ }_{\A}(T)$: 
\begin{widetext}
\bea
\textrm{Tr} \left[ \rho^{n}_{\A}(T) \right] 
&=& 
\!\!\!\!\!\!
\sum_{g^{\ }_{1},\,\ldots,\,g^{\ }_{n}\in G^{\ }_{\A}} 
\left( 
\prod^{n}_{l=1} \: 
  \frac{Z^{\textrm{tot}}_{J}(g^{\ }_{l})} 
       {Z^{\textrm{tot}}_{J}(\openone)} 
\right) 
\sum^{\ }_{\alpha^{\ }_{1},\,\ldots,\,\alpha^{\ }_{n}} 
\left( 
\prod^{n}_{l=1} \: 
  \frac{e^{\beta\lambda^{\ }_{A}M^{\ }_{s}(\alpha^{\ }_{l})}_{\ }} 
       {Z^{\ }_{s}} 
\right) 
\langle \alpha^{\ }_{1,\A} \vert 
    g^{\ }_{1,\A} 
  \vert \alpha^{\ }_{2,\A} \rangle 
  \langle \alpha^{\ }_{2,\A} \vert 
    g^{\ }_{2,\A} 
  \vert \alpha^{\ }_{3,\A} \rangle 
\ldots 
  \langle \alpha^{\ }_{n,\A} \vert 
    g^{\ }_{n,\A} 
\vert \alpha^{\ }_{1,\A} \rangle 
. 
\nonumber \\ 
\label{eq: Tr(rho^n_A(T)) f.c. -- 0}
\eea
\end{widetext}
Each expectation value above imposes that the two configurations 
$\alpha^{\ }_{l+1}$ and $\alpha^{\ }_{l}$, $l=1,\ldots,n$ (with the 
identification $n+1 \equiv 1$), can be mapped 
one onto the other over subsystem $\A$, via the plaquette flipping 
operation $g^{\ }_{l,\A}$. 
This is possible only if the set 
$g^{\ }_{1},\,\ldots,\,g^{\ }_{n}\in G^{\ }_{\A}$ satisfies the condition 
$\prod^{n}_{l=1}\,g^{\ }_{l,\A} = \openone^{\ }_{\A}$, i.e., 
$\prod^{n}_{l=1}\,g^{\ }_{l} = \openone$. 
Therefore, we can decompose each element $g^{\ }_{l}$ into a product 
$g^{\ }_{l} = \tilde{g}^{\ }_{l} \tilde{g}^{\ }_{l+1}$, 
where $\tilde{g}^{\ }_{l} \in G^{\ }_{\A}$, 
$l = 1,\,\ldots,\,n$ with periodic boundary conditions $n+1 \equiv 1$ 
(the fact that this decomposition is highly non-unique is immaterial to the 
calculations below): 
\begin{widetext}
\bea
\textrm{Tr} \left[ \rho^{n}_{\A}(T) \right] 
&=& 
\!\!\!\!\!\!
\sum_{g^{\ }_{1},\,\ldots,\,g^{\ }_{n}\in G^{\ }_{\A}} 
\left( 
\prod^{n}_{l=1} \: 
  \frac{Z^{\textrm{tot}}_{J}(g^{\ }_{l})} 
       {Z^{\textrm{tot}}_{J}(\openone)} 
\right) 
\sum^{\ }_{\alpha^{\ }_{1},\,\ldots,\,\alpha^{\ }_{n}} 
\left( 
\prod^{n}_{l=1} \: 
  \frac{e^{\beta\lambda^{\ }_{A}M^{\ }_{s}(\alpha^{\ }_{l})}_{\ }} 
       {Z^{\ }_{s}} 
\right) 
\nonumber \\ 
&\times& \vphantom{\sum^\sum}
\langle 0 \vert 
  \prod^{n}_{l=1} g^{\ }_{l} 
\vert 0 \rangle 
\:
\langle \alpha^{\ }_{1,\A} \vert 
    \tilde{g}^{\ }_{1,\A} \tilde{g}^{\ }_{2,\A} 
  \vert \alpha^{\ }_{2,\A} \rangle 
  \langle \alpha^{\ }_{2,\A} \vert 
    \tilde{g}^{\ }_{2,\A} \tilde{g}^{\ }_{3,\A} 
  \vert \alpha^{\ }_{3,\A} \rangle 
\ldots 
  \langle \alpha^{\ }_{n,\A} \vert 
    \tilde{g}^{\ }_{n,\A} \tilde{g}^{\ }_{1,\A} 
\vert \alpha^{\ }_{1,\A} \rangle 
\nonumber \\ 
&=& 
\!\!\!\!\!\!
\sum_{g^{\ }_{1},\,\ldots,\,g^{\ }_{n}\in G^{\ }_{\A}} 
\left( 
\prod^{n}_{l=1} \: 
  \frac{Z^{\textrm{tot}}_{J}(g^{\ }_{l})} 
       {Z^{\textrm{tot}}_{J}(\openone)} 
\right) 
\sum^{\ }_{\alpha^{\ }_{1},\,\ldots,\,\alpha^{\ }_{n}} 
\left( 
\prod^{n}_{l=1} \: 
  \frac{e^{\beta\lambda^{\ }_{A}M^{\ }_{s}(\alpha^{\ }_{l})}_{\ }} 
       {Z^{\ }_{s}} 
\right) 
\langle 0 \vert 
  \prod^{n}_{l=1} g^{\ }_{l} 
\vert 0 \rangle 
\: 
\langle \alpha^{\ }_{1,\A} \vert \alpha^{\ }_{2,\A} \rangle 
  \langle \alpha^{\ }_{2,\A} \vert \alpha^{\ }_{3,\A} \rangle 
\ldots 
  \langle \alpha^{\ }_{n,\A} \vert \alpha^{\ }_{1,\A} \rangle 
, 
\nonumber \\ 
\label{eq: Tr(rho^n_A(T)) f.c. -- 0bis}
\eea
\end{widetext}
where we used the fact that the magnetization $M^{\ }_{s}(\alpha)$ 
of state $\vert \alpha \rangle$ is the same as 
$M^{\ }_{s}(g \alpha)$ of state 
$g \vert \alpha \rangle$, for any 
$g \in G$, to do away with the $\tilde{g}^{\ }_{l}$ via relabeling of 
the states 
$
\vert \alpha^{\ }_{l} \rangle 
\to 
\tilde{g}^{\ }_{l} 
\vert \alpha^{\ }_{l} \rangle
$. 

We can further simplify the notation by defining the function 
$
\delta\left( \alpha^{\ }_{\A}, \beta^{\ }_{\A} \right)
= 
\langle \alpha^{\ }_{\A} \vert \beta^{\ }_{\A} \rangle
$, 
and the above equation can be rewritten as 
\begin{widetext}
\bea
\textrm{Tr} \left[ \rho^{n}_{\A}(T) \right] 
&=& 
\sum_{g^{\ }_{1},\,\ldots,\,g^{\ }_{n}\in G^{\ }_{\A}} 
\left( 
\prod^{n}_{l=1} \: 
  \frac{Z^{\textrm{tot}}_{J}(g^{\ }_{l})} 
       {Z^{\textrm{tot}}_{J}(\openone)} 
\right) 
\langle 0 \vert 
  \prod^{n}_{l=1} g^{\ }_{l} 
\vert 0 \rangle 
\: \times \!\!\!\!\!
\sum^{\ }_{\alpha^{\ }_{1},\,\ldots,\,\alpha^{\ }_{n}} 
\left( 
\prod^{n}_{l=1} \: 
  \frac{e^{\beta\lambda^{\ }_{A}M^{\ }_{s}(\alpha^{\ }_{l})}_{\ }} 
       {Z^{\ }_{s}} 
\right) 
\prod^{n-1}_{l=1} 
  \delta \left( \alpha^{\ }_{l,\A}, \alpha^{\ }_{l+1,\A} \right) 
\nonumber \\ 
&=& 
\vphantom{\sum^\sum}
\mathcal{Z}^{(P)}(n) \times \mathcal{Z}^{(S)}(n) 
. 
\label{eq: Tr(rho^n_A(T)) f.c. -- 1}
\eea
\end{widetext}
Notice that the product 
$
\prod^{n-1}_{l=1} 
  \delta ( \alpha^{\ }_{l,\A}, \alpha^{\ }_{l+1,\A} )
$
implies 
$\delta ( \alpha^{\ }_{1,\A}, \alpha^{\ }_{n,\A} )$, 
which is therefore redundant and has been omitted in the previous equation. 
In the notation of Eq.~(\ref{eq: Tr(rho^n_A(T)) f.c. -- 1}), it becomes 
evident that the \emph{star} ($S$) contribution, i.e., involving only the star 
coupling constant $\lambda^{\ }_{A}$, and the \emph{plaquette} ($P$) 
contribution, i.e., involving only the plaquette coupling constant 
$\lambda^{\ }_{B}$, decouple and factorize into two separate terms, 
$\mathcal{Z}^{(S)}(n)$ and $\mathcal{Z}^{(P)}(n)$. 
In particular, $\mathcal{Z}^{(P)}(n=1) = \mathcal{Z}^{(S)}(n=1) = 1$. 

Using the replica trick, we can compute the von Neumann entropy:
\bea
S^{\ }_{\textrm{VN}}(\A;T) 
&=& 
- \lim^{\ }_{n \to 1} 
  \partial^{\ }_{n} \textrm{Tr} \left[ \rho^{n}_{\A} \right] 
= 
- \lim^{\ }_{n \to 1} 
  \partial^{\ }_{n} \left[ \mathcal{Z}^{(P)}(n) \; \mathcal{Z}^{(S)}(n) \right] 
\nonumber \\ 
&=& 
- \mathcal{Z}^{(S)}(1) \lim^{\ }_{n \to 1} 
  \partial^{\ }_{n} \mathcal{Z}^{(P)}(n) 
- \mathcal{Z}^{(P)}(1) \lim^{\ }_{n \to 1} 
  \partial^{\ }_{n} \mathcal{Z}^{(S)}(n) 
\nonumber \\ 
&=& 
- \lim^{\ }_{n \to 1} 
  \partial^{\ }_{n} \mathcal{Z}^{(P)}(n) 
- \lim^{\ }_{n \to 1} 
  \partial^{\ }_{n} \mathcal{Z}^{(S)}(n) 
\nonumber \\ 
&=& 
S^{(P)}_{\textrm{VN}}(\A ; T/\lambda^{\ }_{B}) 
+ 
S^{(S)}_{\textrm{VN}}(\A ; T/\lambda^{\ }_{A}) 
. 
\label{eq: p+s decomposition}
\eea
Thus, from the factorizability in Eq.~(\ref{eq: Tr(rho^n_A(T)) f.c. -- 1}) 
above, it follows that the von Neumann entropy has two
additive contributions from the star and plaquette terms
that can then be computed separately.~\cite{footnote:separability} 

One can check that Eqs.~(\ref{eq: Tr(rho^n_A(T)) f.c. -- 1}) 
and~(\ref{eq: p+s decomposition}) satisfy indeed 
the $T \to 0$ limit discussed in Sec.~\ref{sec: zero temperature gauge}, as 
well as the known $T \to \infty$ limit 
(see Appendix~\ref{app: checks against known limits}). 

Notice that, although in this
paper we are concerned with 3D systems, the derivation is independent of the 
dimensionality, and this result holds true for
${\mathbb Z}_2$ models in {\it any number of dimensions}. 

Because the von Neumann entropy is separable as the sum of the two
independent contributions from star and plaquette terms, so is the
topological entropy, which is a linear combination of the entanglement 
entropies for the partitions shown in Fig.~\ref{fig: bipartition schemes}: 
\bea
S^{\ }_{\textrm{topo}}(T) 
&=& 
S^{(S)}_{\textrm{topo}}(T/\lambda^{\ }_{A}) 
+
S^{(P)}_{\textrm{topo}}(T/\lambda^{\ }_{B}) 
.
\eea

We now turn to the separate analysis of the two contributions. 
%
%

\subsubsection{\label{sec: S^S_topo}
The star contribution $S^{(S)}_{\textrm{topo}}(T/\lambda^{\ }_{A})$
              }
The computation of this contribution is very similar to the one 
in Ref.~\cite{Castelnovo2007} for the 2D Kitaev model, where the limit 
$\lambda_B\to\infty$ was explicitly considered. 

In order to illustrate this analogy, let us define the following entropy 
differentials: 
\bea
\Delta S^{\ }_{\textrm{VN}}(\A;T) 
&\equiv&
S^{\ }_{\textrm{VN}}(\A;T) - S^{\ }_{\textrm{VN}}(\A;0)
\nonumber \\
&=& 
\Delta S^{(S)}_{\textrm{VN}}(\A;T/\lambda^{\ }_{A}) 
+
\Delta S^{(P)}_{\textrm{VN}}(\A;T/\lambda^{\ }_{B}) 
,
\nonumber \\ 
\eea
and 
\bea
\Delta S^{\ }_{\textrm{topo}}(T) 
&\equiv&
S^{\ }_{\textrm{topo}}(T) - S^{\ }_{\textrm{topo}}(0)
\nonumber \\
&=& 
\Delta S^{(S)}_{\textrm{topo}}(T/\lambda^{\ }_{A}) 
+
\Delta S^{(P)}_{\textrm{topo}}(T/\lambda^{\ }_{B}) 
,
\eea
where 
%
%
%
\bea
\Delta S^{(S)}_{\textrm{VN}}(\A;T/\lambda^{\ }_{A}) 
&\equiv&
S^{(S)}_{\textrm{VN}}(\A;T/\lambda^{\ }_{A}) 
-
S^{(S)}_{\textrm{VN}}(\A;0) 
\nonumber \\ 
\Delta S^{(S)}_{\textrm{topo}}(T/\lambda^{\ }_{A}) 
&\equiv&
S^{(S)}_{\textrm{topo}}(T/\lambda^{\ }_{A}) 
-
S^{(S)}_{\textrm{topo}}(0) 
,
\nonumber 
\eea
and
\bea
\Delta S^{(P)}_{\textrm{VN}}(\A;T/\lambda^{\ }_{B}) 
&\equiv&
S^{(P)}_{\textrm{VN}}(\A;T/\lambda^{\ }_{B}) 
-
S^{(P)}_{\textrm{VN}}(\A;0) 
\nonumber \\ 
\Delta S^{(P)}_{\textrm{topo}}(T/\lambda^{\ }_{B}) 
&\equiv&
S^{(P)}_{\textrm{topo}}(T/\lambda^{\ }_{B}) 
-
S^{(P)}_{\textrm{topo}}(0) 
.
\nonumber 
\eea
%
%
%

Notice that for $\lambda_B\to\infty$, 
$\Delta S^{(P)}_{\textrm{VN}}(\A;T/\lambda^{\ }_{B}) =0$ 
and 
$\Delta S^{(P)}_{\textrm{topo}}(T/\lambda^{\ }_{B}) =0$. 
Thus, one obtains that
\bea
\label{S_VN-lambdaB-infty}
\Delta S^{(S)}_{\textrm{VN}}(\A;T/\lambda^{\ }_{A}) 
&=&
\Delta S^{\ }_{\textrm{VN}}(\A;T) \Big|_{\lambda_B\to\infty}
\\ 
\label{S_topo-lambdaB-infty}
\Delta S^{(S)}_{\textrm{topo}}(T/\lambda^{\ }_{A}) 
&=&
\Delta S^{\ }_{\textrm{topo}}(T) \Big|_{\lambda_B\to\infty}
.
\eea

Moreover, in the limit $\lambda_B\to\infty$ and choosing to work in the 
$\sigma^{z}_{\ }$ basis, one can show that both the 
group structure of $G$ and the collective operations in $G^{\ }_{\A}$ are 
very much the same in 2D and in 3D. 
For example, the group $G$ is generated by all but one star operators, and 
the subgroup $G^{\ }_{\A}$ is generated by all star operators in $\A$ 
with the addition of all but one collective operations that obtain as 
products of star operators belonging to each component of $\B$ times the ones 
along the corresponding boundary. 
As a result, the topologically non-trivial bipartitions 1 and 4 in 2D 
correspond to bipartitions 1 and 8 in 3D. 
All calculations generalize straightforwardly to 3D, and one 
can derive the expressions for $\Delta S^{(S)}_{\textrm{VN}}$ and for 
$\Delta S^{(S)}_{\textrm{topo}}$ in a finite system at finite 
temperature. 
The actual values for $S^{(S)}_{\textrm{VN}}$ and $S^{(S)}_{\textrm{topo}}$ 
are then fixed by matching, say, the known $T \to 0$ limits. 

{}From the 2D results in Ref.~\cite{Castelnovo2007}, we infer that the star 
contribution to the 3D topological entropy is fragile, in the sense that it 
vanishes in the thermodynamic limit at any finite temperature. 
Namely, the behavior is singular in that the limits of $T\to 0$ and infinite 
size do not commute. 
If the thermodynamic limit is taken first, 
\begin{equation}
\label{S^S_topo-thermo-limit}
\Delta S^{(S)}_{\textrm{topo}}(T/\lambda^{\ }_{A}) 
=
\begin{cases}
0 & \text{$T=0$} \\
-\ln 2 & \text{$T >0$}.
\end{cases}
\end{equation}
Thus, in the thermodynamic limit, the star contribution to the topological 
entropy evaporates at any infinitesimal temperature. 

(The finite temperature and finite size expressions for the star 
contributions to the von Neumann and topological entropies are shown in 
Appendix~\ref{app: S^S_topo}.) 
%
%

\subsubsection{\label{sec: S^P_topo} 
The plaquette contribution $S^{(P)}_{\textrm{VN}}(\A ; T/\lambda^{\ }_{B})$
              }
Similarly to the above, one obtains for the plaquette contributions
\bea
\Delta S^{(P)}_{\textrm{VN}}(\A;T/\lambda^{\ }_{B}) 
&=&
\Delta S^{\ }_{\textrm{VN}}(\A;T) \Big|_{\lambda_A\to\infty}
\\ 
\Delta S^{(P)}_{\textrm{topo}}(T/\lambda^{\ }_{B}) 
&=&
\Delta S^{\ }_{\textrm{topo}}(T) \Big|_{\lambda_A\to\infty}
.
\eea
Because of the very different nature of the 2D and 3D group structures
when using the $\sigma^{x}_{\ }$ basis, the computation of
the plaquette contribution in 3D is not a trivial extension of that in
2D, and it thus requires some work.  The calculations are shown in
detail in Appendix~\ref{app: S^P_topo}, while only the
results are summarized here for conciseness and clarity. 

The behavior of $\Delta S^{(P)}_{\textrm{topo}}(T/\lambda^{\ }_{B})$ 
as a function of temperature, in the thermodynamic limit, is 
\begin{equation}
\Delta S^{(P)}_{\textrm{topo}}(T/\lambda^{\ }_{B}) =
\begin{cases}
0 & \text{$T < T_c$} \\
-\ln 2 & \text{$T > T_c$},
\end{cases}
\label{eq:S_P-result}
\end{equation}
where the critical temperature is associated with a 3D Ising transition 
and can be located at $T_c = 1.313346(3) \lambda_B$. 
%
%

\section{\label{sec: discussion}
Discussion
        }
We can now put all the pieces together, and argue for the persistence of
topological order at finite temperatures in the 3D Kitaev model. 
Adding the contributions from stars and plaquettes, 
which we have shown to be exactly separable, 
the topological entropy of the system is 
\begin{equation}
S^{\rm 3D}_{\textrm{topo}}(T) =
\begin{cases}
2\ln 2 & \text{$T=0$} \\
\ln 2 & \text{$0< T < T_c$} \\
0 & \text{$T > T_c$} 
. 
\end{cases}
\end{equation}
This is to be contrasted to the 2D case,~\cite{Castelnovo2007} 
\begin{equation}
S^{\rm 2D}_{\textrm{topo}}(T) =
\begin{cases}
2\ln 2 & \text{$T=0$} \\
0 & \text{$T>0 $} 
, 
\end{cases}
\end{equation}
where the topological order is fragile, subsiding for any 
finite $T$ (when the thermodynamic limit is taken first). 

In 3D the order survives up to a transition temperature that is
determined by the coupling constant $\lambda_B$ associated with the
plaquette degrees of freedom alone. The topological order in the
system, as measured by the topological entropy, is thus the same as 
in the case where $\lambda_A = 0$, that is, in a purely classical model. 
In this sense, the order at finite $T$ is classical in 
origin.~\cite{footnote: gauge order parameter} 

Our results show that the extension of the notion of topological
order to classical systems applies beyond the hard constrained
limit already discussed in Ref.~\onlinecite{Castelnovo2006} in two 
dimensions. In the 3D example discussed here, the order persists for 
non-infinite couplings $\lambda_A$, $\lambda_B$. 

Having obtained the result that topological order in the 3D toric code 
survives thermal fluctuations, in a classical sense, up to a finite critical 
temperature, we now turn to a discussion of what this type of order implies. 

At zero temperature, topological sectors can be discerned according to
the eigenvalues $I_\alpha=\pm 1$ of the loop operators $\Gamma^{\
}_{\alpha}$, where $\alpha=1,2,3$, as in Eq.~(\ref{eq: topo ops z}). 
The eight ground states $|I\rangle$ in the different topological
sectors can be labeled by integers $I=0,\dots, 2^3-1$ (made up of
three bits, $I\equiv I_1 I_2 I_3$, $I_\alpha=0,1$).

Suppose to prepare, at an initial time $t=t_i$, a superposition of
states
\bea
\vert \Psi(t_i) \rangle 
&=& 
\sum_{I=0}^{2^3-1}
\sqrt{p_I}\;\vert I \rangle 
\;,
\label{eq:super-p-bit}
\eea
then raise the temperature to some value $0<T<T_c$, and bring it
back to $T=0$ at some time $t_f$. The final $T=0$ state will again be,
assuming thermodynamic equilibrium is reached, a superposition of the eight 
topologically degenerate ground states. 

Following the discussion in Section~\ref{sec: finite temperature},
for temperatures below $T_c$, one can take a winding loop and deform
it past thermal defects, and read off the same eigenvalue of the
topological operator as the path is deformed. The information stored
in all winding loops that {\it do not} cross a thermal defect does not
disappear, as long as there is a way to pass a winding loop that
avoids defects. Therefore, as long as the system temperature is not
raised above $T_c$, upon returning to $T=0$ at $t_f$, the system should 
return to the same topological sector that it was originally prepared in 
at time $t_i$. 

Thus, the state at $t_f$ is a superposition
\bea
\vert \Psi(t_f) \rangle 
&=& 
\sum_{I=0}^{2^3-1}
\sqrt{p_I}\;e^{i\varphi_I}\;\vert I \rangle 
\;,
\label{eq:super-p-bit2}
\eea
where phases $\varphi_I$ are accumulated during the thermal
cycle. These phases, unless locked together by some specific
mechanism, shall be randomized by the thermal bath. However, the
magnitude of the amplitudes remains $\sqrt{p_I}$, for
$I=0,\dots,2^3-1$, as there have been no transitions between different 
topological sectors, if the system was never heated above $T_c$.

Hence, the only (accessible) information preserved under the time evolution 
from $t_i$ to $t_f$ is that the relative probabilities
of find the state in sector $I$ equals $p_I$. The
state in Eq.~(\ref{eq:super-p-bit2}) realizes a pbit, or probabilistic
bit.~\cite{pbit refs} It is not a qubit because of the thermal dephasing 
between the states $\vert I \rangle$. Although still a 
quantum superposition of a sort, in that it has probability $p_I$ of being in
sector $I$, it cannot be told apart by any type of measurement from a
classical probabilistic system with the same probabilities $p_I$. The
stability of the system against local measurements only tells us that
the state is not projected onto a sector {\it until} a non-local
measurement is carried out. This effect is a {\it non-measurable}
difference between the state in Eq.~(\ref{eq:super-p-bit}) and a
classical probabilistic state: whether the projection occurs {\it
before} (as in the classical state) or {\it after} (as in the pbit)
the measurement is not detectable. 
%
%

\section{\label{sec: conclusions}
Conclusions
        }
In this paper we have shown that topological order exists in the 3D
toric code at finite temperatures, up to a critical temperature
$T_c = 1.313346(3) \lambda_B$ which
is set by one of the couplings (that associated to the plaquette terms
in the Hamiltonian). This is in sharp contrast to what happens in the
2D toric code, where in the thermodynamic limit the order subsides for
any infinitesimal temperature.

We first presented simple heuristic arguments for this result. These
arguments are based on the observation that eigenvalues of operators
defined as products of spin operators along winding loops can be used
to determine the order even in the presence of (thermally activated)
local defects, because loops can be deformed around such obstacles in
3D, leaving unchanged the eigenvalues of such loop operators. This is
to be contrasted to the 2D case, where one cannot move a loop around a
point, and thus the eigenvalues of non-local loop operators are
unequal on opposite sides of the point defect.

We subsequently substantiated the heuristic arguments by means of an
exact calculation of the von Neumann and topological entropies in the
system as a function of temperature. In carrying out this exact
calculation, we derived a generic result that applies to toric codes
defined in any number of spatial dimensions: that the von Neumann
entropy is separable as a sum of two terms, one associated with stars
alone (and a function of the dimensionless ratio $T/\lambda_A$) and
another associated with plaquettes alone (and a function of the
dimensionless ratio $T/\lambda_B$). The same separability follows
naturally for the topological entropy, $S^{\ }_{\textrm{topo}}(T) =
S^{(S)}_{\textrm{topo}}(T/\lambda^{\ }_{A}) +
S^{(P)}_{\textrm{topo}}(T/\lambda^{\ }_{B}) $. We then showed that, in
the thermodynamic limit, the star contribution
$S^{(S)}_{\textrm{topo}}(T/\lambda^{\ }_{A})$ vanishes for any $T\ne
0$, while the plaquette contribution
$S^{(P)}_{\textrm{topo}}(T/\lambda^{\ }_{B}) $ remains constant for
$T/\lambda_B < 1.313346(3)$, and vanishes for temperatures above this 
scale. 

Because the critical temperature is set by $\lambda_B$ and not
$\lambda_A$, one can argue that the topological entropy remains
non-zero when $\lambda_A\to 0$. The resulting Hamiltonian is purely
classical, and thus one can argue that the nature of the finite $T$
topological order must be classical as well.

Finally, we discussed the nature of the information that can be stored
robustly in the system because of the topological order at finite
$T$. We argued that the resilient information stored in the 3D system
realizes a pbit. 

We end with a note on an interesting situation that should occur in
systems where both $\mathbb{Z}_2$ gauge defect types are confined. In
3D only one of the defect types is confined, the topological entropy
drops from $2\ln 2$ at $T=0$ to $\ln 2$ for $0<T<T_c$, and only the
probabilities of being in a given topological sector are preserved
(magnitude square of the amplitudes, but not the relative phases). 
If instead both defect types are confined, the notion of sectors in both 
the $\sigma^x$ and $\sigma^z$ basis is retained, and this implies (as
discussed briefly in Appendix~\ref{app: confined-confined}) that, if
the system is prepared in a given superposition at zero temperature
and its temperature is raised and again lowered to zero without ever
going above $T_c$, the system returns to the same original quantum
state (a ``boomerang'' effect). 
%
%

\section*{Acknowledgments}

We are indebted to Xiao-Gang Wen for attracting our attention towards the 
possibility of a finite-temperature topological phase transition in the 
three dimensional Kitaev model, and to Michael Levin, John Cardy, 
Eduardo Fradkin, and Roderich Moessner for several insightful discussions. 
This work was supported in part by 
EPSRC Grant No. GR/R83712/01 (C.~Castelnovo). 
%
%

\appendix

\section{\label{app: confined-confined}
The confined-confined case
        }
In this appendix, we briefly discuss how the nature of the topological 
protection at finite temperature changes when both types of thermal defects 
in a $\mathbb{Z}_2$ gauge theory are confined at low temperature ($T < T_c$). 

For concreteness and simplicity, let us consider a modification of the 2D 
toric code, where some \emph{ad hoc} energy terms have been introduced 
that confine both electric and magnetic thermal defects (without inquiring 
on the nature of these terms. As mentioned in 
Sec.~\ref{sec: finite temperature}, this scenario should be realized in the 
4D case without need of any additional term). 

The $T=0$ ground state (GS) wavefunction in a given topological sector is 
uniquely specified by the ($\pm$) eigenvalues of two independent Wilson toric 
cycles, i.e., winding loop operators. In the $\sigma^z$ basis, it is 
sufficient to consider the product of all $\hat{\sigma}^z_i$ operators along a 
horizontal ($\hat{\T}^z_h$) and a vertical ($\hat{\T}^z_v$) winding loop, 
respectively. 
Similarly, in the $\sigma^x$ basis, using loop operators in the dual lattice, 
$\hat{\T}^x_h$ and $\hat{\T}^x_v$. These loop operators satisfy the algebra 
$\{\hat{\T}^x_h,\hat{\T}^z_v\}=0$ and $\{\hat{\T}^x_v,\hat{\T}^z_h\}=0$.

Let us choose to work in the $\sigma^z$ basis, and define $\vert a,b \rangle$, 
$a = \pm$, to be the normalized GS wavefunctions that are also eigenvectors 
of $\hat{\T}^z_h$ and $\hat{\T}^z_v$, 
\bea
\hat{\T}^z_h \: \vert a,b \rangle = a \, \vert a,b \rangle
\nonumber \\ 
\hat{\T}^z_v \: \vert a,b \rangle = b \, \vert a,b \rangle
. 
\nonumber
\eea
Let us prepare the system in a given superposition of such basis states, 
\beq
\vert \Psi_\textrm{in} \rangle 
= 
\sum_{a,b = \pm} \psi_{a,b} \, \vert a,b \rangle 
, 
\eeq
where $\sum_{a,b = \pm} \vert \psi_{a,b} \vert^2 = 1$, and consider 
coupling the system to a thermal bath so that the temperature can be varied 
from $T_\textrm{in} = 0$, via $0 < T < T_c$, back to $T_\textrm{fi} = 0$, 
as discussed in Sec.~\ref{sec: finite temperature}.

Trivially, the final state of the system must again be a ground state, and 
therefore it can be written as 
\beq
\vert \Psi_\textrm{fi} \rangle 
= 
\sum_{a,b = \pm} \tilde{\psi}_{a,b} \, \vert a,b \rangle 
. 
\eeq
Moreover, so long as the temperature was never raise beyond the deconfining 
transition at $T_c$, the coupling to the thermal bath cannot have transferred 
any amplitude between any of the topological sectors. 
Hence the following topological quantities must be conserved: 
\bea
\langle \Psi_\textrm{in} \vert
\hat{\T}^{z/x}_{h/v} 
\vert \Psi_\textrm{in} \rangle 
= 
\langle \Psi_\textrm{fi} \vert
\hat{\T}^{z/x}_{h/v} 
\vert \Psi_\textrm{fi} \rangle 
. 
\label{eq: conf-conf conservation rules}
\eea

For simplicity, consider the case where 
\bea
\psi_{+,+} 
&=& 
\cos (\theta / 2) 
\nonumber \\ 
\psi_{-,+} 
&=& 
\sin (\theta / 2) \: e^{i \phi} 
, 
\nonumber
\eea
where $\theta \in (0,\pi)$ and $\phi \in (-\pi, \pi)$, and all others vanish. 
After a little algebra, one can show that the conditions in 
Eq.~(\ref{eq: conf-conf conservation rules}) require that the only 
non-vanishing terms in the final GS wavefunction are 
\bea
\tilde{\psi}_{+,+} 
&=& 
\cos (\tilde{\theta} / 2) 
\nonumber \\ 
\tilde{\psi}_{-,+} 
&=& 
\sin (\tilde{\theta} / 2) \: e^{i \tilde{\phi}} 
, 
\nonumber 
\eea
and they satisfy the relations 
\bea
\cos (\theta) &=& \cos (\tilde{\theta}) 
\\ 
\sin (\theta) \cos (\phi) &=& \sin (\tilde{\theta}) \cos (\tilde{\phi}) 
. 
\eea
That is, $\theta = \tilde{\theta}$ and $\phi = \pm \tilde{\phi}$. 

The ambiguity in the sign of $\phi$ is immediately resolved if we further 
require, as expected below $T_c$, that also the expectation values of the 
products $i\,\hat{\T}^{z}_{h} \, \hat{\T}^{x}_{v}$ and
$i\,\hat{\T}^{z}_{v} \, \hat{\T}^{x}_{h}$ are conserved, leading to the 
relation 
\bea
\sin (\theta) \sin (\phi) &=& \sin (\tilde{\theta}) \sin (\tilde{\phi}) 
. 
\eea

Therefore, the quantum topological order in this system is fully protected 
from thermal fluctuations, so long as $T < T_c$, in the sense that the 
system is bound to come back to the same exact initial state upon cooling 
back to zero temperature. 
%
%

\section{\label{app: checks against known limits}
Check against known limits
        }
As a check of the steps leading to Eq.~(\ref{eq: Tr(rho^n_A(T)) f.c. -- 1}) 
and~(\ref{eq: p+s decomposition}), let us verify that the known 
limits are indeed recovered. 

For $T=0$ (i.e., for $J=0$) we have that 
$ 
e^{\beta\lambda^{\ }_{A} M^{\ }_{s}(\alpha^{\ }_{l})}_{\ } 
\,/\, 
Z^{\ }_{s} 
\to 
\delta^{\ }_{M^{\ }_{s}(\alpha^{\ }_{l}),N} \,/\, 2^3 \vert G \vert
$, 
while 
$ 
Z^{\textrm{tot}}_{J}(g) 
= 
Z^{\textrm{tot}}_{J}(\openone)
$, 
$\forall\,g$. 
In the notation introduced below Eq.~(\ref{eq: T = 0 limit of alpha term}), 
this restricts the summation over $\alpha^{\ }_{l}$ to 
states of the form 
$\vert \alpha^{\ }_{l} \rangle = g^{\prime}_{\ } \vert 0^{\ }_{k} \rangle$, 
with $g^{\prime}_{\ } \in G$ and $k=1,\,\ldots,\,2^3$ labeling the states 
obtained from $\vert 0 \rangle$ by the action of the non local $\Gamma$ 
operators in Eq.~(\ref{eq: topo ops z}). 
Namely, the states $\vert 0^{\ }_{k} \rangle$ are of the form 
$\Gamma^{m_1}_{1} \Gamma^{m_2}_{2} \Gamma^{m_3}_{3} \vert 0 \rangle$, 
for all possible choices of $m_1,m_2,m_3 = 0,1$. 
Eq.~(\ref{eq: Tr(rho^n_A(T)) f.c. -- 1}) reduces then to 
\begin{widetext}
\bea
\textrm{Tr} \left[ \rho^{n}_{\A}(T) \right] 
&=& 
d^{n-1}_{\A}
\: \times \:
\frac{1}{2^{3n}_{\ }\,\vert G \vert^{n}_{\ }} 
\sum^{\ }_{\alpha^{\ }_{1},\,\ldots,\,\alpha^{\ }_{n}} 
\left( 
\prod^{n}_{l=1} \: 
  \delta^{\ }_{M^{\ }_{s}(\alpha^{\ }_{l}),N} 
\right) 
\prod^{n-1}_{l=1} 
  \delta \left( \alpha^{\ }_{l,\A}, \alpha^{\ }_{l+1,\A} \right) 
\nonumber \\ 
&=& 
d^{n-1}_{\A}
\: \times \:
\frac{1}{2^{3n}_{\ }\,\vert G \vert^{n}_{\ }} 
\sum^{\ }_{g^{\prime}_1,\,\ldots ,\,g^{\prime}_n \in G} \; 
\sum^{\ }_{k^{\ }_{1},\,\ldots,\,k^{\ }_{n}} \; 
  \prod^{n-1}_{l=1} 
    \delta\left[ (g^{\prime}_l 0^{\ }_{k_l})^{\ }_{\A}, 
                 (g^{\prime}_{l+1} 0^{\ }_{k_{l+1}})^{\ }_{\A} \right] 
\nonumber \\ 
&=& 
d^{n-1}_{\A}
\: \times \:
\frac{1}{2^{3n}_{\ }\,\vert G \vert^{n}_{\ }} 
2^{3n}_{\ } 
\sum^{\ }_{g^{\prime}_1,\,\ldots ,\,g^{\prime}_n \in G} \; 
  \prod^{n-1}_{l=1} 
    \delta\left[ (g^{\prime}_l 0)^{\ }_{\A}, 
                 (g^{\prime}_{l+1} 0)^{\ }_{\A} \right] 
\nonumber \\ 
&=& 
d^{n-1}_{\A}
\: \times \:
\frac{1}{\vert G \vert^{n}_{\ }} 
\vert G \vert 
d^{n-1}_{\B} 
\nonumber \\ 
&=& 
d^{n-1}_{\A}
\: \times \:
\left( \frac{d^{\ }_{\B}}{\vert G \vert} \right)^{n-1}_{\ } 
= 
\left( \frac{d^{\ }_{\A} d^{\ }_{\B}}{\vert G \vert} \right)^{n-1}_{\ } 
, 
\label{eq: low T limit}
\eea
\end{widetext}
where we used the fact that, for the cases of interest, subsystem $\A$ is 
finite and the non local operators $\Gamma$ can always be chosen so as to 
traverse only subsystem $\B$, 
$
\delta\left[ (g^{\prime}_l 0^{\ }_{k_l})^{\ }_{\A}, 
             (g^{\prime}_{l+1} 0^{\ }_{k_{l+1}})^{\ }_{\A} \right]
\equiv 
\delta\left[ (g^{\prime}_l 0)^{\ }_{\A}, 
             (g^{\prime}_{l+1} 0)^{\ }_{\A} \right]
$. 
This in turn implies that $g^{\prime}_l g^{\prime}_{l+1} \in G^{\ }_{\B}$, 
and the constrained summation over 
$g^{\prime}_{1},\ldots,g^{\prime}_{n} \in G$ 
can be replaced by an unconstrained summation over 
$g^{\prime}_{1} \in G$, 
$g^{\prime\prime}_{2},\ldots,g^{\prime\prime}_{n} \in G^{\ }_{\B}$
(where $g^{\prime\prime}_{l+1}\equiv g^{\prime}_l g^{\prime}_{l+1}$, 
for $l=2,\dots,n$).
Eq.~(\ref{eq: low T limit}) is indeed the same as in the 2D case at 
zero-temperature.~\cite{Castelnovo2007} 

In this limit, the von Neumann entropy is given  by 
\beq
S^{\ }_{\textrm{VN}}(\A;T=0) 
= 
- \lim^{\ }_{n \to 1} 
  \partial^{\ }_{n} \textrm{Tr} \left[ \rho^{n}_{\A} \right] 
= 
- \ln \left( \frac{d^{\ }_{\A} d^{\ }_{\B}}{\vert G \vert} \right), 
\eeq
and the topological entropy by $S^{\ }_{\textrm{topo}} = 2 \ln 2$ as discussed in 
Sec.~\ref{sec: zero temperature gauge} (for the full bipartition scheme 1-8). 

For $T \to \infty$ (i.e., for $J \to \infty$), we have 
$
Z^{\textrm{tot}}_{J}(g) 
/ 
Z^{\textrm{tot}}_{J}(\openone) 
\to 
\delta(g - \openone)
$, 
all $\alpha$ are equally weighed, 
and Eq.~(\ref{eq: Tr(rho^n_A(T)) f.c. -- 1}) reduces to 
%
%
\bea
\textrm{Tr} \left[ \rho^{n}_{\A} \right] 
&=& 
1 
\: \times \:
\frac{1}{Z^{n}_{s}}
\sum^{\ }_{\alpha^{\ }_{1},\,\ldots,\,\alpha^{\ }_{n}} 
\prod^{n-1}_{l=1} 
  \delta\left( \alpha^{\ }_{l,\A}, \alpha^{\ }_{l+1,\A} \right) 
\nonumber \\ 
&=& 
1 
\: \times \:
\frac{1}{2^{3Nn}_{\ }} 
2^{3N}_{\ } 
2^{\Sigma^{\ }_{\B} (n-1)}_{\ } 
\nonumber \\ 
&=& 
1 
\: \times \:
\left( 
  \frac{1}{2^{\Sigma^{\ }_{\A}}_{\ }} 
\right)^{n-1}_{\ }
, 
\label{eq: high T limit}
\eea
%
%
where $\Sigma^{\ }_{\A}$ ($\Sigma^{\ }_{\B}$) is the number of $\sigma$ spin 
degrees of freedom in $\A$ ($\B$), and 
$\Sigma^{\ }_{\A} + \Sigma^{\ }_{\B} = 3N$. 
Here we used the fact that 
$\delta\left( \alpha^{\ }_{l,\A}, \alpha^{\ }_{l+1,\A} \right)$ 
involves only subsystem $\A$, hence $\Sigma^{\ }_{\A}$ spins are summed over 
only once, while there are $n$ independent copies of the remaining 
$\Sigma^{\ }_{\B}$ spins. 

This result leads to 
\bea
S^{\ }_{\textrm{VN}}(\A;T\to\infty) 
&=& 
- \lim^{\ }_{n \to 1} 
  \partial^{\ }_{n} \textrm{Tr} \left[ \rho^{n}_{\A} \right] 
\nonumber \\
&=& 
\ln \left( 2^{\Sigma^{\ }_{\A}}_{\ } \right) 
= 
\Sigma^{\ }_{\A} \ln 2
, 
\eea
which is indeed the classical entropy of a collection of $\Sigma^{\ }_{\A}$ 
free Ising spins. 
The topological entropy vanishes in this limit, since the contributions from 
the different bipartitions cancel out exactly (recall that the 
total number of spins in $\A$ for bipartitions 2 and 3 is the same as for 
bipartitions 1 and 4, and similarly for 6,7 and 5,8). 

Notice that, in our chosen factorization scheme in 
Eq.~(\ref{eq: Tr(rho^n_A(T)) f.c. -- 1}), the plaquette term does not yield 
any contribution to the von Neumann entropy at infinite temperature, 
while at zero temperature the plaquette term contribution equals 
$- \ln d^{\ }_{\A}$, and the star term contribution is 
$- \ln ( d^{\ }_{\B} / \vert G \vert )$. 
%
%

\section{\label{app: S^S_topo}
The star contribution 
        }
Here we present the expressions for the star contribution to the
entropies for finite temperatures and finite system sizes. As we
argued in the Sec.~\ref{sec: S^S_topo}, the star contribution to the
entropies can computed using
Eqs.~(\ref{S_VN-lambdaB-infty},\ref{S_topo-lambdaB-infty}), which
relate them to entropies evaluated for a hard constrained system
where $\lambda_B\to\infty$. The calculation in this limit is done most
conveniently in the $\sigma^{z}_{\ }$ basis, very much along
the lines of the calculation carried out for 2D systems in
Ref.~\onlinecite{Castelnovo2007}. Paralleling the steps of the
computation for 2D systems, one obtains for the 3D case that
\begin{widetext}
\bea
\Delta S^{\ }_{\textrm{VN}}(\A;T) \Big|_{\lambda_B\to\infty} 
&=& 
\ln \cosh\left( \frac{K^{\ }_{A}}{2} N \right) 
- 
\N^{(s)}_{\A} \left( x \ln x \right) 
  \frac{\cosh\left( \frac{K^{\ }_{A}}{2} (N - 1) \right)}
       {\cosh\left( \frac{K^{\ }_{A}}{2} N \right)}
- 
\N^{(s)}_{\A} \left( y \ln y \right) 
  \frac{\sinh\left( \frac{K^{\ }_{A}}{2} (N - 1) \right)}
       {\cosh\left( \frac{K^{\ }_{A}}{2} N \right)}
\nonumber \\ 
&& 
- 
\sum^{\ }_{i} 
\left( \tilde{x}^{\ }_{i} \ln \tilde{x}^{\ }_{i} \right) 
  \frac{\cosh\left( \frac{K^{\ }_{A}}{2} (N - \N^{(s)}_{\overline{\A}^{\ }_{i}}) \right)}
       {\cosh\left( \frac{K^{\ }_{A}}{2} N \right)} 
- 
\sum^{\ }_{i} 
\left( \tilde{y}^{\ }_{i} \ln \tilde{y}^{\ }_{i} \right) 
  \frac{\sinh\left( \frac{K^{\ }_{A}}{2} (N - \N^{(s)}_{\overline{\A}^{\ }_{i}}) \right)}
       {\cosh\left( \frac{K^{\ }_{A}}{2} N \right)} 
\label{eq: delta S_VN limit}
, 
\eea
\end{widetext}
where $K^{\ }_{A} = -\ln [\tanh(\lambda^{\ }_{A} / T)]$, 
$
\N^{(s)}_{\overline{\A}^{\ }_{i}} 
\equiv 
\N^{(s)}_{\B^{\ }_{i}} 
+ 
\N^{(s)}_{\A\B^{\ }_{i}} 
$ 
is the total number of star operators acting on the $i$th component of 
subsystem $\B$ (either entirely in $\B^{\ }_{i}$, or at its boundary 
$\A\B^{\ }_{i}$), and 
\begin{subequations}
\bea
x = \cosh\left( \frac{K^{\ }_{A}}{2} \right)
&\quad& 
y = \sinh\left( \frac{K^{\ }_{A}}{2} \right)
\\ 
\tilde{x}^{\ }_{i} 
= 
\cosh\left( \frac{K^{\ }_{A}}{2} \N^{(s)}_{\overline{\A}^{\ }_{i}} \right)
&\quad& 
\tilde{y}^{\ }_{i} 
= 
\sinh\left( \frac{K^{\ }_{A}}{2} \N^{(s)}_{\overline{\A}^{\ }_{i}} \right) 
. 
\eea
\end{subequations}
Notice that only the last two terms in Eq.~(\ref{eq: delta S_VN limit}) 
yield a topological contribution in our bipartition scheme, since 
$
\N^{(s)}_{1\A} 
-
\N^{(s)}_{2\A} 
-
\N^{(s)}_{3\A} 
+ 
\N^{(s)}_{4\A} 
= 
0
$ 
and likewise for bipartitions 5-8. 
Therefore, 
\begin{widetext}
\bea
\Delta S^{(S)}_{\textrm{topo}}(T/\lambda^{\ }_{A},N) 
&=&
\sum^{2}_{i=1} 
\left( \tilde{x}^{(1)}_{i} \ln \tilde{x}^{(1)}_{i} \right) 
  \frac{\cosh\left( \frac{K^{\ }_{A}}{2} (N - \N^{(s)}_{1\overline{\A}^{\ }_{i}}) \right)}
       {\cosh\left( \frac{K^{\ }_{A}}{2} N \right)} 
+ 
\sum^{2}_{i=1} 
\left( \tilde{y}^{(1)}_{i} \ln \tilde{y}^{(1)}_{i} \right) 
  \frac{\sinh\left( \frac{K^{\ }_{A}}{2} (N - \N^{(s)}_{1\overline{\A}^{\ }_{i}}) \right)}
       {\cosh\left( \frac{K^{\ }_{A}}{2} N \right)} 
\nonumber \\ 
&& 
- 
\left( \tilde{x}^{(2)}_{\ } \ln \tilde{x}^{(2)}_{\ } \right) 
  \frac{\cosh\left( \frac{K^{\ }_{A}}{2} (N - \N^{(s)}_{2\overline{\A}}) \right)}
       {\cosh\left( \frac{K^{\ }_{A}}{2} N \right)} 
- 
\left( \tilde{y}^{(2)}_{\ } \ln \tilde{y}^{(2)}_{\ } \right) 
  \frac{\sinh\left( \frac{K^{\ }_{A}}{2} (N - \N^{(s)}_{2\overline{\A}}) \right)}
       {\cosh\left( \frac{K^{\ }_{A}}{2} N \right)} 
\nonumber \\ 
&& 
- 
\left( \tilde{x}^{(3)}_{\ } \ln \tilde{x}^{(3)}_{\ } \right) 
  \frac{\cosh\left( \frac{K^{\ }_{A}}{2} (N - \N^{(s)}_{3\overline{\A}}) \right)}
       {\cosh\left( \frac{K^{\ }_{A}}{2} N \right)} 
- 
\left( \tilde{y}^{(3)}_{\ } \ln \tilde{y}^{(3)}_{\ } \right) 
  \frac{\sinh\left( \frac{K^{\ }_{A}}{2} (N - \N^{(s)}_{3\overline{\A}}) \right)}
       {\cosh\left( \frac{K^{\ }_{A}}{2} N \right)}  
\nonumber \\ 
&& 
+
\left( \tilde{x}^{(4)}_{\ } \ln \tilde{x}^{(4)}_{\ } \right) 
  \frac{\cosh\left( \frac{K^{\ }_{A}}{2} (N - \N^{(s)}_{4\overline{\A}}) \right)}
       {\cosh\left( \frac{K^{\ }_{A}}{2} N \right)} 
+ 
\left( \tilde{y}^{(4)}_{\ } \ln \tilde{y}^{(4)}_{\ } \right) 
  \frac{\sinh\left( \frac{K^{\ }_{A}}{2} (N - \N^{(s)}_{4\overline{\A}}) \right)}
       {\cosh\left( \frac{K^{\ }_{A}}{2} N \right)} 
\nonumber \\ 
&&
+ 
\left( \tilde{x}^{(5)}_{\ } \ln \tilde{x}^{(5)}_{\ } \right) 
  \frac{\cosh\left( \frac{K^{\ }_{A}}{2} (N - \N^{(s)}_{5\overline{\A}^{\ }_{\ }}) \right)}
       {\cosh\left( \frac{K^{\ }_{A}}{2} N \right)} 
+ 
\left( \tilde{y}^{(5)}_{\ } \ln \tilde{y}^{(5)}_{\ } \right) 
  \frac{\sinh\left( \frac{K^{\ }_{A}}{2} (N - \N^{(s)}_{5\overline{\A}^{\ }_{\ }}) \right)}
       {\cosh\left( \frac{K^{\ }_{A}}{2} N \right)} 
\nonumber \\ 
&& 
- 
\left( \tilde{x}^{(6)}_{\ } \ln \tilde{x}^{(6)}_{\ } \right) 
  \frac{\cosh\left( \frac{K^{\ }_{A}}{2} (N - \N^{(s)}_{6\overline{\A}}) \right)}
       {\cosh\left( \frac{K^{\ }_{A}}{2} N \right)} 
- 
\left( \tilde{y}^{(6)}_{\ } \ln \tilde{y}^{(6)}_{\ } \right) 
  \frac{\sinh\left( \frac{K^{\ }_{A}}{2} (N - \N^{(s)}_{6\overline{\A}}) \right)}
       {\cosh\left( \frac{K^{\ }_{A}}{2} N \right)} 
\nonumber \\ 
&& 
- 
\left( \tilde{x}^{(7)}_{\ } \ln \tilde{x}^{(7)}_{\ } \right) 
  \frac{\cosh\left( \frac{K^{\ }_{A}}{2} (N - \N^{(s)}_{7\overline{\A}}) \right)}
       {\cosh\left( \frac{K^{\ }_{A}}{2} N \right)} 
- 
\left( \tilde{y}^{(7)}_{\ } \ln \tilde{y}^{(7)}_{\ } \right) 
  \frac{\sinh\left( \frac{K^{\ }_{A}}{2} (N - \N^{(s)}_{7\overline{\A}}) \right)}
       {\cosh\left( \frac{K^{\ }_{A}}{2} N \right)}  
\nonumber \\ 
&& 
+
\left( \tilde{x}^{(8)}_{\ } \ln \tilde{x}^{(8)}_{\ } \right) 
  \frac{\cosh\left( \frac{K^{\ }_{A}}{2} (N - \N^{(s)}_{8\overline{\A}}) \right)}
       {\cosh\left( \frac{K^{\ }_{A}}{2} N \right)} 
+ 
\left( \tilde{y}^{(8)}_{\ } \ln \tilde{y}^{(8)}_{\ } \right) 
  \frac{\sinh\left( \frac{K^{\ }_{A}}{2} (N - \N^{(s)}_{8\overline{\A}}) \right)}
       {\cosh\left( \frac{K^{\ }_{A}}{2} N \right)} 
, 
\label{eq: delta S_topo star} 
\eea
\end{widetext}
where we used the fact that subsystem $\B$ has always one component except 
for bipartition 1, where it has two components.

With the expression above for $\Delta
S^{(S)}_{\textrm{topo}}(T/\lambda^{\ }_{A},N)$, one can determine the
topological entropy contribution from the star operators as a function of
temperature and system sizes. In particular, let us look at two
particular limits: that of the zero temperature limit taken first, and
that of the thermodynamic limit taken first.

For $T\to 0$ first, $K_A\to 0$, and one can easily check that all
terms in Eq.~(\ref{eq: delta S_topo star}) vanish, which is expected as
the difference $\Delta S^{(S)}_{\textrm{topo}}(T/\lambda^{\ }_{A},N)$
is, by definition, zero at $T=0$.

Now, when the thermodynamic limit is taken first, i.e., when the
sizes $N$ and all of $\N^{(s)}_{1\overline{\A}^{\ }_{i}}$ (for
$i=1,2$) and $\N^{(s)}_{p\overline{\A}^{\ }_{i}}$, $p=2,\ldots,8$ are
taken to infinity at fixed $K_A$, each term in the expression in 
Eq.~(\ref{eq: delta S_topo star}) gives $\mp \ln 2$ (with the sign
determined by whether the partition is added or
subtracted). Bipartition 1 gives $-2\ln2$ (its contribution is doubled
because $1\B$ has two disconnected components) and it is
added to bipartitions 4,5, and 8, which give $-\ln 2$ each;
bipartitions 2,3,6, and 7 are subtracted, and each of them gives $+\ln
2$. Altogether, we obtain $\Delta S^{(S)}_{\textrm{topo}}(T/\lambda^{\
}_{A},N\to \infty)=-\ln 2$, for any temperature $T$. Therefore, we
obtain in the thermodynamic limit the result used in
Eq.~(\ref{S^S_topo-thermo-limit}).

One can finally add the zero temperature contributions, to obtain 
\bea
S^{(S)}_{\textrm{VN}}(T/\lambda^{\ }_{A}) 
&=&
\Delta S^{(S)}_{\textrm{VN}}(T/\lambda^{\ }_{A}) 
- 
\frac{d^{\ }_{\B}}{\vert G \vert} 
, 
\label{eq: S_VN star} 
\eea
and 
\bea
S^{(S)}_{\textrm{topo}}(T/\lambda^{\ }_{A}) 
&=&
\Delta S^{(S)}_{\textrm{topo}}(T/\lambda^{\ }_{A}) 
+ 
\ln 
\frac{d^{\ }_{1\B}\,d^{\ }_{4\B}\,d^{\ }_{5\B}\,d^{\ }_{8\B}}
     {d^{\ }_{2\B}\,d^{\ }_{3\B}\,d^{\ }_{6\B}\,d^{\ }_{7\B}}
\nonumber \\ 
&=& 
\Delta S^{(S)}_{\textrm{topo}}(T/\lambda^{\ }_{A}) 
+ 
\ln 2 
. 
\label{eq: S_topo star} 
\eea
%
%
%

\section{\label{app: S^P_topo}
The plaquette contribution
        }
As anticipated in Sec.~\ref{sec: S^P_topo}, the plaquette contribution in 
3D is very different from the 2D case, and we need to carry out the 
calculations explicitly. 

Consider the expression for $\mathcal{Z}^{(P)}$, 
\bea
\mathcal{Z}^{(P)}(n) 
&=& 
\!\!\!\!\!
\sum_{g^{\ }_{1},\,\ldots,\,g^{\ }_{n}\in G^{\ }_{\A}} 
\left( 
\prod^{n}_{l=1} \: 
  \frac{Z^{\textrm{tot}}_{J}(g^{\ }_{l})} 
       {Z^{\textrm{tot}}_{J}(\openone)} 
\right) 
\langle 0 \vert 
  \prod^{n}_{l=1} g^{\ }_{l} 
\vert 0 \rangle 
, 
\qquad 
\label{eq: Tr(rho^n_A(T)) f.c. -- 2}
\eea
where
\beq 
Z^{\textrm{tot}}_{J}(g) 
= 
\sum_{\{S_i\}} 
  \exp \left( J \sum_{\langle ij\rangle}\eta_{ij}(g)\,S_iS_j \right) 
\eeq
is the partition function of the 3D random-bond Ising model (summed over 
all possible boundary conditions), whose randomness is controlled by 
$g$ according to Eq.~(\ref{eq: random bond g f.c.}). Namely, 
$\eta_{ij}(g)=\pm 1$ depending on whether the plaquette perpendicular to the
bond $\langle ij \rangle$ is flipped in configuration $g$ 
($\eta_{ij}=-1$) or not ($\eta_{ij}=+1$). 

Recall that the group $G$, and therefore its subgroup $G^{\ }_{\A}$, is 
defined modulo the identities 
$\prod^{\ }_{\textrm{closed membrane}} B^{\ }_{p} = \openone$. 
In the language of the randomness realizations $\{ \eta_{ij} \}$, this amounts 
to summing over gauge \emph{in}equivalent configurations. 
In fact, any $\eta_{ij}$ and $\eta^{\prime}_{ij}$ that 
differ by the product of plaquettes around closed surfaces are related by 
\beq
\eta^{\ }_{ij} 
= 
\eta^{\prime}_{ij} \: \overline{S}_i \overline{S}_j , 
\qquad \qquad 
\exists \: \{ \overline{S}^{\ }_{i} \} . 
\eeq
Specifically, $\{ \overline{S}^{\ }_{i} \}$ corresponds to either of the two 
spin configurations that exhibit the closed surfaces in question as their only 
antiferromagnetic boundary (the two configurations are related by an overall 
$Z^{\ }_{2}$ symmetry). 
Recall that the product of plaquettes belonging to an infinite crystal plane 
is also an allowed gauge transformation, and all possible boundary conditions 
(periodic or antiperiodic in each direction) should be taken into account 
when enumerating all configurations 
$\{ \overline{S}^{\ }_{i} \}$. 
In conclusion, every $\eta^{\ }_{ij}(g)$ admits $2^{N+3}_{\ }$ equivalent 
randomness realizations 
$\eta^{\prime}_{ij} = \eta^{\ }_{ij} \overline{S}_i \overline{S}_j$, 
labeled by all possible Ising configurations $\{ \overline{S}_i \}^{N}_{i=1}$ 
(where $\{ \overline{S}_i \}^{N}_{i=1}$ and 
$\{ - \overline{S}_i \}^{N}_{i=1}$ yield the exact same $\eta^{\prime}_{ij}$). 

In the case of a summation over the whole group $G$, one has then the 
identity 
\bea
&& 
\sum^{\ }_{g \in G} 
  \sum_{\{S_i\}} 
    \exp \left( J \sum_{\langle ij\rangle}\eta_{ij}(g)\,S_iS_j \right) 
\equiv 
\nonumber \\ 
&& \qquad 
\equiv 
\frac{1}{2^{N+3}_{\ }}
  \sum^{\ }_{\{ \eta^{\ }_{ij} \}} 
    \sum_{\{S_i\}} 
      \exp \left( J \sum_{\langle ij\rangle}\eta_{ij}\,S_iS_j \right) 
. 
\eea

For the subgroup $G^{\ }_{\A}$, the situation is more convoluted. 
First of all, the operators $g\in G^{\ }_{\A}$ correspond to randomness 
realizations $\{ \eta^{\ }_{ij}(g) \}$ where all the bonds outside $\A$ can 
be gauged to assume the value $+1$. 
Rather than considering all the equivalent configurations as for the whole 
group $G$, it is more convenient to introduce a 
restricted set of randomness realizations $\{\eta^{(\A)}_{ij}\}$ where 
$\eta^{(\A)}_{ij}$ is constrained to assume the value $+1$ 
whenever $\langle ij \rangle \notin \A$. 
Notice that we do not constrain the bonds inside $\A$, and we are therefore 
over-counting all the gauge equivalent configurations with respect to these 
bonds. 
The number of equivalent realizations in the restricted 
subgroup can be counted as seen in Sec.~\ref{sec: zero temperature gauge}, 
and repeated hereafter for convenience. 
All cubic unit cells entirely contained in $\A$ are independent generators 
of gauge transformations. Also, if $\A$ contains crystal planes, there are 
up to three additional generators. Finally, we have one extra generator per 
connected component of $\B$ (i.e., entirely surrounded by $\A$), but for one 
of them. 
Thus, the total number of gauge equivalent configurations is now 
$2^{\N^{(c)}_{\A} + m^{\textrm{(c.p.)}}_{\A} + (m^{\ }_{\B} - 1)}_{\ }$, 
where again 
$\N^{(c)}_{\A}$ is the number of cubic unit cells entirely contained in $\A$, 
$m^{\textrm{(c.p.)}}_{\A}$ is the number of independent crystal planes in 
$\A$ ($m^{\textrm{(c.p.)}}_{\A} = 0$, $m^{\textrm{(c.p.)}}_{\B} = 3$ for all 
cases of interest), 
and $m^{\ }_{\B}$ is the number of connected components of $\B$. 

As a result, one obtains: 
\bea
&&
\sum^{\ }_{g^{\prime}_{\ } \in G^{\ }_{\A}} 
  \sum_{\{S_i\}} 
    \exp 
      \left( 
        J \sum_{\langle ij\rangle}\eta_{ij}(g^{\prime}_{\ })\,S_i S_j 
      \right) 
\equiv 
\nonumber \\ 
&& 
\qquad
\equiv 
\frac{1}{2^{\N^{(c)}_{\A} + (m^{\ }_{\B} - 1)}_{\ }}
  \sum^{\ }_{\{ \eta^{(\A)}_{ij} \}} 
    \sum_{\{S_i\}}
      \left[ 
        \exp 
          \left( 
            J \sum_{\langle ij\rangle \in \A} \eta^{(\A)}_{ij}\,S_i S_j 
          \right) 
\right. 
\nonumber \\ 
&& 
\qquad
\times
\left. 
        \exp 
          \left( 
	    J \sum_{\langle ij\rangle \notin \A}\,S_i S_j 
          \right) 
      \right] 
. 
\nonumber \\
\label{eq: g to eta f.c.}
\eea
%
%

Having done so, the summation over $\{ \eta^{(\A)}_{ij} \}$ is now 
unconstrained, namely the bond variables $\eta^{(\A)}_{ij} = \pm 1$ are 
generated by freely flipping any of the plaquettes in $\A$, starting from 
the configuration with all $\eta^{(\A)}_{ij} = +1$ 
(which we refer to in the following as 
$\eta^{0}_{\ } \equiv \{ \eta^{0}_{ij} \}$, the ferromagnetic configuration). 
Notice that this accounts only for the bipartitions where the plaquette 
operators in $\A$ are sufficient to generate the whole group $G^{\ }_{\A}$ 
(bipartitions 1,2,3 and 6,7,8). 
As discussed in Sec.~\ref{sec: zero temperature gauge}, this is not always 
the case and additional collective operations may be needed to generate 
$G^{\ }_{\A}$ (bipartitions 4,5). 
The summation encompasses then all configurations obtained by flipping 
plaquettes in $\A$ starting from $\{ \eta^{0}_{ij} \}$, 
\emph{and} starting from the configurations derived from the ferromagnetic 
one via the action of each of the independent collective operations. 
For concreteness, in bipartitions 4 and 5 there is only one collective 
operation in $\A$, illustrated in the bottom panel of 
Fig.~\ref{fig: collective ops 2}. 
In this case, the configurations $\{ \eta^{(\A)}_{ij} \}$ are obtained by 
flipping plaquettes in $\A$ starting from the ferromagnetic configuration 
$\eta^{0}_{\ }$, and starting from the configuration with all 
$\eta^{(\A)}_{ij} = +1$, except for those inside the blue thick line in the 
bottom panel of Fig.~\ref{fig: collective ops 2} (i.e., plaquettes in $\B$ 
or at the boundary), where $\eta^{(\A)}_{ij} = -1$. 
(We will refer to this configuration in the following as 
$\eta^{1}_{\ } \equiv \{ \eta^{1}_{ij} \}$). 
If we label 
$\{ \tilde{\eta}^{(\A)}_{\ } \equiv \{ \tilde{\eta}^{(\A)}_{ij} \} \}$ 
the set of all configurations obtained from the ferromagnetic one via the 
action of the plaquette operators in $\A$ alone, 
the summation in Eq.~(\ref{eq: g to eta f.c.}) runs over 
$
\eta^{0}_{\ } \{ \tilde{\eta}^{(\A)}_{\ } \} 
\cup 
\eta^{1}_{\ } \{ \tilde{\eta}^{(\A)}_{\ } \}
$, 
where the product of two configurations represents the new configuration 
with variables given by the site-by-site product of the two original 
variables 
$\eta^{0}_{ij} \tilde{\eta}^{(\A)}_{ij}$ ($\equiv \tilde{\eta}^{(\A)}_{ij}$), 
and 
$\eta^{1}_{ij} \tilde{\eta}^{(\A)}_{ij}$. 

We can then apply the identity in Eq.~(\ref{eq: g to eta f.c.}) to simplify 
our expression in Eq.~(\ref{eq: Tr(rho^n_A(T)) f.c. -- 2}). 
The condition that a term is non-vanishing, namely 
$
\langle 0^{\ }_{\A} \vert 
  g^{\prime}_{1,\A} \:\ldots\: g^{\prime}_{n,\A} 
\vert 0^{\ }_{\A} \rangle 
= 
1
$, 
translates into the condition that 
\beq
\prod_{l=1}^n \eta^{(\A,l)}_{ij}(g_l) = {\tilde S}_i {\tilde S}_j 
\quad \qquad 
\forall\,\langle ij\rangle, \;\; \exists \{ {\tilde S}_i \} , 
\label{eq: rho^n constraint f.c. 1}
\eeq
i.e., the product of all $\eta^{(\A,l)}_{ij}(g_l)$, $l=1,\,\ldots,\,n$ is 
gauge equivalent to $\eta^{0}_{\ }$ 
(equivalently $g^{\prime}_{1,\A} \:\ldots\: g^{\prime}_{n,\A} = \openone$). 
The very same nature of a collective operation in $\A$ requires that such 
operation cannot be completed to an identity (a closed membrane) by means 
of plaquette operators in $\A$ alone. 
Therefore the above equation holds independently for the collective 
operations and for the $\tilde{\eta}^{(\A)}_{\ }$ configurations. 
Namely, it imposes that the number of collective operations appearing in 
$\{ \eta^{(\A,l)}_{ij} \}^{n}_{l=1}$ is even, and that 
\beq
\prod_{l=1}^n \tilde{\eta}^{(\A,l)}_{ij}(g_l) = {\tilde S}_i {\tilde S}_j 
\quad \qquad 
\forall\,\langle ij\rangle, \;\; \exists \{ {\tilde S}_i \} . 
\label{eq: rho^n constraint f.c. 2}
\eeq
Trivially, Eq.~(\ref{eq: rho^n constraint f.c. 1}) and 
Eq.~(\ref{eq: rho^n constraint f.c. 2}) become equivalent if no collective 
operations are present in $\A$. 

Notice that ${\tilde S}_i {\tilde S}_j \equiv 1$ for all 
$\langle ij \rangle \notin \A$: all possible $\{ {\tilde S}_i \}$ 
configurations must be ferromagnetically ordered outside $\A$. 
If $m_\B$ is the number of connected components in $\B$, then the 
ferromagnetic order holds across each component separately, and from one 
component to the next the overall sign of the $\tilde{S}$ spins may change. 
An overall sign change of the spins $\tilde{S}$ is immaterial, as one can 
see from Eq.~(\ref{eq: rho^n constraint f.c. 2}), and therefore one needs to 
introduce a corresponding factor of $1/2$ when summing over 
$\{ \tilde{S}_i \}$. 

Eq.~(\ref{eq: Tr(rho^n_A(T)) f.c. -- 2}) becomes then 
\begin{widetext}
\bea
\mathcal{Z}^{(P)}(n) 
&=& 
\left( \frac{1}{Z^{\textrm{tot}}_{J}(\openone)} \right)^{n}_{\ } 
\frac{1}{2^{[\N^{(c)}_{\A} + m^{\ }_{\B} - 1]\, n}_{\ }} 
  \frac{1}{2} 
  \sum^{\ }_{\{ \tilde{S}_i \}} 
  \mathop{\sum^{\ }_{\left\{ \{ \eta^{(\A,l)}_{ij} \} \right\}^{n}_{l=1}}} 
         _{\prod_{l=1}^n \eta^{(\A,l)}_{ij} = {\tilde S}_i {\tilde S}_j} 
    \prod^{n}_{l=1} \:
     \left[ 
      \sum_{\left\{ S^{(l)}_i \right\}} 
	\left( 
          \prod^{\ }_{\langle ij\rangle} \: 
            \exp\left( J \eta^{(\A,l)}_{ij}\,S^{(l)}_i S^{(l)}_j \right) 
	\right) 
     \right] 
\nonumber \\ 
&=& 
\left( 
  \frac{1}
       {Z^{\textrm{tot}}_{J}(\openone) \: 
        2^{\N^{(c)}_{\A} + m^{\ }_{\B} - 1}_{\ }} 
\right)^{n}_{\ } 
  \frac{1}{2} 
  \sum^{\ }_{\{ \tilde{S}_i \}} 
  \mathop{\sum^{\ }_{\left\{ \{ \eta^{(\A,l)}_{ij} \} \right\}^{n}_{l=1}}}
         _{\prod_{l=1}^n \eta^{(\A,l)}_{ij} = {\tilde S}_i {\tilde S}_j} 
  \sum^{\ }_{\left\{ \{ S^{(l)}_{i} \} \right\}^{n}_{l=1}} 
	\left[ 
          \prod^{\ }_{\langle ij\rangle} \: 
            \exp\left( 
	      J \sum^{n}_{l=1} \eta^{(\A,l)}_{ij}\,S^{(l)}_i S^{(l)}_j 
	    \right) 
	\right] 
\nonumber \\ 
&=& 
\left( 
  \frac{1}
       {Z^{\textrm{tot}}_{J}(\openone) \: 
        2^{\N^{(c)}_{\A} + m^{\ }_{\B} - 1}_{\ }} 
\right)^{n}_{\ } 
  \frac{1}{2} 
  \sum^{\ }_{\{ \tilde{S}_i \}} 
  \sum^{\ }_{\left\{ \{ S^{(l)}_{i} \} \right\}^{n}_{l=1}} 
  \prod^{\ }_{\langle ij\rangle} \: 
    \left[ 
      \mathop{\sum^{\ }_{\left\{ \eta^{(\A,l)}_{ij} \right\}^{n}_{l=1}}}
             _{\prod_{l=1}^n \eta^{(\A,l)}_{ij} = {\tilde S}_i {\tilde S}_j} 
            \exp\left( 
	      J \sum^{n}_{l=1} \eta^{(\A,l)}_{ij}\,S^{(l)}_i S^{(l)}_j 
            \right)
    \right] 
\nonumber \\ 
&=& 
\left( 
  \frac{1}
       {Z^{\textrm{tot}}_{J}(\openone) \: 
        2^{\N^{(c)}_{\A} + m^{\ }_{\B} - 1}_{\ }} 
\right)^{n}_{\ } 
  \frac{1}{2} 
  \sum^{\ }_{\{ \tilde{S}_i \}} 
  \sum^{\ }_{\left\{ \{ S^{(l)}_{i} \} \right\}^{n}_{l=1}} 
\nonumber \\ 
&& 
\times 
\sum^{\textrm{(even)}}_{\{ \overline{\eta}^{(l)}_{\ } \}^{n}_{l=1}} 
  \prod^{\ }_{\langle ij\rangle\in\A} \: 
    \left[ 
      \mathop{\sum^{\ }_{\left\{ \tilde{\eta}^{(\A,l)}_{ij} \right\}^{n}_{l=1}}}
             _{\prod_{l=1}^n \tilde{\eta}^{(\A,l)}_{ij} = {\tilde S}_i {\tilde S}_j} 
            \exp\left(
	      J \sum^{n}_{l=1} \overline{\eta}^{(l)}_{ij} 
	        \tilde{\eta}^{(\A,l)}_{ij}\,S^{(l)}_i S^{(l)}_j 
            \right) 
    \right] 
  \prod^{\ }_{\langle ij\rangle\notin\A} \: 
    \left[ 
      \exp\left( 
        J \sum^{n}_{l=1} \overline{\eta}^{(l)}_{ij} S^{(l)}_i S^{(l)}_j 
      \right) 
    \right] 
, 
\label{eq: Tr(rho^n_A(T)) f.c. -- 3}
\eea
%
%
where $\sum^{\textrm{(even)}}_{\{ \overline{\eta}^{(l)}_{\ } \}^{n}_{l=1}}$ 
runs over all $n$tuples 
$
\{ \overline{\eta}^{(l)}_{\ } \in (\eta^{0}_{\ }, \eta^{1}_{\ }) \}^{n}_{l=1}
$ 
with an even number of $\eta^{1}_{\ }$ terms. 
Notice that the summation 
%
%
\bea 
\mathop{\sum^{\ }_{\left\{ \tilde{\eta}^{(\A,l)}_{ij} \right\}^{n}_{l=1}}}
       _{\prod_{l=1}^n \tilde{\eta}^{(\A,l)}_{ij} = {\tilde S}_i {\tilde S}_j} 
\!\!\!\!\!\!\!\!\!\!\!\!
         \exp\left(
	   J \sum^{n}_{l=1} \overline{\eta}^{(l)}_{ij} 
	     \tilde{\eta}^{(\A,l)}_{ij}\,S^{(l)}_i S^{(l)}_j 
         \right) 
= 
\mathcal{Z}^{\ }_{n} 
  \left( 
    \{J \, \overline{\eta}^{(l)}_{ij} S^{(l)}_i S^{(l)}_j \}; \,
    \tilde{S}_i \tilde{S}_j 
  \right) 
\label{eq: chain part funct f.c.}
\eea
\end{widetext}
where 
$
\mathcal{Z}^{\ }_{n} 
  \left( 
    \{J \, \overline{\eta}^{(l)}_{ij} S^{(l)}_i S^{(l)}_j \}; \,
    \tilde{S}_i \tilde{S}_j 
  \right)
$
can be interpreted as the partition function of an Ising chain of degrees 
of freedom $\{\tilde{\eta}^{(\A,l)}_{ij}\}^{n}_{l=1}$ in a random field of 
local strength $J \, \overline{\eta}^{(l)}_{ij} S^{(l)}_i S^{(l)}_j$, and 
subject to the condition that the product of all Ising spins 
$\prod^{n}_{l=1} \tilde{\eta}^{(\A,l)}_{ij}$ equals $\tilde{S}_i \tilde{S}_j$. 
By means of the change of variables 
$\tilde{\eta}^{(\A,l)}_{ij} = m^{(\A,l)}_{ij} m^{(\A,l+1)}_{ij}$, this becomes 
the partition function of a nearest-neighbor Ising chain with periodic or 
antiperiodic boundary conditions (BC) depending on the sign of 
$\tilde{S}_i \tilde{S}_j = \pm 1$ 
(i.e., $m^{(\A,n+1)}_{ij} = m^{(\A,1)}_{ij} \tilde{S}_i \tilde{S}_j$): 
\bea
\mathcal{Z}^{\ }_{n} 
&\equiv&
\mathcal{Z}^{\ }_{n} 
  \left( 
    \{J \, \overline{\eta}^{(l)}_{ij} S^{(l)}_i S^{(l)}_j \}; \,
    \tilde{S}_i \tilde{S}_j 
  \right)
\nonumber \\ 
&=& 
\frac{1}{2} 
\mathop{\sum^{\ }_{\left\{ m^{(\A,l)}_{ij} \right\}^{n}_{l=1}}}
       _{\textrm{BC} = {\tilde S}_i {\tilde S}_j} 
         \exp\left( 
	   J \sum^{n}_{l=1} \, \overline{\eta}^{(l)}_{ij} 
	     m^{(\A,l)}_{ij} m^{(\A,l+1)}_{ij} \, 
	     S^{(l)}_i S^{(l)}_j 
         \right)
. 
\nonumber \\
\eea
This in turn can be computed exactly, 
\bea
2 \mathcal{Z}^{\ }_{n} 
&=& 
\left( 2\cosh J \right)^n
+
\left[
\left(\prod_{l=1}^n \, \overline{\eta}^{(l)}_{ij} S^{(l)}_i S^{(l)}_j \right) 
\;{\tilde S}_i {\tilde S}_j
\right]
\left( 2\sinh J \right)^n 
\nonumber \\ 
&=& 
\left( 2\cosh J \right)^n
+
\left[
\left(\prod_{l=1}^n S^{(l)}_i S^{(l)}_j \right)\;{\tilde S}_i {\tilde S}_j
\right]
\left( 2\sinh J \right)^n 
. 
\nonumber \\
\eea
%
%
We also used the fact that $\overline{\eta}^{(l)}_{ij} = + 1$ if 
$\langle ij \rangle \in \A$ by construction. (Notice that this convenient 
choice does not introduce any limitations. In general, the number of times 
when a $-1$ appears in the $l=1,\,\ldots,\,n$ sequence of 
$\overline{\eta}^{(l)}_{ij}$ values must be even, and therefore 
$\prod_{l=1}^n \, \overline{\eta}^{(l)}_{ij} = +1$, $\forall\,i,j$). 

For convenience of notation, let us consider the following change of 
summation variables 
\beq
{\tilde S}_i \to \theta_i 
= 
\left( \prod_{l=1}^n S^{(l)}_i \right)\;{\tilde S}_i 
\eeq
so that we can write 
$
{\cal Z}_n = \frac{1}{2} \, e^{A_n}\,e^{B_n\theta_i \theta_j} 
$, 
with $A_n$ and $B_n$ defined as 
\begin{subequations}
\bea
&& e^{A_n+B_n}=\left(2\cosh J\right)^n+\left(2\sinh J\right)^n \\
&& e^{A_n-B_n}=\left(2\cosh J\right)^n-\left(2\sinh J\right)^n
\eea
\label{eq: A_n B_n f.c.}
\end{subequations}

Given that $\prod_{l=1}^n S^{(l)}_i = \pm 1$, for all sites $i$ whose adjacent 
bonds $\langle ij \rangle$ are \emph{solely} in $\A$, the summation over 
$\{ {\tilde S}_i = \pm 1 \}$ and the summation over $\{ \theta_i = \pm 1 \}$ 
are unconstrained. 
The case is different for the sites $i$ that have an adjacent bond not in 
$\A$. The correlation across such bond is in fact ferromagnetic by 
construction, and, if $\B$ has only one connected component, 
the spin ${\tilde S}_i$ has the same sign as all other spins not entirely 
surrounded by bonds in $\A$. 
Consequently, all the boundary spins ${\tilde S}_i$ have the same sign, 
and the values of the associated spins $\theta_i$ are determined uniquely 
by the product $\prod_{l=1}^n S^{(l)}_i$. 
If $m_\B$ is the number of connected components in $\B$, then the 
ferromagnetic order holds across each component separately, and from one 
component to the next the overall sign of the $\tilde{S}$ spins may change. 
This is accounted for by summing over boundary sign variables $q_r = \pm 1$, 
$r = 1,\ldots,m_\B$, assigned to each boundary $\partial_r$ defined 
as the set of sites that have adjacent bonds both 
in $\A$ and in the $r$th component of $\overline{\A}$. 

In the end, Eq.~(\ref{eq: Tr(rho^n_A(T)) f.c. -- 3}) becomes 
\begin{widetext}
\bea
\mathcal{Z}^{(P)}(n) 
&=& 
\left( 
  \frac{1}
       {Z^{\textrm{tot}}_{J}(\openone) \: 
        2^{\N^{(c)}_{\A} + m^{\ }_{\B} - 1}_{\ }} 
\right)^{n}_{\ } 
  \frac{1}{2} 
  \sum^{\ }_{\{ \tilde{S}_i \}} 
  \sum^{\ }_{\left\{ \{ S^{(l)}_{i} \} \right\}^{n}_{l=1}} 
  \prod^{\ }_{\langle ij\rangle\in\A} \: 
    \mathcal{Z}^{\ }_{n} 
      \left( 
        \{J \, S^{(l)}_i S^{(l)}_j \}; \,
        \tilde{S}_i \tilde{S}_j 
      \right) 
\sum^{\textrm{(even)}}_{\{ \overline{\eta}^{(l)}_{\ } \}^{n}_{l=1}} 
  \prod^{\ }_{\langle ij\rangle\notin\A} \: 
    \left[ 
      \exp\left(
        J \sum^{n}_{l=1} \overline{\eta}^{(l)}_{ij} S^{(l)}_i S^{(l)}_j 
      \right)
    \right] 
\nonumber \\ 
&=& 
\left( 
  \frac{1}
       {Z^{\textrm{tot}}_{J}(\openone) \: 
        2^{\N^{(c)}_{\A} + m^{\ }_{\B} - 1}_{\ }} 
\right)^{n}_{\ } 
\frac{1}{2} 
  \sum^{\ }_{\left\{ \{ S^{(l)}_{i} \} \right\}^{n}_{l=1}} 
  \sum^{\ }_{\{ \theta_i \}} 
  \prod^{\ }_{\langle ij\rangle \in \A} 
    \frac{1}{2} e^{A_n}\,e^{B_n \theta_i \theta_j}
\sum^{\textrm{(even)}}_{\{ \overline{\eta}^{(l)}_{\ } \}^{n}_{l=1}} 
  \prod^{\ }_{\langle ij\rangle\notin\A} \: 
    \left[ 
      \exp\left(
        J \sum^{n}_{l=1} \overline{\eta}^{(l)}_{ij} S^{(l)}_i S^{(l)}_j 
      \right) 
    \right] 
\nonumber \\ 
&\times&
\sum_{\{ q_r = \pm 1 \}^{m_\B}_{r=1}} \: 
  \prod^{m_\B}_{r=1} \; 
    \prod_{i \in \partial_r} 
      \delta\left( \theta_i \prod^n_{l=1} S^{(l)}_i = q_r \right) 
\nonumber \\ 
&=& 
\left( 
  \frac{1}
       {Z^{\textrm{tot}}_{J}(\openone) \: 
        2^{\N^{(c)}_{\A} + m^{\ }_{\B} - 1}_{\ }} 
\right)^{n}_{\ } 
\frac{e^{\N^{(p)}_{\A} A_n}_{\ }}{2^{\N^{(p)}_{\A}}_{\ }}
\frac{1}{2} 
  \sum^{\ }_{\left\{ \{ S^{(l)}_{i} \} \right\}^{n}_{l=1}} 
  \sum^{\ }_{\{ \theta_i \}} 
  \prod^{\ }_{\langle ij\rangle \in \A} 
    e^{B_n \theta_i \theta_j}
\sum^{\textrm{(even)}}_{\{ \overline{\eta}^{(l)}_{\ } \}^{n}_{l=1}} 
  \prod^{\ }_{\langle ij\rangle\notin\A} \: 
    \left[ 
      \exp\left(
        J \sum^{n}_{l=1} \overline{\eta}^{(l)}_{ij} S^{(l)}_i S^{(l)}_j 
      \right) 
    \right] 
\nonumber \\ 
&\times&
\sum_{\{ q_r = \pm 1 \}^{m_\B}_{r=1}} \: 
  \prod^{m_\B}_{r=1} \; 
    \prod_{i \in \partial_r} 
      \delta\left( \theta_i \prod^n_{l=1} S^{(l)}_i = q_r \right) 
. 
\label{eq: Tr(rho^n_A(T)) f.c. -- 4} 
\eea
\end{widetext}

Notice that $\overline{\eta}^{(l)}_{ij} = +1$ if the plaquette 
$\langle ij \rangle$ does not belong to the collective operation, 
and that whenever $\langle ij \rangle$ belongs to the collective operation 
the value of 
$\overline{\eta}^{(l)}_{ij} = \pm 1$ is the same for all $\langle ij \rangle$. 
(We restrict here for simplicity to the case where there is at most one 
collective operation in $\A$. In order to extend to the general case one needs 
to repeat the derivation for each collective operation separately.) 

Notice also that the sum over $S^{(l)}_i$ that are entirely 
surrounded by bonds in $\A$ is unconstrained, and it contributes a trivial 
factor $2^{\N^{(c)}_{\A} n}_{\ }$ to the sum over the remaining spins. 
In the following, we use this simplification and all summations over 
$S^{(l)}_i$ are intended as constrained only to the remaining spins 
(for convenience, we do not increase the already complex notation). 

Let us focus on the boundary condition 
\beq
\sum_{\{ q_r = \pm 1 \}^{m_\B}_{r=1} } \: 
  \prod^{m_\B}_{r=1} 
    \prod_{i \in \partial_r} 
      \delta\left( \theta_i \prod^n_{l=1} S^{(l)}_i = q_r \right) 
. 
\label{eq: boundary condition}
\eeq
Given that the $\theta$ and the $S$ spins can assume only the values $\pm 1$, 
then the quantity $\theta_i + \sum^n_{l=1} S^{(l)}_i$ can only assume the 
values $n+1, n-1, n-3, \ldots, -(n-1), -(n+1)$.~\cite{footnote: Cardy} 
In particular, the product $\theta_i \prod^n_{l=1} S^{(l)}_i$ is positive 
whenever said summation equals $n+1, n-3, n-7 \ldots$, and it is 
negative otherwise. 
We can therefore rewrite the delta function in the above equation as 
\bea
&& 
\delta\left( \theta_i \prod^n_{l=1} S^{(l)}_i = q_r \right) 
= 
\nonumber \\ 
&& 
\qquad\qquad
\sum^{\lfloor (n+q_r)/2 \rfloor}_{p=0} 
  \delta\left( 
    \theta_i + \sum^n_{l=1} S^{(l)}_i = n+q_r-4p
  \right) 
\nonumber 
\eea
where $\lfloor \cdot \rfloor$ stands for the integer part of its argument. 
In other words, the sum $\theta_i + \sum^n_{l=1} S^{(l)}_i$ must equal 
$n+q \;(\textrm{mod} \: 4)$, or 
\beq
\theta_i + \sum^n_{l=1} S^{(l)}_i - (n+q_r) = 0 \;\;(\textrm{mod} \: 4) 
. 
\eeq

Using the function 
\bea
f(x) 
&=& 
\frac{1}{4} 
  \sum^{3}_{k=0} 
    \exp\left( i \frac{\pi}{2} k x \right) 
\nonumber \\
&=& 
\begin{cases}
1 & \textrm{if} \;\; x = 0 \;\;(\textrm{mod} \: 4)
\\
0 & \textrm{if} \;\; x = 1,2,3 \;\;(\textrm{mod} \: 4)
\end{cases}
, 
\eea
we can finally write the delta function as 
\bea
&& 
\delta\left( \theta_i \prod^n_{l=1} S^{(l)}_i = q_r \right) 
= 
\nonumber \\ 
&& 
\qquad\qquad
\frac{1}{4} 
\sum^{\ }_{k_i} 
  \exp\left[ 
    i \frac{\pi}{2} k_i 
      \left( \theta_i + \sum^n_{l=1} S^{(l)}_i - (n+q_r) \right) 
  \right] 
. 
\nonumber \\ 
\label{eq: delta rewritten}
\eea

Substituting into Eq.~(\ref{eq: Tr(rho^n_A(T)) f.c. -- 4}), we obtain 
\begin{widetext}
\bea
\mathcal{Z}^{(P)}(n) 
&=& 
\left( 
  \frac{1}
       {Z^{\textrm{tot}}_{J}(\openone) \: 
        2^{\N^{(c)}_{\A} + m^{\ }_{\B} - 1}_{\ }} 
\right)^{n}_{\ } 
\frac{e^{\N^{(p)}_{\A} A_n}_{\ }}{2^{\N^{(p)}_{\A}}_{\ }}
\: 
2^{\N^{(c)}_{\A} n}_{\ } 
\frac{1}{2}
  \sum^{\ }_{\left\{ \{ S^{(l)}_{i} \} \right\}^{n}_{l=1}} 
  \sum^{\ }_{\{ \theta_i \}} 
  \prod^{\ }_{\langle ij\rangle \in \A} 
    e^{B_n \theta_i \theta_j}
\sum^{\textrm{(even)}}_{\{ \overline{\eta}^{(l)}_{\ } \}^{n}_{l=1}} 
  \prod^{\ }_{\langle ij\rangle\notin\A} \: 
    \left[ 
      \exp\left(
        J \sum^{n}_{l=1} \overline{\eta}^{(l)}_{ij} S^{(l)}_i S^{(l)}_j 
      \right) 
    \right] 
\nonumber \\ 
&\times&
\sum_{\{ q_r = \pm 1 \}^{m_\B}_{r=1}} \: 
  \prod^{m_\B}_{r=1} 
    \prod_{i \in \partial_r} 
\frac{1}{4} 
\sum^{\ }_{k_i} 
  \exp\left[ 
    i \frac{\pi}{2} k_i 
      \left( \theta_i + \sum^n_{l=1} S^{(l)}_i - (n+q_r) \right) 
  \right] 
\nonumber \\ 
&=& 
\left( 
  \frac{1}
       {Z^{\textrm{tot}}_{J}(\openone) \: 
        2^{m^{\ }_{\B} - 1}_{\ }} 
\right)^{n}_{\ } 
\frac{e^{\N^{(p)}_{\A} A_n}_{\ }}{2^{\N^{(p)}_{\A}}_{\ }}
\frac{1}{2} 
\frac{1}{4^{\N^{\ }_{\partial}}} 
\sum_{\{ q_r = \pm 1 \}^{m_\B}_{r=1}} \: 
\sum^{\ }_{\{ k_i \}^{\N^{\ }_{\partial}}_{i=1} } 
\nonumber \\ 
&\times&
  \sum^{\ }_{\{ \theta_i \}} 
  \prod^{\ }_{\langle ij\rangle \in \A} 
    e^{B_n \theta_i \theta_j}
\sum^{\textrm{(even)}}_{\{ \overline{\eta}^{(l)}_{\ } \}^{n}_{l=1}} 
\; 
 \sum^{\ }_{\{ S^{(1)}_{i} \}} 
    \left[ 
      \exp\left(
        J \sum^{\ }_{\langle ij\rangle\notin\A} 
	  \overline{\eta}^{(1)}_{ij} S^{(1)}_i S^{(1)}_j 
      \right) 
    \right] 
\exp\left[ 
  i \frac{\pi}{2} \sum^{m_\B}_{r=1} \sum_{i \in \partial_r} k_i 
    \left( \theta_i + S^{(1)}_i - 1 - q_r \right) 
\right] 
\nonumber \\ 
&\times&
 \sum^{\ }_{\left\{ \{ S^{(l)}_{i} \} \right\}^{n}_{l=2}} 
    \left[ 
      \exp\left(
        J \sum^{\ }_{\langle ij\rangle\notin\A} \: 
	  \sum^{n}_{l=2} \overline{\eta}^{(l)}_{ij} S^{(l)}_i S^{(l)}_j 
      \right) 
    \right] 
\exp\left\{ 
  i \frac{\pi}{2} \sum_{i \in \partial} k_i 
    \left[ \sum^n_{l=2} \left( S^{(l)}_i -1 \right) \right] 
\right\} 
, 
\label{eq: Tr(rho^n_A(T)) f.c. -- 5} 
\eea
\end{widetext}
where $\partial$ and $\N^{\ }_{\partial}$ are, respectively, the full set and the 
total number of boundary sites, i.e., sites that have adjacent bonds both 
in $\A$ and outside $\A$. 
In the language introduced earlier, 
$
\N^{(c)}_{\ } 
= 
\N^{(c)}_{\A} + \N^{(c)}_{\overline{\A}} 
= 
\N^{(c)}_{\A} + \N^{(c)}_{\B} + \N^{\ }_{\partial}
$, 
and therefore 
$\N^{(c)}_{\overline{\A}} = \N^{(c)}_{\B} + \N^{\ }_{\partial}$. 

Note that the last line in Eq.~(\ref{eq: Tr(rho^n_A(T)) f.c. -- 5}) does 
not depend on the $S^{(1)}$ or $\theta$ spins. 
If we introduce the partition functions 
\bea
Z^{\overline{\A},+}_{\{ k_i \}} 
&=& 
\sum^{\ }_{\{ S^{\ }_{i} \}} 
  \exp\left[
    J \sum^{\ }_{\langle ij\rangle\notin\A} S^{\ }_i S^{\ }_j 
    + 
    i \frac{\pi}{2} 
      \sum_{i \in \partial} k_i \left( S^{\ }_i - 1 \right)
  \right] 
\nonumber 
\eea
\bea
Z^{\overline{\A},-}_{\{ k_i \}} 
&=& 
\sum^{\ }_{\{ S^{\ }_{i} \}} 
  \exp\left[
    J \sum^{\ }_{\langle ij\rangle\notin\A} \eta^{1}_{ij} S^{\ }_i S^{\ }_j 
    + 
    i \frac{\pi}{2} 
      \sum_{i \in \partial} k_i \left( S^{\ }_i - 1 \right) 
  \right] 
, 
\nonumber 
\eea
we can carry out the summation over the even number of collective operations 
$\{ \overline{\eta}^{(l)}_{\ } \}^{n}_{l=1}$ explicitly, and arrive at 
\begin{widetext}
\bea
\mathcal{Z}^{(P)}(n) 
&=& 
\left( 
  \frac{1}
       {Z^{\textrm{tot}}_{J}(\openone) \: 
        2^{m^{\ }_{\B} - 1}_{\ }} 
\right)^{n}_{\ } 
\frac{e^{\N^{(p)}_{\A} A_n}_{\ }}{2^{\N^{(p)}_{\A}}_{\ }}
\frac{1}{2} 
\sum_{\{ q_r = \pm 1 \}^{m_\B}_{r=1}} \: 
\frac{1}{4^{\N^{\ }_{\partial}}} 
\sum^{\ }_{\{ k_i \}^{\N^{\ }_{\partial}}_{i=1} } 
\nonumber \\ 
&\times&
\sum^{\ }_{\{ \theta_i \}} 
  \exp\left( 
    B_n \sum^{\ }_{\langle ij\rangle \in \A} \theta_i \theta_j
  \right) 
\sum^{\ }_{\{ S^{(1)}_{i} \}} 
\exp\left[ 
  i \frac{\pi}{2} \sum^{m_\B}_{r=1} \sum_{i \in \partial_r} k_i 
    \left( \theta_i + S^{(1)}_i - 1 - q_r\right)
\right] 
\nonumber \\ 
&\times&
\frac{1}{2}
\left\{ 
\left[ 
  \exp\left(
    J \sum^{\ }_{\langle ij\rangle\notin\A} 
      S^{(1)}_i S^{(1)}_j 
  \right) 
 + 
  \exp\left(
    J \sum^{\ }_{\langle ij\rangle\notin\A} 
      \eta^{1}_{ij} S^{(1)}_i S^{(1)}_j 
  \right) 
\right]
\left(
  Z^{\overline{\A},+}_{\{ k_i \}} + Z^{\overline{\A},-}_{\{ k_i \}}
\right)^{n-1}
\right. 
\nonumber \\ 
&&
\;\;\;
\left. 
+ 
\left[ 
  \exp\left(
    J \sum^{\ }_{\langle ij\rangle\notin\A} 
      S^{(1)}_i S^{(1)}_j 
  \right) 
 - 
  \exp\left(
    J \sum^{\ }_{\langle ij\rangle\notin\A} 
      \eta^{1}_{ij} S^{(1)}_i S^{(1)}_j 
  \right) 
\right]
\left(
  Z^{\overline{\A},+}_{\{ k_i \}} - Z^{\overline{\A},-}_{\{ k_i \}}
\right)^{n-1}
\right\}
. 
\label{eq: Tr(rho^n_A(T)) f.c. -- 6} 
\eea
%
%

We are finally in the position to take the derivative with respect to $n$, 
and to compute the von Neumann entropy of the bipartition 
%
%
\begin{subequations}
\bea
S^{(P)}_{\textrm{VN}}(\A;T/\lambda^{\ }_{B})
&=& 
- \lim^{\ }_{n \to 1} 
  \partial^{\ }_{n} \mathcal{Z}^{(P)}(n) 
\nonumber \\ 
&=& 
- \lim^{\ }_{n \to 1} \partial^{\ }_{n} 
\left\{ 
\left( 
  \frac{1}
       {Z^{\textrm{tot}}_{J}(\openone) \: 
        2^{m^{\ }_{\B} - 1}_{\ }} 
\right)^{n}_{\ } 
\frac{e^{\N^{(p)}_{\A} A_n}_{\ }}{2^{\N^{(p)}_{\A}}_{\ }}
\right\} 
\sum^{\ }_{\{ \theta_i \}} 
  \exp\left( 
    B_1 \sum^{\ }_{\langle ij\rangle \in \A} \theta_i \theta_j
  \right)
\nonumber \\ 
&\times&
\sum^{\ }_{\{ S^{(1)}_{i} \}} 
    \left[ 
      \exp\left(
        J \sum^{\ }_{\langle ij\rangle\notin\A} 
	  S^{(1)}_i S^{(1)}_j 
      \right) 
    \right] 
\frac{1}{2}
\sum^{\ }_{\{ q_r \}^{m_\B}_{r=1}} 
\frac{1}{4^{\N^{\ }_{\partial}}} 
\sum^{\ }_{\{ k_i \}^{\N^{\ }_{\partial}}_{i=1} } 
\exp\left[ 
  i \frac{\pi}{2} \sum^{m_\B}_{r=1} \sum_{i \in \partial_r} k_i 
    \left( \theta_i + S^{(1)}_i - 1 - q_r\right)
\right] 
\label{eq: S_VN plaquette -- 1a}
\\ 
&-&
\left( 
  \frac{1}
       {Z^{\textrm{tot}}_{J}(\openone) \: 2^{m_\B - 1}_{\ }} 
\right) 
\frac{e^{\N^{(p)}_{\A} A_1}_{\ }}{2^{\N^{(p)}_{\A}}_{\ }}
\lim_{n \to 1} \partial_n 
\left[ 
  \sum^{\ }_{\{ \theta_i \}} 
    \exp\left( 
      B_n \sum^{\ }_{\langle ij\rangle \in \A} \theta_i \theta_j
    \right)
\right] 
\nonumber \\ 
&\times&
\sum^{\ }_{\{ S^{(1)}_{i} \}} 
    \left[ 
      \exp\left(
        J \sum^{\ }_{\langle ij\rangle\notin\A} 
	  S^{(1)}_i S^{(1)}_j 
      \right) 
    \right] 
\frac{1}{2}
\sum^{\ }_{\{ q_r \}^{m_\B}_{r=1}} 
\frac{1}{4^{\N^{\ }_{\partial}}} 
\sum^{\ }_{\{ k_i \}^{\N^{\ }_{\partial}}_{i=1} } 
\exp\left[ 
  i \frac{\pi}{2} \sum^{m_\B}_{r=1} \sum_{i \in \partial_r} k_i 
    \left( \theta_i + S^{(1)}_i - 1 - q_r\right)
\right] 
\label{eq: S_VN plaquette -- 1b}
\\ 
&-&
\left( 
  \frac{1}
       {Z^{\textrm{tot}}_{J}(\openone) \: 2^{m_\B - 1}_{\ }} 
\right) 
\frac{e^{\N^{(p)}_{\A} A_1}_{\ }}{2^{\N^{(p)}_{\A}}_{\ }}
\sum^{\ }_{\{ \theta_i \}} 
  \exp\left(
    B_1 \sum^{\ }_{\langle ij\rangle \in \A} \theta_i \theta_j
  \right)
\nonumber \\ 
&\times&
\sum^{\ }_{\{ S^{(1)}_{i} \}} 
\frac{1}{2}
\sum^{\ }_{\{ q_r \}^{m_\B}_{r=1}} 
\frac{1}{4^{\N^{\ }_{\partial}}} 
\sum^{\ }_{\{ k_i \}^{\N^{\ }_{\partial}}_{i=1} } 
\exp\left[ 
  i \frac{\pi}{2} \sum^{m_\B}_{r=1} \sum_{i \in \partial_r} k_i 
    \left( \theta_i + S^{(1)}_i - 1 - q_r\right)
\right] 
\nonumber \\ 
&\times&
\frac{1}{2}
\left\{ 
\left[ 
  \exp\left(
    J \sum^{\ }_{\langle ij\rangle\notin\A} 
      S^{(1)}_i S^{(1)}_j 
  \right) 
  + 
  \exp\left(
    J \sum^{\ }_{\langle ij\rangle\notin\A} 
      \eta^{1}_{ij} S^{(1)}_i S^{(1)}_j 
  \right) 
\right]
\ln 
\left(
  Z^{\overline{\A},+}_{\{ k_i \}} + Z^{\overline{\A},-}_{\{ k_i \}}
\right) 
\right. 
\nonumber \\ 
&&
\left. 
+ 
\left[ 
  \exp\left(
    J \sum^{\ }_{\langle ij\rangle\notin\A} 
      S^{(1)}_i S^{(1)}_j 
  \right) 
  - 
  \exp\left(
    J \sum^{\ }_{\langle ij\rangle\notin\A} 
      \eta^{1}_{ij} S^{(1)}_i S^{(1)}_j 
  \right) 
\right]
\ln 
\left(
  Z^{\overline{\A},+}_{\{ k_i \}} - Z^{\overline{\A},-}_{\{ k_i \}}
\right)
\right\}
. 
\label{eq: S_VN plaquette -- 1c}
\eea
\label{eq: S_VN plaquette -- 1}
\end{subequations}
\end{widetext}

The summation over $\{ k_i \}$ can be carried out explicitly both in the 
first~(\ref{eq: S_VN plaquette -- 1a}) and 
second~(\ref{eq: S_VN plaquette -- 1b}) contribution to 
Eq.~(\ref{eq: S_VN plaquette -- 1}). 
This leads to a delta function that identifies 
$\theta_i = q_r \, S^{(1)}_i$, $i \in \partial_r$ and $r = 1,\ldots,m_\B$. 
One can verify that the factor $q_r$ is actually immaterial, and 
the $\theta$ and $S^{(1)}$ terms in the above equation can be 
gathered into a single partition function 
\bea
&& 
\sum^{\ }_{\{ \theta_i \}} 
  \exp\left(
    B_1 \sum^{\ }_{\langle ij\rangle \in \A} \theta_i \theta_j
  \right)
\sum^{\ }_{\{ S^{(1)}_{i} \}} 
  \exp\left(
    J \sum^{\ }_{\langle ij\rangle\notin\A} 
      S^{(1)}_i S^{(1)}_j 
  \right) 
\nonumber \\ 
&& \qquad 
= 
\sum^{\ }_{\{ S^{\ }_{i} \}} 
  \exp\left(
    J \sum^{\ }_{\langle ij\rangle} 
      S^{\ }_i S^{\ }_j 
  \right) 
\equiv 
Z^{\textrm{tot}}_{J}(\openone) 
, 
\eea
where we used the fact that $B_1 = J$ 
(see Eq.~(\ref{eq: rho_n derivative tools f.c. 1}) below). 
The summation over $\{ q_r = \pm 1 \}^{m_\B}_{r=1}$ becomes then trivial, 
yielding an overall factor $2^{m_\B}_{\ }$. 

In the third contribution~(\ref{eq: S_VN plaquette -- 1c}), 
each summation over $q_r = \pm 1$ yields a factor 
$2 \cos(\pi \sum_{i \in \partial_r} k_i / 2)$, 
which vanishes unless $\sum k_i$ is even. 
Thus, we can constrain the summation over 
$\{ k_i = 0,\ldots,3 \}_{i \in \partial_r}$ to 
satisfy this condition, and we can drop the terms 
$
\exp\left[ 
  i \frac{\pi}{2} \sum_{i \in \partial_r} k_i 
    \left( 1 - q_r \right)
\right]
$, 
since $1-q_r$ is even and the term is identically one. 
The summation over $\{ q_r = \pm 1 \}^{m_\B}_{r=1}$ becomes again trivial. 
In particular, 
\begin{widetext}
\bea
\exp\left[ 
  i \frac{\pi}{2} \sum_{i \in \partial} k_i 
    \left( \theta_i + S^{(1)}_i \right) 
\right]
= 
\exp\left\{ 
  i \frac{\pi}{2} \sum_{i \in \partial} k_i 
    \left[ \left( \theta_i - 1 \right) + \left( S^{(1)}_i - 1 \right) \right] 
\right\}
\eea
\end{widetext}
for the same reasoning, and we can write the $\theta$ and $S^{(1)}$ terms in 
a more compact form using the definition of 
$Z^{\overline{\A},\pm}_{\{ k_i \}}$, and introducing the notation 
\bea
Z^{\overline{\B},+}_{\{ k_i \}} 
&=& 
\sum^{\ }_{\{ \theta^{\ }_{i} \}} 
  \exp\left[
    J \sum^{\ }_{\langle ij\rangle\in\A} \theta^{\ }_i \theta^{\ }_j 
    + 
    i \frac{\pi}{2} \sum_{i \in \partial} k_i 
      \left( \theta^{\ }_i - 1 \right) 
  \right] 
. 
\nonumber \\ 
\eea
(The labeling $\overline{\B}$ instead of $\A$ is used here as a reminder 
that the summation over $\{ \theta^{\ }_{i} \}$ includes both spins 
surrounded only by bonds in $\A$, and spins on the boundary $\partial$. 
Therefore, the total number of $\theta$ spins is 
$\N^{(c)}_{\overline{\B}} = \N^{(c)}_{\A} + \N^{\ }_{\partial}$.) 

These considerations allow us to simplify 
Eq.~(\ref{eq: S_VN plaquette -- 1}) to 
\begin{widetext}
\begin{subequations}
\bea
S^{(P)}_{\textrm{VN}}(\A;T/\lambda^{\ }_{B})
&=& 
- \lim^{\ }_{n \to 1} \partial^{\ }_{n} 
\left\{ 
\left( 
  \frac{1}
       {Z^{\textrm{tot}}_{J}(\openone) \: 2^{m^{\ }_{\B} - 1}_{\ }} 
\right)^{n-1}_{\ } 
\frac{e^{\N^{(p)}_{\A} A_n}_{\ }}{2^{\N^{(p)}_{\A}}_{\ }}
\right\} 
\label{eq: S_VN plaquette -- 2a}
\\ 
&-&
\left( 
  \frac{1}
       {Z^{\textrm{tot}}_{J}(\openone)} 
\right) 
\frac{e^{\N^{(p)}_{\A} A_1}_{\ }}{2^{\N^{(p)}_{\A}}_{\ }}
\; 
\lim_{n \to 1} \partial_n 
\left[ 
  \sum^{\ }_{\{ S_i \}} 
    \exp\left( 
      B_n \sum^{\ }_{\langle ij\rangle \in \A} S_i S_j 
      + 
      J \sum^{\ }_{\langle ij\rangle\notin\A} S^{\ }_i S^{\ }_j 
    \right) 
\right] 
\label{eq: S_VN plaquette -- 2b}
\\ 
&-&
\left( 
  \frac{1}
       {Z^{\textrm{tot}}_{J}(\openone)} 
\right) 
\frac{e^{\N^{(p)}_{\A} A_1}_{\ }}{2^{\N^{(p)}_{\A}}_{\ }}
\frac{1}{4^{\N^{\ }_{\partial}}} 
\nonumber \\ 
&\times&
\sum^{\textrm{(even)}}_{\{ k_i \}^{\N^{\ }_{\partial}}_{i=1} } 
\frac{Z^{\overline{\B},+}_{\{ k_i \}}}{2}
\left\{ 
\left( 
  Z^{\overline{\A},+}_{\{ k_i \}}
  + 
  Z^{\overline{\A},-}_{\{ k_i \}}
\right)
\ln 
\left(
  Z^{\overline{\A},+}_{\{ k_i \}} + Z^{\overline{\A},-}_{\{ k_i \}}
\right) 
+ 
\left( 
  Z^{\overline{\A},+}_{\{ k_i \}}
  - 
  Z^{\overline{\A},-}_{\{ k_i \}}
\right)
\ln 
\left(
  Z^{\overline{\A},+}_{\{ k_i \}} - Z^{\overline{\A},-}_{\{ k_i \}}
\right)
\right\}
. 
\label{eq: S_VN plaquette -- 2c}
\eea
\label{eq: S_VN plaquette -- 2}
\end{subequations}
\end{widetext}

In order to proceed further, let us first study some of the terms in 
Eq.~(\ref{eq: S_VN plaquette -- 2}) separately. 
{}From Eqs.~(\ref{eq: A_n B_n f.c.}) we have that 
\begin{subequations} 
\bea
A_n 
&=& 
\frac{1}{2} \ln 
\left\{\vphantom{\sum}
\left[\vphantom{\sum}
  \left(2\cosh J\right)^n+\left(2\sinh J\right)^n
\right] 
\right.
\nonumber \\ 
&& 
\quad\;\;
\times
\left.
\left[\vphantom{\sum}
  \left(2\cosh J\right)^n-\left(2\sinh J\right)^n
\right] 
\right\}
\nonumber \\ 
&=& 
\frac{1}{2} \ln 
\left\{
\left(2\cosh J\right)^{2n}-\left(2\sinh J\right)^{2n} 
\right\}
\eea
\bea
B_n 
&=& 
\frac{1}{2} \ln 
\frac{\left(2\cosh J\right)^n+\left(2\sinh J\right)^n}
     {\left(2\cosh J\right)^n-\left(2\sinh J\right)^n} 
\\ 
A_1 
&=& 
\ln 2 
\\ 
B_1 
&=& 
\frac{1}{2} \ln \frac{1+\tanh{J}}{1-\tanh{J}} 
= 
J 
\eea
\bea 
\left. \frac{d}{dn} A_n \right\vert_{n=1} 
&=& 
\ln 2 
+ 
\cosh^{2}{J} \: \ln\left(\cosh{J}\right) 
- 
\sinh^{2}{J} \: \ln\left(\sinh{J}\right) 
\nonumber \\ 
\\
\left. \frac{d}{dn} B_n \right\vert_{n=1} 
&=& 
\sinh{J}\,\cosh{J}\:\ln\frac{\sinh{J}}{\cosh{J}} . 
\eea
\label{eq: rho_n derivative tools f.c. 1}
\end{subequations} 
Notice that 
$\left. \frac{d}{dn} A_n \right\vert_{n=1} \to \ln2$ for $J \to 0$, 
$\left. \frac{d}{dn} A_n \right\vert_{n=1} \sim J+1/2+\mathcal{O}(e^{-2J})$ 
for $J \to \infty$,  
and that 
$\left. \frac{d}{dn} B_n \right\vert_{n=1} \to 0$ for $J \to 0$, 
$\left. \frac{d}{dn} B_n \right\vert_{n=1} \to -1/2 + \mathcal{O}(e^{-2J})$ 
for $J \to \infty$. 

{}We can also carry out the derivative in 
Eq.~(\ref{eq: S_VN plaquette -- 2b}): 
\begin{widetext}
\begin{subequations} 
\bea
\lim_{n \to 1} \partial_n 
\left[ 
  \sum^{\ }_{\{ S_i \}} 
    \exp\left( 
      B_n \sum^{\ }_{\langle ij\rangle \in \A} S_i S_j 
      + 
      J \sum^{\ }_{\langle ij\rangle\notin\A} S^{\ }_i S^{\ }_j 
    \right) 
\right] 
&=& 
\left. 
  \frac{d}{dn} B_n 
\right\vert^{\ }_{n=1} 
\; 
\sum^{\ }_{\{ S_i \}} 
  \left( \sum^{\ }_{\langle ij\rangle\in\A} S_i S_j \right) 
    \exp\left(
      B_1 \sum^{\ }_{\langle ij\rangle\in\A} S_i S_j + 
       J \sum^{\ }_{\langle ij\rangle\notin\A} S_i S_j 
    \right) 
\nonumber \\ 
&=& 
\sinh{J}\,\cosh{J}\:\ln\frac{\sinh{J}}{\cosh{J}} 
\; 
\sum^{\ }_{\{ S_i \}} 
  \left( \sum^{\ }_{\langle ij\rangle\in\A} S_i S_j \right) 
    \exp\left( J \sum^{\ }_{\langle ij\rangle} S_i S_j \right) 
\nonumber \\ 
&=& 
\sinh{J}\,\cosh{J}\:\ln\frac{\sinh{J}}{\cosh{J}} 
\; 
\langle 
  E^{\ }_{\A} 
\rangle^{\ }_{Z^{\textrm{tot}}_{J}(\openone)} 
\; 
Z^{\textrm{tot}}_{J}(\openone) 
\\ 
\langle 
  E^{\ }_{\A} 
\rangle^{\ }_{Z^{\textrm{tot}}_{J}(\openone)} 
&\equiv& 
\frac{
\sum^{\ }_{\{ S_i \}} 
  \left( \sum^{\ }_{\langle ij\rangle\in\A} S_i S_j \right) 
    \exp\left( J \sum^{\ }_{\langle ij\rangle} S_i S_j \right) 
     }
     {Z^{\textrm{tot}}_{J}(\openone)}, 
\eea
\label{eq: rho_n derivative tools f.c. 2}
\end{subequations} 
%
%
where $E^{\ }_{\A}$ is the extensive energy of the bonds in $\A$ (in units 
of $J$), in the Ising model described by the equilibrium partition function 
$Z^{\textrm{tot}}_{J}(\openone)$. 

The last calculation we still need is 
\bea
&&
\left. 
  \frac{d}{dn}  e^{\N^{(p)}_{\A} A_n}_{\ } 
\right\vert^{\ }_{n=1} 
= 
\N^{(p)}_{\A} 2^{\N^{(p)}_{\A}}_{\ } 
\left[\vphantom\int 
  \ln 2 
  + 
  \cosh^{2}{J} \: \ln\left(\cosh{J}\right) 
  - 
  \sinh^{2}{J} \: \ln\left(\sinh{J}\right) 
\right] 
. 
\nonumber \\
\label{eq: rho_n derivative tools f.c. 3}
\eea

Combining all the results in Eqs.~(\ref{eq: rho_n derivative tools f.c. 1}), 
(\ref{eq: rho_n derivative tools f.c. 2}) 
and~(\ref{eq: rho_n derivative tools f.c. 3}), 
Eq.~(\ref{eq: S_VN plaquette -- 2}) reduces to 
%
%
\begin{subequations}
\bea
S^{(P)}_{\textrm{VN}}(\A;T/\lambda^{\ }_{B})
&=& 
\ln \left(\vphantom\int 2^{m_\B - 1}_{\ } \right)
+ 
\ln \, Z^{\textrm{tot}}_{J}(\openone)
- 
\N^{(p)}_{\A} 
  \left[\vphantom\int 
    \ln 2 
    + 
    \cosh^{2}{J} \: \ln\left(\cosh{J}\right) 
    - 
    \sinh^{2}{J} \: \ln\left(\sinh{J}\right) 
  \right] 
\label{eq: S_VN plaquette -- 3a}
\\ 
&-&
\sinh{J}\,\cosh{J}\:\ln\frac{\sinh{J}}{\cosh{J}} 
\; 
\langle 
  E^{\ }_{\A} 
\rangle^{\ }_{Z^{\textrm{tot}}_{J}(\openone)} 
\label{eq: S_VN plaquette -- 3b}
\\ 
&-&
\frac{1}
     {Z^{\textrm{tot}}_{J}(\openone)} 
\frac{1}{4^{\N^{\ }_{\partial}}} 
\sum^{\textrm{(even)}}_{\{ k_i \}^{\N^{\ }_{\partial}}_{i=1} } 
\frac{Z^{\overline{\B},+}_{\{ k_i \}}}{2} 
\left[\vphantom\int 
\left( 
  Z^{\overline{\A},+}_{\{ k_i \}} 
  + 
  Z^{\overline{\A},-}_{\{ k_i \}} 
\right)
\ln 
\left(
  Z^{\overline{\A},+}_{\{ k_i \}} + Z^{\overline{\A},-}_{\{ k_i \}}
\right) 
+ 
\left( 
  Z^{\overline{\A},+}_{\{ k_i \}} 
  - 
  Z^{\overline{\A},-}_{\{ k_i \}} 
\right)
\ln 
\left(
  Z^{\overline{\A},+}_{\{ k_i \}} - Z^{\overline{\A},-}_{\{ k_i \}}
\right)
\right]
. 
\nonumber \\ 
\label{eq: S_VN plaquette -- 3c}
\eea
\label{eq: S_VN plaquette -- 3}
\end{subequations}
\end{widetext}

Recall that $\sum k_i$ is even, and therefore $\sum k_i S_i$ 
is also even, irrespective of the values of the spins $\{ S_i = \pm 1 \}$. 
In particular, 
\bea
&& 
\exp\left[ 
  i \frac{\pi}{2} \sum^{\ }_{i \in \partial} 
    k_i \left( S^{\ }_{i} - 1 \right) 
\right] 
= 
\nonumber \\ 
&& 
\qquad\qquad
= 
\prod_{i  \in \partial}
\left[ 
e^{-i \frac{\pi}{2} k_i}_{\ } 
  \cos \frac{\pi}{2} k_i 
  + 
  i e^{-i \frac{\pi}{2} k_i}_{\ } 
  \left( \sin \frac{\pi}{2} k_i \right) S^{\ }_{i}
\right] 
\nonumber \\ 
&& 
\qquad\qquad
= 
\prod_{i  \in \partial}
\left[ 
  \delta_{k_i \; \textrm{even}} 
  + 
  S^{\ }_{i} \, \delta_{k_i \; \textrm{odd}} 
\right] 
, 
\eea
and both $Z^{\A,+}_{\{ k_i \}}$ and $Z^{\overline{\A},+}_{\{ k_i \}}$ can be 
rewritten as 
\bea
Z^{\A,+}_{\{ k_i \}} 
&=& 
\sum^{\ }_{\{ S^{\ }_{i} \}} 
  \left( \prod^{k_i \; \textrm{odd}}_{i \in \partial} S^{\ }_{i} \right) 
  \exp\left(
    J \sum^{\ }_{\langle ij\rangle\in\A} S^{\ }_i S^{\ }_j 
  \right) 
\nonumber \\ 
&=& 
Z^{\A,+}_{\ } \; 
\left\langle
\prod^{k_i \; \textrm{odd}}_{i \in \partial} S^{\ }_{i}
\right\rangle
\\ 
Z^{\overline{\A},+}_{\{ k_i \}} 
&=& 
\sum^{\ }_{\{ S^{\ }_{i} \}} 
  \left( \prod^{k_i \; \textrm{odd}}_{i \in \partial} S^{\ }_{i} \right) 
  \exp\left(
    J \sum^{\ }_{\langle ij\rangle\notin\A} S^{\ }_i S^{\ }_j 
  \right) 
\nonumber \\ 
&=& 
Z^{\overline{\A},+}_{\ } \; 
\left\langle
\prod^{k_i \; \textrm{odd}}_{i \in \partial} S^{\ }_{i}
\right\rangle
, 
\eea
where 
$
Z^{\A,+}_{\ }
= 
\sum^{\ }_{\{ S^{\ }_{i} \}} 
  \exp\left(
    J \sum^{\ }_{\langle ij\rangle\in\A} S^{\ }_i S^{\ }_j 
  \right) 
$ 
and 
$
Z^{\overline{\A},+}_{\ }
= 
\sum^{\ }_{\{ S^{\ }_{i} \}} 
  \exp\left(
    J \sum^{\ }_{\langle ij\rangle\notin\A} S^{\ }_i S^{\ }_j 
  \right) 
$. 
Similarly for $Z^{\A,-}_{\{ k_i \}}$ and $Z^{\overline{\A},-}_{\{ k_i \}}$. 
Thus, all these quantities can be interpreted as correlation functions of 
boundary spins located at the odd entries of the set $\{ k_i \}$ times a 
partition function. 
Note that the constraint $\sum_{i \in \partial_r} k_i$ even, $\forall\,r$, 
requires that the number of such odd entries is also even separately on each 
boundary component $r=1,\ldots,m_\B$. 

If we are interested in computing the topological entropy of the system, 
it is convenient to decompose the last term in 
Eq.~(\ref{eq: S_VN plaquette -- 3}) so that 
\begin{widetext}
\begin{subequations}
\bea
S^{(P)}_{\textrm{VN}}(\A;T/\lambda^{\ }_{B})
&=& 
\ln \left(\vphantom\int 2^{m_\B - 1}_{\ } \right)
+ 
\ln \, Z^{\textrm{tot}}_{J}(\openone)
- 
\N^{(p)}_{\A} 
  \left[\vphantom\int 
    \ln 2 
    + 
    \cosh^{2}{J} \: \ln\left(\cosh{J}\right) 
    - 
    \sinh^{2}{J} \: \ln\left(\sinh{J}\right) 
  \right] 
\label{eq: S_VN plaquette -- 4a}
\\ 
&-&
\sinh{J}\,\cosh{J}\:\ln\frac{\sinh{J}}{\cosh{J}} 
\; 
\langle 
  E^{\ }_{\A} 
\rangle^{\ }_{Z^{\textrm{tot}}_{J}(\openone)} 
\label{eq: S_VN plaquette -- 4b}
\\ 
&-&
\frac{1}{4^{\N^{\ }_{\partial}}} 
\sum^{\textrm{(even)}}_{\{ k_i \}^{\N^{\ }_{\partial}}_{i=1} } 
\frac{Z^{\overline{\B},+}_{\{ k_i \}} \, Z^{\overline{\A},+}_{\{ k_i \}}} 
     {Z^{\textrm{tot}}_{J}(\openone)} 
\ln Z^{\overline{\A},+}_{\{ k_i \}} 
\label{eq: S_VN plaquette -- 4c}
\\ 
&-&
\frac{1}{4^{\N^{\ }_{\partial}}} 
\sum^{\textrm{(even)}}_{\{ k_i \}^{\N^{\ }_{\partial}}_{i=1} } 
\frac{Z^{\overline{\B},+}_{\{ k_i \}} \, Z^{\overline{\A},+}_{\{ k_i \}}} 
     {Z^{\textrm{tot}}_{J}(\openone)} 
\frac{1}{2} 
\left[\vphantom\int 
\left( 
  1 
  + 
  \frac{Z^{\overline{\A},-}_{\{ k_i \}}}{Z^{\overline{\A},+}_{\{ k_i \}}} 
\right)
\ln 
\left(
  1 
  + 
  \frac{Z^{\overline{\A},-}_{\{ k_i \}}}{Z^{\overline{\A},+}_{\{ k_i \}}}
\right) 
+ 
\left( 
  1 
  - 
  \frac{Z^{\overline{\A},-}_{\{ k_i \}}}{Z^{\overline{\A},+}_{\{ k_i \}}} 
\right)
\ln 
\left(
  1 
  - 
  \frac{Z^{\overline{\A},-}_{\{ k_i \}}}{Z^{\overline{\A},+}_{\{ k_i \}}}
\right)
\right]
. 
\nonumber \\ 
\label{eq: S_VN plaquette -- 4d}
\eea
\label{eq: S_VN plaquette -- 4}
\end{subequations}
\end{widetext}

The result in Eq.~(\ref{eq: S_VN plaquette -- 4}) holds for 
$n^{\ }_{\A} = 1$ (i.e., there is only one collective operation in $\A$). 
In order to compute the topological entropy of the system with the 
bipartition scheme in Sec.~\ref{sec: zero temperature gauge}, we also need 
to consider the case where $n^{\ }_{\A} = 0$. 
Repeating the derivation above, from Eq.~(\ref{eq: Tr(rho^n_A(T)) f.c. -- 5}) 
to Eq.~(\ref{eq: S_VN plaquette -- 4}), in the absence of collective 
operations leads rather straightforwardly to the result that 
\begin{widetext}
\bea
\mathcal{Z}^{(P)}(n) 
&=& 
\left( 
  \frac{1}
       {Z^{\textrm{tot}}_{J}(\openone) \: 2^{m^{\ }_{\B} - 1}_{\ }} 
\right)^{n}_{\ } 
\frac{e^{\N^{(p)}_{\A} A_n}_{\ }}{2^{\N^{(p)}_{\A}}_{\ }}
\frac{1}{2} 
\sum_{\{ q_r \}^{m_\B}_{r=1}} \: 
\frac{1}{4^{\N^{\ }_{\partial}}} 
\sum^{\ }_{\{ k_i \}^{\N^{\ }_{\partial}}_{i=1} } 
\sum^{\ }_{\{ \theta_i \}} 
  \exp\left( 
    B_n \sum^{\ }_{\langle ij\rangle \in \A} \theta_i \theta_j
  \right) 
\nonumber \\ 
&\times&
\sum^{\ }_{\{ S^{(1)}_{i} \}} 
\exp\left[ 
  i \frac{\pi}{2} \sum^{m_\B}_{r=1} \sum_{i \in \partial_r} k_i 
    \left( \theta_i + S^{(1)}_i - 1 - q_r \right)
\right] 
  \left[ 
    \exp\left(
      J \sum^{\ }_{\langle ij\rangle\notin\A} 
	S^{(1)}_i S^{(1)}_j 
    \right) 
  \right] 
\left(
  Z^{\overline{\A},+}_{\{ k_i \}} 
\right)^{n-1} 
\label{eq: Tr(rho^n_A(T)) f.c. -- 6 no coll.} 
\eea
and
\begin{subequations}
\bea
S^{(P)}_{\textrm{VN}}(\A;T/\lambda^{\ }_{B})
&=& 
\ln \left(\vphantom\int 2^{m_\B - 1}_{\ } \right)
+ 
\ln \, Z^{\textrm{tot}}_{J}(\openone)
- 
\N^{(p)}_{\A} 
  \left[\vphantom\int 
    \ln 2 
    + 
    \cosh^{2}{J} \: \ln\left(\cosh{J}\right) 
    - 
    \sinh^{2}{J} \: \ln\left(\sinh{J}\right) 
  \right] 
\label{eq: S_VN plaquette -- 4a no coll.}
\\ 
&-&
\sinh{J}\,\cosh{J}\:\ln\frac{\sinh{J}}{\cosh{J}} 
\; 
\langle 
  E^{\ }_{\A} 
\rangle^{\ }_{Z^{\textrm{tot}}_{J}(\openone)} 
\label{eq: S_VN plaquette -- 4b no coll.}
\\ 
&-&
\frac{1}{4^{\N^{\ }_{\partial}}} 
\sum^{\textrm{(even)}}_{\{ k_i \}^{\N^{\ }_{\partial}}_{i=1} } 
\frac{Z^{\overline{\B},+}_{\{ k_i \}} \, Z^{\overline{\A},+}_{\{ k_i \}}} 
     {Z^{\textrm{tot}}_{J}(\openone)} 
\ln Z^{\overline{\A},+}_{\{ k_i \}} 
%
%
%
\label{eq: S_VN plaquette -- 4c no coll.}
\eea
\label{eq: S_VN plaquette -- 4 no coll.}
\end{subequations}
\end{widetext}
Notice that Eq.~(\ref{eq: S_VN plaquette -- 4 no coll.}) differs from 
Eq.~(\ref{eq: S_VN plaquette -- 4}) only in that it lacks its last 
contribution~(\ref{eq: S_VN plaquette -- 4d}). 

We can finally compute the plaquette contribution to the topological entropy 
$S^{(P)}_{\textrm{topo}}(T/\lambda^{\ }_{B})$, using the full bipartition 
scheme. All the terms that do not carry a topological contribution cancel. 
Namely, as discussed in Sec.~\ref{sec: zero temperature gauge}, 
\begin{subequations}
\bea
\N^{(p)}_{1\A} + \N^{(p)}_{4\A} &=& \N^{(p)}_{2\A} + \N^{(p)}_{3\A} 
, 
\eea
and on similar grounds 
\bea
\langle 
  E^{\ }_{1\A} 
\rangle^{\ }_{Z^{\textrm{tot}}_{J}(\openone)} 
+ 
\langle 
  E^{\ }_{4\A} 
\rangle^{\ }_{Z^{\textrm{tot}}_{J}(\openone)} 
&=& 
\langle 
  E^{\ }_{2\A} 
\rangle^{\ }_{Z^{\textrm{tot}}_{J}(\openone)} 
+ 
\langle 
  E^{\ }_{3\A} 
\rangle^{\ }_{Z^{\textrm{tot}}_{J}(\openone)} 
. 
\nonumber \\
\eea
\end{subequations}
Likewise for bipartitions 5-8. 
Recall also that $m_\B = 1$ and $n_\A = 0$ for all bipartitions, except 
bipartitions 4 and 5 (which have $m_\B = 1$ and $n_\A = 1$), and bipartition 
1, (which has $m_\B = 2$ and $n_\A = 0$). 
Using Eq.~(\ref{eq: S_VN plaquette -- 4}) and 
Eq.~(\ref{eq: S_VN plaquette -- 4 no coll.}) accordingly, we obtain 
\begin{widetext}
\bea
S^{(P)}_{\textrm{topo}}(T/\lambda^{\ }_{B}) 
&=& 
\ln
\left( 
  2^{-m^{\ }_{1\B}+m^{\ }_{2\B}+m^{\ }_{3\B}-m^{\ }_{4\B}}_{\ } 
\right) 
\nonumber \\ 
&+&
\frac{1}{4^{\N^{\ }_{\partial}}} 
\sum^{\textrm{(even)}}_{\{ k_i \}^{\N^{\ }_{\partial}}_{i=1} } 
\left\{
\frac{Z^{\overline{1\B},+}_{\{ k_i \}} \, Z^{\overline{1\A},+}_{\{ k_i \}}} 
     {Z^{\textrm{tot}}_{J}(\openone)} 
\ln Z^{\overline{1\A},+}_{\{ k_i \}} 
-
\frac{Z^{\overline{2\B},+}_{\{ k_i \}} \, Z^{\overline{2\A},+}_{\{ k_i \}}} 
     {Z^{\textrm{tot}}_{J}(\openone)} 
\ln Z^{\overline{2\A},+}_{\{ k_i \}} 
-
\frac{Z^{\overline{3\B},+}_{\{ k_i \}} \, Z^{\overline{3\A},+}_{\{ k_i \}}} 
     {Z^{\textrm{tot}}_{J}(\openone)} 
\ln Z^{\overline{3\A},+}_{\{ k_i \}} 
+
\frac{Z^{\overline{4\B},+}_{\{ k_i \}} \, Z^{\overline{4\A},+}_{\{ k_i \}}} 
     {Z^{\textrm{tot}}_{J}(\openone)} 
\ln Z^{\overline{4\A},+}_{\{ k_i \}} 
\right\}
\nonumber \\ 
&+& 
\frac{1}{4^{\N^{\ }_{\partial}}} 
\sum^{\textrm{(even)}}_{\{ k_i \}^{\N^{\ }_{\partial}}_{i=1} } 
\frac{Z^{\overline{4\B},+}_{\{ k_i \}} \, Z^{\overline{4\A},+}_{\{ k_i \}}} 
     {Z^{\textrm{tot}}_{J}(\openone)} 
\frac{1}{2} 
\left[\vphantom\int 
\left( 
  1 
  + 
  \frac{Z^{\overline{4\A},-}_{\{ k_i \}}}{Z^{\overline{4\A},+}_{\{ k_i \}}} 
\right)
\ln 
\left(
  1 
  + 
  \frac{Z^{\overline{4\A},-}_{\{ k_i \}}}{Z^{\overline{4\A},+}_{\{ k_i \}}}
\right) 
+ 
\left( 
  1 
  - 
  \frac{Z^{\overline{4\A},-}_{\{ k_i \}}}{Z^{\overline{4\A},+}_{\{ k_i \}}} 
\right)
\ln 
\left(
  1 
  - 
  \frac{Z^{\overline{4\A},-}_{\{ k_i \}}}{Z^{\overline{4\A},+}_{\{ k_i \}}}
\right)
\right]
\nonumber \\ 
&+& 
(\textrm{partitions}\;5\:\ldots\:8) 
. 
\eea
\end{widetext}
Using the fact that 
$
m^{\ }_{1\B}-m^{\ }_{2\B}-m^{\ }_{3\B}+m^{\ }_{4\B} 
= 
1 
$, 
that 
$
m^{\ }_{5\B}-m^{\ }_{6\B}-m^{\ }_{7\B}+m^{\ }_{8\B} 
= 
0
$, 
and that 
$
Z^{\overline{4\A},\pm}_{\{ k_i \}}(J) 
\equiv 
Z^{\overline{5\A},\pm}_{\{ k_i \}}(J)
$ 
since bipartitions 4 and 5 are in fact identical, 
one arrives to the result 
\begin{widetext}
\bea
S^{(P)}_{\textrm{topo}}(T/\lambda^{\ }_{B}) 
&=& 
- \ln 2 
\nonumber \\ 
\nonumber \\ 
&+&
\frac{1}{4^{\N^{\ }_{\partial}}} 
\sum^{\textrm{(even)}}_{\{ k_i \}^{\N^{\ }_{\partial}}_{i=1} } 
\left\{
\frac{Z^{\overline{1\B},+}_{\{ k_i \}} \, Z^{\overline{1\A},+}_{\{ k_i \}}} 
     {Z^{\textrm{tot}}_{J}(\openone)} 
\ln Z^{\overline{1\A},+}_{\{ k_i \}} 
-
\frac{Z^{\overline{2\B},+}_{\{ k_i \}} \, Z^{\overline{2\A},+}_{\{ k_i \}}} 
     {Z^{\textrm{tot}}_{J}(\openone)} 
\ln Z^{\overline{2\A},+}_{\{ k_i \}} 
-
\frac{Z^{\overline{3\B},+}_{\{ k_i \}} \, Z^{\overline{3\A},+}_{\{ k_i \}}} 
     {Z^{\textrm{tot}}_{J}(\openone)} 
\ln Z^{\overline{3\A},+}_{\{ k_i \}} 
+
\frac{Z^{\overline{4\B},+}_{\{ k_i \}} \, Z^{\overline{4\A},+}_{\{ k_i \}}} 
     {Z^{\textrm{tot}}_{J}(\openone)} 
\ln Z^{\overline{4\A},+}_{\{ k_i \}} 
\right\}
\nonumber \\ 
&+&
\frac{1}{4^{\N^{\ }_{\partial}}} 
\sum^{\textrm{(even)}}_{\{ k_i \}^{\N^{\ }_{\partial}}_{i=1} } 
\left\{
\frac{Z^{\overline{5\B},+}_{\{ k_i \}} \, Z^{\overline{5\A},+}_{\{ k_i \}}} 
     {Z^{\textrm{tot}}_{J}(\openone)} 
\ln Z^{\overline{5\A},+}_{\{ k_i \}} 
-
\frac{Z^{\overline{6\B},+}_{\{ k_i \}} \, Z^{\overline{6\A},+}_{\{ k_i \}}} 
     {Z^{\textrm{tot}}_{J}(\openone)} 
\ln Z^{\overline{6\A},+}_{\{ k_i \}} 
-
\frac{Z^{\overline{7\B},+}_{\{ k_i \}} \, Z^{\overline{7\A},+}_{\{ k_i \}}} 
     {Z^{\textrm{tot}}_{J}(\openone)} 
\ln Z^{\overline{7\A},+}_{\{ k_i \}} 
+
\frac{Z^{\overline{8\B},+}_{\{ k_i \}} \, Z^{\overline{8\A},+}_{\{ k_i \}}} 
     {Z^{\textrm{tot}}_{J}(\openone)} 
\ln Z^{\overline{8\A},+}_{\{ k_i \}} 
\right\}
\nonumber \\ 
&+& 
\frac{1}{4^{\N^{\ }_{\partial}}} 
\sum^{\textrm{(even)}}_{\{ k_i \}^{\N^{\ }_{\partial}}_{i=1} } 
\frac{Z^{\overline{4\B},+}_{\{ k_i \}} \, Z^{\overline{4\A},+}_{\{ k_i \}}} 
     {Z^{\textrm{tot}}_{J}(\openone)} 
\left[\vphantom\int 
\left( 
  1 
  + 
  \frac{Z^{\overline{4\A},-}_{\{ k_i \}}}{Z^{\overline{4\A},+}_{\{ k_i \}}} 
\right)
\ln 
\left(
  1 
  + 
  \frac{Z^{\overline{4\A},-}_{\{ k_i \}}}{Z^{\overline{4\A},+}_{\{ k_i \}}}
\right) 
+ 
\left( 
  1 
  - 
  \frac{Z^{\overline{4\A},-}_{\{ k_i \}}}{Z^{\overline{4\A},+}_{\{ k_i \}}} 
\right)
\ln 
\left(
  1 
  - 
  \frac{Z^{\overline{4\A},-}_{\{ k_i \}}}{Z^{\overline{4\A},+}_{\{ k_i \}}}
\right)
\right]
. 
\label{eq: S_topo plaquette}
\eea
\end{widetext}

This expression can be cast in a more useful way by noticing the
following. Factors like
\begin{eqnarray}
{\cal P}^{\,p}_{\{ k_i \}}&&\equiv
\frac{1}{4^{\N_{\partial^p}}} 
\frac{Z^{\overline{p\B},+}_{\{ k_i \}} \, Z^{\overline{p\A},+}_{\{ k_i \}}} 
     {Z^{\textrm{tot}}_{J}(\openone)} 
\label{prob-P}
\\
&&=\frac{1}{4^{\N^{\ }_{\partial^p}}} 
\frac{Z^{\overline{p\B},+} \, Z^{\overline{p\A},+}}
{Z^{\textrm{tot}}_{J}(\openone)} \;
\left\langle
\prod^{k_i \; \textrm{odd}}_{i \in \partial^p} \theta^{\ }_{i}
\right\rangle
\left\langle
\prod^{k_i \; \textrm{odd}}_{i \in \partial^p} S^{\ }_{i}
\right\rangle
\nonumber\\
&&\ge 0,
\nonumber
\end{eqnarray}
for each of the partitions $p=1,\dots,8$. This is because the
expectation values of the products of spins is always non-negative
because the interactions are {\it ferromagnetic} (this can be shown
explicitly in a high temperature expansion, for example). Recall that
the set $\{ k_i \}$ contains always an even number of odd $k_i$'s.

Moreover, one can check that
\begin{widetext}
\begin{eqnarray}
\sum^{\textrm{(even)}}_{\{ k_i \}^{\N^{\ }_{\partial^p}}_{i=1} } 
{\cal P}^{\,p}_{\{ k_i \}}
&&=
\frac{1}{Z^{\textrm{tot}}_{J}(\openone)} 
\sum^{\ }_{\{ \theta^{\ }_{i} \}} 
\sum^{\ }_{\{ S^{\ }_{i} \}} 
  \exp\left(
    J \sum^{\ }_{\langle ij\rangle\in p\A} \theta^{\ }_i \theta^{\ }_j 
  \right) 
  \exp\left(
    J \sum^{\ }_{\langle ij\rangle\not\in p\A} S^{\ }_i S^{\ }_j 
  \right) 
\;
\frac{1}{4^{\N^{\ }_{\partial^p}}} 
\sum^{\textrm{(even)}}_{\{ k_i \}^{\N^{\ }_{\partial^p}}_{i=1} } 
\exp\left[ 
  i \frac{\pi}{2} \sum_{i \in \partial^p} k_i 
    \left( \theta_i + S_i -2 \right) 
\right]
\nonumber\\
&&=
\frac{1}{Z^{\textrm{tot}}_{J}(\openone)} 
\sum^{\ }_{\{ \theta^{\ }_{i} \}} 
\sum^{\ }_{\{ S^{\ }_{i} \}} 
  \exp\left(
    J \sum^{\ }_{\langle ij\rangle\in p\A} \theta^{\ }_i \theta^{\ }_j 
  \right) 
  \exp\left(
    J \sum^{\ }_{\langle ij\rangle\not\in p\A} S^{\ }_i S^{\ }_j 
  \right) 
\;
\frac{1}{4^{\N^{\ }_{\partial^p}}} 
\sum^{\textrm{(even)}}_{\{ k_i \}^{\N^{\ }_{\partial^p}}_{i=1} } 
\exp\left[ 
  i \frac{\pi}{2} \sum_{i \in \partial^p} k_i 
    \left( \theta_i + S_i -1 \right) 
\right]
\nonumber \\ 
&& 
\qquad\qquad\qquad
\times 
\left[ 
  \cos \left( \frac{\pi}{2} \sum_{i \in \partial^p} k_i \right) 
  - i 
  \sin \left( \frac{\pi}{2} \sum_{i \in \partial^p} k_i \right)
\right]
\nonumber\\
&&=
\frac{1}{Z^{\textrm{tot}}_{J}(\openone)} 
\sum^{\ }_{\{ \theta^{\ }_{i} \}} 
\sum^{\ }_{\{ S^{\ }_{i} \}} 
  \exp\left(
    J \sum^{\ }_{\langle ij\rangle\in p\A} \theta^{\ }_i \theta^{\ }_j 
  \right) 
  \exp\left(
    J \sum^{\ }_{\langle ij\rangle\not\in p\A} S^{\ }_i S^{\ }_j 
  \right) 
\;
\frac{1}{2} \sum_{q = \pm 1} 
\frac{1}{4^{\N^{\ }_{\partial^p}}} 
\sum^{\ }_{\{ k_i \}^{\N^{\ }_{\partial^p}}_{i=1} } 
\exp\left[ 
  i \frac{\pi}{2} \sum_{i \in \partial^p} k_i 
    \left( \theta_i + S_i -1 -q \right) 
\right]
\nonumber\\
&&=
\frac{1}{Z^{\textrm{tot}}_{J}(\openone)} 
\sum^{\ }_{\{ \theta^{\ }_{i} \}} 
\sum^{\ }_{\{ S^{\ }_{i} \}} 
  \exp\left(
    J \sum^{\ }_{\langle ij\rangle\in\A} \theta^{\ }_i \theta^{\ }_j 
  \right) 
  \exp\left(
    J \sum^{\ }_{\langle ij\rangle\not\in\A} S^{\ }_i S^{\ }_j 
  \right) 
\;
\frac{1}{2} \sum_{q = \pm 1} 
\delta(\theta_i S_i=q)
\nonumber\\
&&=
\frac{1}{Z^{\textrm{tot}}_{J}(\openone)} 
\times Z^{\textrm{tot}}_{J}(\openone)
=1,
\end{eqnarray}
\end{widetext}
and thus the ${\cal P}^{\,p}_{\{ k_i \}}\ge 0$ are probability
weights. 

Similarly, we can define a probability 
\begin{eqnarray}
{\cal P}_{\{ k_i \}}
&&=
\left(
{\cal P}^{\,1}
\;{\cal P}^{\,4}
\;{\cal P}^{\,5}
\;{\cal P}^{\,8}
\right)_{\{ k_i \}}
\\
&&
=
\left(
{\cal P}^{\,2}
\;{\cal P}^{\,3}
\;{\cal P}^{\,6}
\;{\cal P}^{\,7}
\right)_{\{ k_i \}}
\nonumber\\
&&\ge 0,
\end{eqnarray}
where the ${\{ k_i \}}$ are defined on the total boundary of the added
partitions, and we used the fact that partitions $1,4,5,8$ and $2,3,6,7$ have
exactly the same total boundary. We can then define averages with respect to 
this measure, 
\begin{equation}
\langle \cdots \rangle_{\{ k_i \}}\equiv
\sum^{\textrm{(even)}}_{\{ k_i \}^{\N^{\ }_{\partial}}_{i=1} } 
{\cal P}_{\{ k_i \}} \left(\cdots\right)
\, ,
\end{equation}
and Eq.~(\ref{eq: S_topo plaquette}) reduces to 
\begin{widetext}
\begin{subequations}
\bea
S^{(P)}_{\textrm{topo}}(T/\lambda^{\ }_{B}) 
&=& 
- \ln 2 
\nonumber \\ 
\nonumber \\ 
&+&
\left\langle
\ln 
\left(
\frac{
Z^{\overline{1\A},+}_{\{ k_i \}} 
\; Z^{\overline{4\A},+}_{\{ k_i \}} 
\; Z^{\overline{5\A},+}_{\{ k_i \}} 
\; Z^{\overline{8\A},+}_{\{ k_i \}} 
}{
 Z^{\overline{2\A},+}_{\{ k_i \}} 
\; Z^{\overline{3\A},+}_{\{ k_i \}} 
\; Z^{\overline{6\A},+}_{\{ k_i \}} 
\; Z^{\overline{7\A},+}_{\{ k_i \}} 
}
\right)
\right\rangle_{\{ k_i \}}
\label{eq: S_topo plaquette-aveA} \\ 
&+& 
\left\langle
\left( 
  1 
  + 
  \frac{Z^{\overline{4\A},-}_{\{ k_i \}}}{Z^{\overline{4\A},+}_{\{ k_i \}}} 
\right)
\ln 
\left(
  1 
  + 
  \frac{Z^{\overline{4\A},-}_{\{ k_i \}}}{Z^{\overline{4\A},+}_{\{ k_i \}}}
\right) 
+ 
\left( 
  1 
  - 
  \frac{Z^{\overline{4\A},-}_{\{ k_i \}}}{Z^{\overline{4\A},+}_{\{ k_i \}}} 
\right)
\ln 
\left(
  1 
  - 
  \frac{Z^{\overline{4\A},-}_{\{ k_i \}}}{Z^{\overline{4\A},+}_{\{ k_i \}}}
\right)
\right\rangle_{\{ k_i \}}
. 
\label{eq: S_topo plaquette-aveB}
\eea
\end{subequations}
\end{widetext}

We can finally analyze this expression as a function of temperature. 
Recall that 
$J=-(1/2) \ln[\tanh(\beta\lambda^{\ }_{B})]$, so that $J\to 0$ when $T\to 0$, 
and the disordered Ising phase occurs for 
$T < T_c \simeq 1.313346(3) \lambda_B$. 
Below the Ising transition at $J = J_c \simeq 0.2216544(3)$, 
one can use a high-temperature loop expansion to estimate the ratio of 
$Z^{\overline{4\A},-}_{\{ k_i \}}$ over $Z^{\overline{4\A},+}_{\{ k_i \}}$. 

\begin{figure}[ht]
\vspace{0.4 cm}
\begin{center}
\includegraphics[width=0.95\columnwidth]{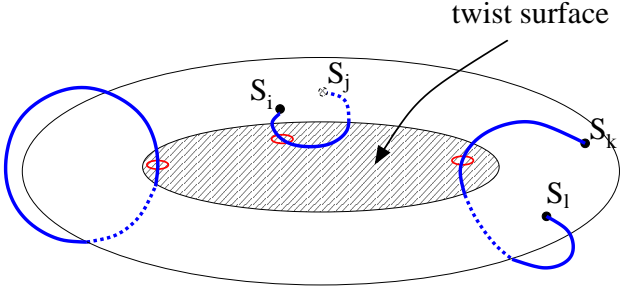}
\end{center}
\caption{
\label{fig: loop expansion}
(Color online) -- 
Qualitative examples of terms in the loop expansion that appear with different 
signs in $Z^{\overline{4\A},-}_{\{ k_i \}}$ and 
$Z^{\overline{4\A},+}_{\{ k_i \}}$: 
closed loops that wind around the donut shape, 
and open strings that connect boundary spins $S_i$ (which appear in the 
high-temperature expansion whenever the corresponding $k_i$ is odd). 
}
\end{figure}

The high-temperature expansion contains either closed loops, or open
strings that terminate at the boundary, because an $S_i$ is inserted
for each site $i$ where $k_i$ is odd.
The corresponding expansions for $Z^{\overline{4\A},-}_{\{ k_i \}}$
over $Z^{\overline{4\A},+}_{\{ k_i \}}$
differ only by loop terms that intersect 
the twist surface (generated by the collective operation in 
Fig.~\ref{fig: collective ops 2} bottom) an odd number of times. These terms 
appear indeed with opposite sign in the two expansions. 
This can be achieved only by closed loops that wind around the donut shape, 
and by open strings that connect boundary spins $S_i$ among those identified 
by the set of odd $k_i$'s (see Fig.~\ref{fig: loop expansion}). 

\begin{figure}[ht]
\vspace{0.4 cm}
\begin{center}
\includegraphics[width=0.95\columnwidth]{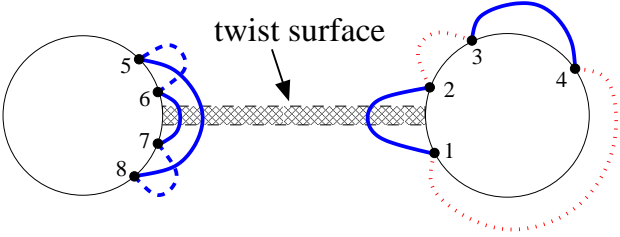}
\end{center}
\caption{
\label{fig: loop expansion detail}
(Color online) -- 
Schematic, projected illustration of open strings between boundary 
spins. The location of the spins $1,2,\ldots,8$ are given by the sites where 
$k_i$ is odd (recall that their total number must be even). 
One can verify that the parity of the number of intersections with the twist 
surface is fixed by the choice of the locations $1,2,\ldots,8$, up to 
exponentially small corrections such as the red dotted string in the figure, 
which vanish in the thermodynamic limit of $N_\partial \to \infty$. 
For example, consider the change upon reconnecting spin $5,\ldots,8$ via the 
dashed lines instead of the solid lines. 
(Notice that the case where, say, the points $1,\ldots,4$ are uniformly 
distributed on the boundary is exponentially suppressed by the probability
${\cal P}_{\{ k_i \}}$.) 
}
\end{figure}

In the high temperature limit, long loops are exponentially suppressed
and we can safely neglect the winding loop contributions when the
size of the partition is taken to infinity.
Similarly, out of all 
possible ways of connecting boundary spins in the $k_i$ odd set, only 
`short' strings between spins `close' to the twist surface need be considered, 
as illustrated in Fig.~\ref{fig: loop expansion detail}. 

For $k_i$ points near the twist surface, rearranging the way that
points are paired does not change the parity of the number of
crossings of the twist surface. This is illustrated in 
Fig.~\ref{fig: loop expansion detail}, where reconnecting spins $5,\ldots,8$ via the
dashed lines instead of the solid lines give 0 instead of 2 crossings,
thus not changing the parity. Now, a reconnection that changes the parity
involves drawing long strings. Below the Ising transition, the probability
${\cal P}_{\{ k_i \}}$ keeps the points with odd 
$k_i$ confined in pairs; thus there are ways to connect them together with short 
strings. But changing the parity of the intersections requires re-matching them 
in such a way that connections with sites far away are made, and the total length of 
these strings is of order the system size. This is illustrated in 
Fig.~\ref{fig: loop expansion detail}: for example, reconnecting spins
$1,\ldots,4$ requires 
strings whose total length spans the system size.

Therefore, one can verify that all the loop terms corresponding to a given 
choice of $k_i$'s have the same parity in the number of intersections to the 
twist surface, up to corrections that are exponentially small in the size of 
the bipartition. 
As a result, the ratio 
$Z^{\overline{4\A},-}_{\{ k_i \}} / Z^{\overline{4\A},+}_{\{ k_i \}}$ 
tends to $\pm 1$ in the thermodynamic limit of $N_\partial \to \infty$, and 
the sign is purely determined by the choice of $k_i$. 

Eq.~(\ref{eq: S_topo plaquette-aveB}) is clearly symmetric under the change 
$
Z^{\overline{4\A},-}_{\{ k_i \}} / Z^{\overline{4\A},+}_{\{ k_i \}}
\to 
- 
Z^{\overline{4\A},-}_{\{ k_i \}} / Z^{\overline{4\A},+}_{\{ k_i \}}
$, 
and we finally arrive at the result that at low temperature $T < T_c$, the 
term in Eq.~(\ref{eq: S_topo plaquette-aveB}) gives $2\ln 2$.

In the Ising ordered phase ($T>T_c$ here), on the other hand, the
ratio $Z^{4\overline\A,-}_{\ } / Z^{4\overline\A,+}_{\ } \to 0$
in the thermodynamic limit, because of the energy cost associated with 
the twist in boundary condition (domain wall) in the `$-$' 
partition. Hence, in this case the term in 
Eq.~(\ref{eq: S_topo plaquette-aveB}) gives $0$. 

A similar reasoning gives that the ratios entering 
Eq.~(\ref{eq: S_topo plaquette-aveA}) are equal to 1 
in the thermodynamic limit, and
corrections appear only as the correlation length becomes of the order of the 
size of the bipartitions, i.e., infinite in the 
thermodynamic limit. Thus, in the low
temperature phase, Eq.~(\ref{eq: S_topo plaquette-aveA}) 
gives $\ln 1=0$ for $T<T_c$. 

On the other hand, for $T>T_c$, the partitions order
ferromagnetically, and one must account for the fact that partition
$1\overline\A$ has two disconnected components, and therefore these
two components can order in two ways relative to one another, giving a
factor of 2 in the ratio appearing in 
Eq.~(\ref{eq: S_topo plaquette-aveA}), and hence this terms gives a 
contribution $\ln 2$.

Putting it all together, we obtain that 
\begin{equation}
S^{(P)}_{\textrm{topo}}(T/\lambda^{\ }_{B}) =
\begin{cases}
\ln 2 & \text{$T < T_c$} \\
0 & \text{$T > T_c$},
\end{cases}
\end{equation}
and 
$
\Delta S^{(P)}_{\textrm{topo}}(T/\lambda^{\ }_{B})
=
S^{(P)}_{\textrm{topo}}(T/\lambda^{\ }_{B}) - S^{(P)}_{\textrm{topo}}(0)
$ 
is given by Eq.~(\ref{eq:S_P-result}). 
%
%


\begin{thebibliography}{99}
\bibitem{footnote-basis} 
The entanglement entropy is invariant under a local spin rotation.
\bibitem{footnote:separability}
Note that this decomposition is visible only at finite temperature,
where the full Hamiltonian enters the calculations for the von Neumann
entropy of the system via the density matrix.  At zero temperature,
the two contributions do remain distinct, but they cannot be told
apart as there is no explicit dependence on $\lambda_\A$ and
$\lambda_\B$ in the GS wavefunction.
\bibitem{footnote: gauge order parameter}
Note that the classical model at finite $T$ that obtains by setting 
$\lambda^{\ }_{A} = 0$ is nothing but a classical $\mathbb{Z}_2$ 
gauge theory in three 
dimensions. 
Therefore, our results show that the topological entropy of this classical 
system behaves as a proper (non-local) order parameter that captures its 
finite temperature phase transition.
\end{thebibliography}
\end{document}